\documentclass[submission, Phys]{SciPost}
\hypersetup{
    colorlinks,
    linkcolor={red!50!black},
    citecolor={blue!50!black},
    urlcolor={blue!80!black}
}

\usepackage[bitstream-charter]{mathdesign}
\DeclareSymbolFont{usualmathcal}{OMS}{cmsy}{m}{n}
\DeclareSymbolFontAlphabet{\mathcal}{usualmathcal}

\definecolor{ao(english)}{rgb}{0.0, 0.5, 0.0}

\usepackage{microtype}
\usepackage[utf8]{inputenx}
\usepackage{physics}                            
\usepackage{bm}									
\usepackage{enumerate}							
\usepackage[caption=false]{subfig}
\usepackage{mathtools}							
\usepackage{tensor}								
\usepackage{wrapfig}							
\usepackage{listings}							
\usepackage{xcolor}								
\usepackage{graphicx,color}
\usepackage[capitalize]{cleveref}               
\usepackage{xstring}                            
\usepackage{hyperref}

\newcommand{\cor}[1]{\mathcal{#1}}									
\newcommand{\T}[1]{\text{#1}}										
\newcommand{\eg}{e.g.}
\newcommand{\ie}{i.e.}
\newcommand{\n}{\nonumber}
\newcommand{\ccite}[1]{\IfSubStr{#1}{,}{Refs.~}{Ref.~}\cite{#1}}

\newcommand{\comments}[1]{}   

\newcommand{\rev}[1]{#1}

\usepackage{tikz,tkz-tab}
\usetikzlibrary{matrix,decorations.pathreplacing,calc, positioning,automata,arrows, arrows.meta, petri, topaths, shapes, decorations.markings}
\usepackage{pgfplots}
\newcommand{\littlescheme}{
\begin{tikzpicture}[scale=1, every node/.style={scale=1, inner sep=0pt}]

\node   (0) {~};
\node[right= 200pt of 0]   (9) {~};
\node[right= 3pt of 9]{$\gamma$};

\fill[scipostphys!30!white] (2.7,5pt) rectangle (4.3,-5pt);
\draw[-latex] (0) --  (9) ;
    \draw[densely dotted] ($ (0)+(2.7,25pt)$) -- ($ (0)+(2.7,0pt)$);
    \draw[densely dotted] ($ (0)+(4.3,-13pt)$) -- ($ (0)+(4.3,-32pt)$);

\node[below right = 3pt and 2.6 of 0]{1};
\node[below right = 3pt and 4.2 of 0]{2};

\node[above right = 13pt and -3.5 of 0]{\textcolor{red!50!black}{\textbf{Spectral density}}};
\node[below right = 16pt and -3.5 of 0]{\textcolor{red!50!black}{\textbf{Level statistics}} };

\node[above right = 14pt and 0.5 of 0]{Semi-circle};
\node[above right = 12pt and 3.7 of 0, align=left]{$p_a(\lambda)$};

\node[below right = 17pt and 1.6 of 0]{Wigner-Dyson};
\node[below right = 17pt and 5.1 of 0, align=left]{Poisson};

\end{tikzpicture}
}

\begin{document}

\begin{center}{\Large \textbf{
Replica approach to the generalized Rosenzweig-Porter model\\
}}\end{center}

\begin{center}
Davide Venturelli\textsuperscript{1$\star$},
Leticia F. Cugliandolo\textsuperscript{2,3},
Grégory Schehr\textsuperscript{2} and
Marco Tarzia\textsuperscript{3,4}
\end{center}

\begin{center}
{\bf 1} SISSA -- International School for Advanced Studies and INFN, via Bonomea 265, 34136 Trieste, Italy\\
{\bf 2} Sorbonne Université, Laboratoire de Physique Théorique et Hautes Energies, CNRS UMR 7589, 4 Place Jussieu, Tour 13, 5ème étage, 75252 Paris 05, France\\
{\bf 3} Institut Universitaire de France, 1 rue Descartes, 75231 Paris Cedex 05, France\\
{\bf 4} Sorbonne Université, Laboratoire de Physique Théorique de la Matière Condensée, CNRS UMR 7600, 4 Place Jussieu, Tour 13, 5ème étage, 75252 Paris Cedex 05, France\\
${}^\star$ {\small \sf davide.venturelli@sissa.it}
\end{center}

\begin{center}
\today
\end{center}


\section*{Abstract}
{\bf
The generalized Rosenzweig-Porter model \rev{with real (GOE) off-diagonal entries} arguably constitutes the simplest random matrix ensemble displaying a phase with fractal eigenstates, which we characterize here by using replica methods. We first derive analytical expressions for the average spectral density in the limit in which the size $N$ of the matrix is large but finite. We then focus on the number of eigenvalues in a finite interval and compute its cumulant generating function as well as the level compressibility, i.e., the ratio of the first two cumulants: these are useful tools to describe the local level statistics. In particular, the level compressibility is shown to be described by a universal scaling function, which we compute explicitly, when the system is probed over scales of the order of the Thouless energy. \rev{Interestingly, the same scaling function is found to describe the level compressibility of the complex (GUE) Rosenzweig-Porter model in this regime.}
We confirm our results with numerical tests.
}

\vspace{10pt}
\noindent\rule{\textwidth}{1pt}
\tableofcontents\thispagestyle{fancy}
\noindent\rule{\textwidth}{1pt}
\vspace{10pt}

\section{Introduction}

Quantum non-interacting particles in a disordered potential undergo the
Anderson localization transition as the disorder strength is increased~\cite{anderson1958absence}. In one and two dimensions an infinitesimal amount of disorder is sufficient to localize all eigenstates of the Hamiltonian, while in dimension larger than two a critical value of the disorder strength separates a metallic phase, where the eigenstates are similar to plane
waves and spread over the whole volume uniformly, from an insulating phase, where
the eigenstates are instead exponentially localized around specific points
in space, and thereby occupy a finite $\cor{O}(1)$ portion of the total volume. It
is well established that exactly at the Anderson localization critical
point the wave-functions are \textit{multifractal}~\cite{wegner1980inverse,rodriguez2011multifractal}. 
This means that they are
neither fully delocalized (as in the metallic regime), nor fully localized
(as in the insulating phase), since their support set grows with the system
size but remains a vanishing fraction of the total volume. \rev{The \textit{multi}-fractal character of the wave functions is due to the fact that the $q$th moments of their amplitudes  decay with the size $N$ as $\langle \sum_{i=1}^N | \psi_i |^{2q} \rangle \propto N^{-(q-1)D_q}$, with different 
$q$-dependent exponents $0 \le D_q \le 1$.} 

In the last decade, the Hilbert space localization properties of quantum
disordered many-body systems have attracted much interest. In this context,
the emergence of multifractal states has been discussed as a
key and robust feature of their
phase diagram, and has been invoked to explain some of their unconventional
properties beyond the single-particle limit. In the many-body setting,
multifractal eigenstates that do not cover the whole accessible Hilbert
space 
may lead to the violation of the eigenstate thermalization
hypothesis (ETH)~\cite{srednicki1994chaos,rigol2008thermalization}. \rev{Therefore, they are often called partially delocalized but
\textit{non-ergodic}, in contrast to the \textit{ergodic} fully delocalized eigenstates, which are supposed to satisfy ETH.}

In the context of Many Body Localization (MBL), recent studies indicate that the many-body
eigenstates are in fact multifractal in the whole insulating phase~\cite{mace2019multifractal,de2021rare,gornyi2017spectral,tarzia2020many,Luitz_2015,Serbyn_2017,Tikhonov_2018,Luitz_2020}. Furthermore, the seminal work by Basko, Aleiner and Altshuler~\cite{basko2006metal} predicted the existence of a novel unconventional ``bad
metal'’  regime in between the fully ergodic metallic phase at low disorder
and the insulating one at strong disorder. Following the pioneering ideas
of~\cite{altshuler1997quasiparticle}, the unusual properties of the bad metal regime have \rev{also} been
put in relation with the possible multifractal nature of the many-body
eigenstates. 
Recent investigations of the out-of-equilibrium phase diagram of
the quantum random  energy model~\cite{Faoro_2019,baldwin2018quantum,biroli2021out,parolini2020multifractal,kechedzhi2018efficient} and Josephson junction arrays~\cite{pino2017multifractal,pino2016nonergodic} seem to support
\rev{this scenario.}
The existence of partially extended but non-ergodic
wave-functions is also believed to have relevant practical and conceptual
implications in the efficient population transfer in the context of quantum
computing~\cite{parolini2020multifractal,kechedzhi2018efficient,Smelyanskiy_2020}. Moreover, recent studies of the Sachdev-Ye-Kitaev model in
high-energy physics and quantum  gravity have reported evidence for the
emergence of a non-ergodic, 
\rev{but partially}
extended phase when the model is perturbed by a
single-body term~\cite{Atland_2019,monteiro2021minimal}.

\begin{figure}
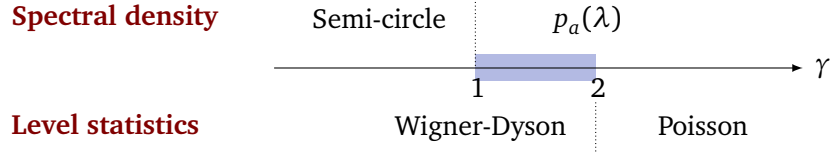

    \centering
    \littlescheme
    \caption{Sketch of the different phases of the RP model, depending on the value of the parameter $\gamma$ in \cref{eq:hamiltonian}. Considering the average spectral density $\rho(\lambda)$, a transition is first observed at $\gamma=1$ which separates a fully delocalized phase where $\rho(\lambda)=\rho_\T{GOE}(\lambda)$ from a partially delocalized phase where $\rho(\lambda)=p_a(\lambda)$ in the thermodynamic limit (see the main text). Focusing instead on the local level statistics, another transition is found at $\gamma=2$ from the partially delocalized phase characterized by the Wigner-Dyson statistics, to an Anderson-localized phase characterized by the Poisson statistics. The shaded region $1<\gamma<2$ indicates the intermediate phase studied in this work.}
    \label{fig:scheme}
\end{figure}

Matrix models have been an invaluable tool to describe and help
understanding complex physical systems, in particular those with quenched
randomness. The physical mechanism at the origin of the above-mentioned
multiftactal eigenstates is one such problem, and specific
matrix models have recently been used as proxies which capture the peculiar
spectral properties associated with them. In this respect, the
Rosenzweig-Porter (RP) random matrix ensemble~\cite{RP_1960}, originally introduced to
reproduce the spectral properties of complex atomic spectra, provides an
archetypal illustration of a system in which a partially extended phase featuring fractal eigenstates (along with other unconventional spectral
properties that will be extensively discussed below) appears in an
intermediate region of the phase diagram between a fully
delocalized phase and a fully Anderson localized phase \rev{(see Fig.~\ref{fig:scheme})}. For this reason the
RP model has been the focus of a strong resurgence of attention over the
last few years~\cite{Kravtsov_2015,vonSoosten_2019,Facoetti_2016,Truong_2016,Bogomolny_2018,DeTomasi_2019,amini2017spread,pino2019ergodic,berkovits2020super}. Although one cannot expect that simple random matrix models could capture all the properties of interacting quantum systems, they
provide natural and powerful tools to understand the deep physical
mechanisms behind some of their features, which are often  elusive to
analytical treatments in more realistic settings.

The Hamiltonian of the RP model $\cor{H}=A+c(N)B$ can be written as the sum of an $N \times N$ diagonal matrix $A$, whose entries $a_i$'s are independent and identically distributed (i.i.d.) random variables drawn from a Gaussian distribution $p_a(a_i)$, and another $N \times N$ random matrix $B$ belonging to the Gaussian orthogonal (or unitary) ensemble (GOE or GUE, respectively). If the variances of the matrix elements $b_{ij}$ are chosen of $\order{1}$, then the width of the spectrum of $B$ is of $\order{\sqrt{N}}$: thus the matrix $A$ (whose spectral width is of $\order{1}$) can produce significant deviations from the GOE/GUE behavior only if $c(N)$ decays sufficiently fast for large $N$.
The properties of this model have been extensively studied by using different techniques, such as a mapping to the Dyson Brownian motion \cite{Pandey_1995}, supersymmetry \cite{Guhr_1996,Guhr_1997}, resolvent methods \cite{Brezin_1996,Kunz_1998}, and first order perturbation theory \cite{Atland_1997}.

Due to the strong surge of attention towards multifractal states in quantum many-body disordered systems, a generalized version of the RP model in which the distribution $p_a(a_i)$ is not necessarily Gaussian has then been introduced in Ref.~\cite{Kravtsov_2015}, and thoroughly investigated by using the techniques recalled above \cite{vonSoosten_2019,Facoetti_2016,Truong_2016,Bogomolny_2018,DeTomasi_2019,amini2017spread,pino2019ergodic,berkovits2020super} -- we will refer to this as the GRP model. In addition, new connections and applications have been pointed out in disordered elastic systems \cite{Krajenbrink_2021}, many-body localization \cite{Faoro_2019,baldwin2018quantum,biroli2021out,tarzia2020many}, quantum gravity \cite{Atland_2019,monteiro2021minimal}, quantum information \cite{parolini2020multifractal,Smelyanskiy_2020,kechedzhi2018efficient}, models of theoretical ecology \cite{Mergny_2021}, and noise reduction in big data matrices \cite{Bouchaud_Les_Houches}.

In this work we revisit the generalized RP model by analyzing some properties of the energy levels and their correlations which have not been investigated in the literature yet. In particular, we perform a thorough study of the finite-$N$ corrections to the average spectral density and compute the level compressibility in the intermediate phase, thereby providing a deeper understanding of the properties of the intermediate regime. Our analysis uses the replica method largely exploited in the analysis of spin glass models~\cite{Mezard_1987}, but not only -- for example, this tool has been recently applied to study the properties of the ground states in a deformed GOE ensemble~\cite{Harukuni_2022}. 
\rev{Our replica study is developed for} the case in which the entries of $\cor{H}$ are real numbers, so that $\cor{H}$ belongs to \rev{a} deformed GOE ensemble \rev{(where the deformation is introduced by the addition of the diagonal random matrix $A$).} \rev{However, and quite surprisingly, we 
show that the exact same behavior of the level compressibility applies to the crossover regime of the Hermitian GRP model 
\cite{Kravtsov_2015} 
(in which the off-diagonal entries are complex).}

In the rest of this introductory section we will present the RP model and some of its salient properties (\cref{subsec:RPmodel}), 
and we will outline our study and main results
(\cref{subsec:outline}).

\subsection{The generalized Rosenzweig-Porter model}
\label{subsec:RPmodel}
We consider the Hamiltonian represented by the $N \times N$ matrix
\begin{equation}
    \cor{H} = A + \frac{\nu}{N^{\gamma/2}} B \, ,
    \label{eq:hamiltonian}
\end{equation}
where the matrix $B$ belongs to the GOE ensemble: its elements are Gaussian random variables with zero mean and unit variance (i.e., 
\rev{$\langle b_{ii}^2 \rangle = 1$ and $\langle b_{ij}^2 \rangle = 1/2$ for $i \neq j$}). 
With this choice, the spectrum of $B$ in the limit $N\to \infty$ converges to a Wigner semicircle supported within $\lambda \in [-\sqrt{2N},\sqrt{2N} ]$, where we denote hereafter with $\lambda$ the eigenvalues of $\cor{H}$. The parameter $\nu$ is of $\order{1}$ and does not scale with $N$. The deformation matrix $A$ is instead diagonal, with independent entries $a_i$'s identically distributed according to a generic distribution $p_a (a)$, hence the name Generalized Rosenzweig-Porter model (GRP). 

Following the analogy with disordered quantum many-body systems, each matrix index can be thought of as a site of the reference Hilbert space, which is connected to every other site with the transition rates distributed according to the Gaussian law. Different phenomenologies are expected depending on the value of the parameter $\gamma$, which renders one of the two matrices $A$ or $B$ subleading with respect to the other in the limit of large $N$. As summarized in~\cref{fig:scheme}, the model features three distinct phases (and two transition points between them): a fully delocalized phase for $\gamma<1$, a fully Anderson localized phase for $\gamma>2$, and an intermediate 
fractal phase for $1<\gamma<2$. 

The 
transition from fully extended to fractal eigenstates at $\gamma=1$ manifests itself as a transition for 
the average spectral density $\rho(\lambda)$, which reproduces the Wigner semicircle law for $\gamma <1$ in the $N\to \infty$ limit, while it reduces to $\rho(\lambda) = p_a(\lambda)$ if $\gamma>1$
and the same $N\to\infty$ limit is taken. At $\gamma=1$, 
$\rho(\lambda)$ interpolates between $p_a(\lambda)$ and the Wigner semicircle as the value of the parameter $\nu$ is increased. This transition becomes sharp (\ie, it occurs at a particular value of $\nu=\nu_c$) provided that $p_a(a)$ has a compact support and vanishes sufficiently fast at its upper edge \cite{Claeys_2018,Krajenbrink_2021}.

The value $\gamma=2$ corresponds instead to a genuine Anderson localization transition.
Indeed, the region $\gamma <1$ is characterized by the Wigner-Dyson statistics, meaning that the eigenvectors of $\cor{H}$ are uniformly delocalized on $N$ sites in the large $N$ limit, and the average level spacing follows the Wigner surmise \cite{Livan_2018}, signaling level repulsion. Conversely, in the region $\gamma > 2$ the eigenvectors are completely localized over $\cor{O}(1)$ sites and the mean level spacing exhibits Poisson statistics. 

The intermediate region with $1 < \gamma < 2$ is particularly interesting, because the average spectral density tends to $p_a(\lambda)$, but the \textit{local} level statistics remains of the Wigner-Dyson type. Here the eigenvectors are known to be delocalized over a large number of sites $N^{D_\gamma}$, which represent, however, a vanishing fraction of the total number of sites $N$ in the thermodynamic limit, their fractal dimension being $D_\gamma = 2-\gamma <1$ \cite{Kravtsov_2015}. 

The simplest and most intuitive way to understand the spectral properties in the intermediate region is provided by the Fermi golden rule.
In the limit in which the off-diagonal matrix $B$ is absent and all the eigenvectors are trivially localized on a single site, one has $\vert \psi_i \rangle =\vert i \rangle$, with corresponding eigenenergies $\lambda_i = a_i$. When the GOE perturbation $\nu N^{-\gamma/2} B$ is turned on, the transition probability per unit time from a state $i$ to another state $j$ can be evaluated perturbatively as
\[
\Gamma_{i \to j} = \frac{2 \pi \rho}{\hbar} \frac{4 \eta}{N} |b_{ij}|^2 \, ,
\]
where $\rho=\rho(\lambda)$ and, for future convenience, we have introduced the combination
\begin{equation}
    \eta \equiv  N^{1-\gamma} \nu^2/4 \, .
    \label{eq:eta}
\end{equation}
Hence, the average escape rate per unit time for a ``particle'' created in site $i$ at time $t=0$ reads
\begin{equation} \label{eq:FGR}
\Gamma = \sum_{j \neq i} \langle \Gamma_{i \to j} \rangle = \frac{2 \pi \rho}{\hbar} 4 \eta \, .
\end{equation}
The quantity $\hbar \Gamma$ can thus be interpreted as the bandwidth $\Delta E$ that can be reached in a time of $\cor{O}(1)$ from a given site $i$:
\rev{
\begin{equation}
\Delta E \sim \hbar \Gamma
\; . \n
\end{equation}
}
This implies that the eigenvectors within this energy window are hybridized by the GOE perturbation. For $1<\gamma<2$ such energy band \rev{decays} with the system size as $\Delta E \propto N^{1 - \gamma}$ 
\rev{but} is much larger than the mean level spacing 
\begin{equation}
    \delta_N \simeq [N \rho]^{-1} \, ,
    \label{eq:mean_level_spacing}
\end{equation}
entailing that the system is not Anderson localized; still, $\Delta E$ remains much smaller than the total bandwidth, which is of $\cor{O}(1)$. This signifies that the particle can only explore a sub-extensive portion of the total volume. 
The Anderson localization transition thus occurs when $\Delta E$ becomes smaller than the mean level spacing, i.e., for $\gamma \ge 2$: this implies that the average escape time from site $i$ \rev{(\ie, $\Delta t \equiv \hbar / \Delta E$)} grows \rev{at least linearly} with $N$, and thus the eigenfunctions remain localized on $\cor{O}(1)$ sites. Conversely, the transition to the fully delocalized phase takes place when $\Delta E$ becomes of the order of the total bandwidth, i.e., for $\gamma \le 1$: this implies that, starting from site $i$, the wave-packet can reach any other site in a time of $\cor{O}(1)$.

In the intermediate phase, $1 < \gamma < 2$, the support set of the eigenvectors (i.e., the number of sites which are hybridized by the perturbation) is simply given by the spreading of the energy interval divided by the average gap between adjacent energy levels, and thus scales as $\Delta E / \delta_N \sim N^{D_\gamma}$, with $D_\gamma = 2 - \gamma$.
The partially extended but fractal eigenstates are therefore linear combinations of a bunch of $N^{D_\gamma}$ localized states associated to nearby energy levels, \ie,
\[
\vert \psi_i \rangle \approx 
\!\! \sum_{\substack{i^\prime \; s.t.\\
|a_i - a_i^\prime| \le \Delta E}} \!\!
c_{i^\prime} \, \vert i^\prime \rangle \, ,
\]
with coefficients $c_{i^\prime}$ of order $N^{-D_\gamma/2}$ to ensure normalization. These eigenstates give 
rise to the so-called \textit{mini-bands} in the local spectrum~\cite{Kravtsov_2015}. The width of the mini-bands sets the energy scale 
\rev{\begin{equation}
    E_T \sim \Delta E \sim N^{D_\gamma - 1} = N^{1 - \gamma}
\, , 
\label{eq:thouless_first_occurrence}
\end{equation}
}
 often called the Thouless energy~\cite{altshuler1986repulsion,cuevas2007two}, within which GOE-like spectral correlations (and in particular level repulsion) have been established. All the moments of the wave-functions' coefficients (the so-called generalized \textit{inverse participation ratios}, IPR) behave as 
 \rev{\begin{equation}
 I_q = \sum_i \vert \langle i \vert \psi \rangle \vert^{2 q} \propto N^{D_\gamma (1-q)}
 \qquad\quad\implies 
 \qquad\quad
 D_q = D_\gamma
\; .
 \end{equation}
 }
 This implies that all the fractal dimensions $D_q$ are degenerate and equal to $D_\gamma$ for all \rev{positive integer} $q$, i.e., that the intermediate phase of the GRP model is fractal but not multifractal.

As discussed above, the emergence of such fractal phase is particularly relevant in many physical contexts. Its existence was first suggested in \ccite{Kravtsov_2015} and then rigorously proven in \ccite{vonSoosten_2019}. In recent years several generalizations of the RP model have been put forward and analysed~\cite{kravtsov2020localization,khaymovich2020fragile,monthus2017multifractality,biroli2021levy,buijsman2022circular,khaymovich2021dynamical}, and many other random matrix ensembles have been shown to have an intermediate partially delocalized phase with similar spectral properties~\cite{sarkar2021mobility,roy2018multifractality,wang2016phase,nosov2019correlation,duthie2022anomalous,kutlin2021emergent,motamarri2021localization,tang2022non,tarzia2022fully,Skvortsov_2022,Cai_2013,DeGottardi_2013,Liu_2015,Das_2022,Ahmed_2022,Lee_2022}. Yet, the GRP setting is still a very useful playground to analyze the properties of fractal states in 
a controlled framework.

\subsection{Outline of this work and summary of the main results}
\label{subsec:outline}
As explained above, the GRP model has been intensively investigated over the past few years with a great variety of analytical and numerical techniques.
In this paper we tackle this model by applying yet another approach, namely the replica formalism~\cite{Mezard_1987}, which allows us to obtain new results on the average spectral density, and  the statistics of the number of energy levels within a finite interval.

In \cref{par:spectral_replica_approach} we start by analyzing the {average density of states} 
$\rho(\lambda)$. 
When the size $N$ of the matrix is large, we find the leading order estimate
\begin{equation}
    \rho(\lambda) =  -\frac{1}{\pi \eta} \lim_{\varepsilon \to 0^+} \Re C(\lambda_\varepsilon) +\order{1/N} \, ,
\label{eq:rho_N_intro}
\end{equation}
where $\lambda_\varepsilon=\lambda-i\varepsilon$, the parameter
$\eta$ depends on $N$ and was introduced in \cref{eq:eta}, and the function $C(\lambda)$ is implicitly defined by the self-consistency equation
\begin{equation}
    C(\lambda) = i \eta \, \cor{G}_a\left[\lambda+2iC(\lambda)\right] \, .
    \label{eq:C(l)_intro}
\end{equation}
Here $\cor{G}_a$ is the resolvent associated to the distribution $p_a(a)$ of the entries of $A$ (c.f. \cref{eq:resolvent}). In general, it is quite difficult to solve explicitly \cref{eq:C(l)_intro} for $C(\lambda)$ for any arbitrary distribution $p_a(a)$. However, we show in \cref{par:exactly_solvable} that it can be solved explicitly for two special cases: (i) when $p_a(a)$ is a Wigner semicircle -- which is expected, since it is stable under free-convolution \cite{voiculescu1992free} --  and (ii) when it is a Cauchy distribution, which is more surprising.

As we show in \cref{par:resolvent}, by taking the limit $N\to \infty$ with $\eta$ kept finite, \cref{eq:rho_N_intro,eq:C(l)_intro} reduce to the free addition formula \cite{voiculescu1992free}, sometimes called the ``Zee formula'' in the physics literature~\cite{Zee_1996}. However, as we show in \cref{par:approximate_solutions}, these equations contain more information and, in the case where $\eta=\eta(N) \ll 1$ defined in \cref{eq:eta} with $1<\gamma<2$,  they allow us  to obtain the leading $1/N$ corrections to the limiting density $\rho(\lambda) = p_a(\lambda)$ in a controlled way, even when the resolvent $\cor{G}_a(z)$ does not admit a closed-form analytic expression. These corrections turn out to be quite difficult to compute using the standard free addition formula, which in principle holds only in the limit $N \to \infty$.

Next, in \cref{par:level_compress} we analyze the behavior of the \textit{level compressibility} $\chi(E)$, which is a simple indicator of the degree of level repulsion and is defined as follows\footnote{It is known in statistics under the name of ``Fano factor''~\cite{fano1947ionization}, and it has been studied recently in physics in the context of extremes and record statistics of time series~\cite{majumdar2019exactly,majumdar2021universal}.}\cite{Mirlin_2000}. We first introduce the empirical density of the (real) eigenvalues $\lambda_i$ of $\cal{H}$, defined as
\begin{equation} \label{eq:def_density}
\rho_N(\lambda) = \frac{1}{N} \sum_{i=1}^N \delta(\lambda - \lambda_i) \, .  
\end{equation}
Note that $\rho_N(\lambda)$ is normalised to unity. Let
\begin{equation}
    I_N[\alpha,\beta] \equiv N \int_\alpha^\beta \dd{\lambda} \rho_N(\lambda)
    \label{eq:levels_number}
\end{equation}
denote the number of eigenvalues lying in the interval 
$ [\alpha,\beta]\subseteq \mathbb{R}$, which is a random variable. Then 
\begin{equation}
    \chi(E) \equiv \frac{\kappa_2(E)}{\kappa_1(E)} = \frac{\expval{I_N^2[-E,E]}-\expval{I_N[-E,E]}^2}{\expval{I_N[-E,E]}} \rev{= \frac{\expval{I_N^2[-E,E]}_c}{\expval{I_N[-E,E]}}} \, ,
    \label{eq:level_compress}
\end{equation}
where $\kappa_1$ and $\kappa_2$ are the first two cumulants of $I_N$. For Poisson statistics, one has $\kappa_2(E) \simeq \kappa_1(E)$ for small $E$, and then $\chi(E)\simeq 1$ (see \cref{app:iid}). On the contrary, for a rigid spectrum like that of the GOE matrix $B$ in \cref{eq:hamiltonian}, the mean number of eigenvalues behaves as $\expval{I_N[-E,E]} \propto \tilde E$, with $\tilde E \equiv N \rho_N(0)\, E$ and where $[N \rho_N(0)]^{-1}$ is the mean level spacing close to $E=0$, while $\expval{I_N^2[-E,E]}_c \propto \ln \tilde E$ for large $\tilde E$. Hence in the GOE case one finds $\chi(E) \to 0$ for $\tilde E \to \infty$, i.e., for $E \gg [N \rho_N(0)]^{-1}$ (but still much smaller than $E\sim\order{1}$).

In \cref{par:symmetric_interval} we 
provide the cumulant generating function of the variable $I_N[\alpha,\beta]$ at leading order for large $N$.
For a symmetric interval $[-E,E]$ and a symmetric distribution $p_a(a)$, the result reads
\begin{equation}
    \cor{F}_{[-E,E]}(s) \equiv  \frac{1}{N} \ln \expval{e^{-s I_N[-E,E]}} = -m s + \ln \expval{e^{-s f(a) }}_a + \order{\eta/N} \, ,
    \label{eq:intro_F}
\end{equation}
where
\begin{equation}
    m= - \frac{2\eta}{\pi} \Im \left[ \cor{G}_a\left(-i \Delta^{-1} \right) \right]^2  \, , \;\;\;\;\;
    f(a) \equiv \frac{1}{\pi} \arctan \left( \frac{ \sin 2\theta}{a^2r^2 + \cos 2\theta} \right) \in [0,1] \, ,
\end{equation}
and $\Delta(E) = r(E) e^{i\theta(E)}$ has to be determined by solving the self-consistency equation
\begin{equation}
     \Delta^{-1}= \varepsilon-iE -2i\eta \cor{G}_a\left(-i \Delta^{-1} \right) \, ,
\end{equation}
before sending $\varepsilon \to 0$.
Again, we show that a closed-form solution can be found in some particular cases. Our results are supported by the comparison with the numerical diagonalization of large random matrices.

The ratio of the first two cumulants of $I_N[-E,E]$ gives the level compressibility $\chi(E)$ introduced in \cref{eq:level_compress}. In agreement with the picture presented above for the region $1<\gamma<2$, and having identified $E_T \propto \eta \propto N^{1-\gamma}$ as the Thouless energy of the system, we explicitly verify in \cref{par:scaling} that $\chi(E)\sim 0$ for $E\ll E_T$ -- but still much larger than the mean level spacing $\delta_N \propto N^{-1}$ -- corresponding to level repulsion, while $\chi(E)$ follows the Poisson statistics for $E\gg E_T$. \rev{However}, in the scaling limit $E = 2\pi p_a(0) \eta \cdot y $ with $N^{-1} \ll \eta \ll 1$, we show that $\chi(E)$ takes the universal form
\begin{equation}
   \chi(E) \simeq \chi_T \left( y= \frac{E}{2\pi p_a(0) \eta} \right) \,, \quad\quad \chi_T(y) = \frac{1}{\pi y}\left[2y \arctan(y) -\ln (1+y^2)   \right] \, ,
    \label{eq:intro_scaling_comp}
\end{equation}
where the scaling function $\chi_T(y)$ is
independent of the specific choice of $p_a(a)$. This function is plotted in \cref{fig:scaling_chi}: it behaves as $\chi_T(y) \simeq y/\pi$ for small $y$, while it tends to $1$ for large $y$. 
Note that the crossover energy scale $2 \pi p_a(0) \eta$ coincides (apart from a factor of $4$) with the width of the mini-bands identified in Eq.~\eqref{eq:FGR} using the Fermi golden rule, which allows us to put the intuitive arguments given above on the structure of spectral correlations on a much firmer basis.

\rev{Interestingly, using results from Refs.~\cite{Kravtsov_2015} and~\cite{Kunz_1998}, we show that the same scaling function 
$\chi_T$ also describes the crossover for Hermitian $B$ matrices -- while the level compressibilities for the real and the Hermitian GRP ensembles differ outside of this energy regime.}
\rev{We do so by relating the level compressibility to the 2-level spectral correlation function of the Hermitian GRP model, previously derived in \cite{Kunz_1998,Kravtsov_2015} by means of the Harish-Chandra-Itzykson-Zuber integral (which notably does not admit a counterpart for matrices with real entries).}

In \cref{par:low_energy} we \rev{finally} inspect, using extensive numerical diagonalization of large random \rev{real} matrices, the low-energy region where $E$ is chosen on the scale of the Thouless energy. 
The scaling form of $\chi(E)$ presented in \cref{eq:intro_scaling_comp} is thus shown to represent a universal crossover between the classical GOE result $\chi(E \sim N^{-1} )\simeq \chi_\T{GOE}(E)$ for energies of the order of the mean level spacing (and much smaller than the Thouless energy, see \cref{chi_GOE_final}), and the model-dependent prediction of \cref{eq:level_compress,eq:intro_F}, valid for energies of the order of the total spectral band-width, \ie, $E\sim \order{1}$. 

\section{Average spectral density}
\label{par:spectral_replica_approach}
Let us begin by considering the density of states of the matrix $\cor{H}$ in \cref{eq:def_density}.
Its mean value can be obtained by means of the Edwards-Jones (E-J) formula \cite{Edwards_1976,Livan_2018}, which we briefly recall here. One starts from the Plemelj-Sokhotski relation: if $f(x)$ is a complex-valued function which is continuous on the real axis, and given $\alpha <0<\beta$, then
\begin{equation}
    \lim_{\varepsilon \to 0^+} \int_{\alpha}^\beta \dd{x} \frac{f(x)}{x\pm i \varepsilon} = \mp i \pi f(0) + \pv{\int_{\alpha}^\beta \dd{x} \, \frac{f(x)}{x}} \, ,
    \label{eq:plemelj}
\end{equation}
where $\pv$ indicates the Cauchy principal value of the integral. From \cref{eq:def_density} we then have
\begin{equation}
    \expval{\rho_N(\lambda)} =  \frac{1}{\pi N} \lim_{\varepsilon \to 0^+} \Im \expval{ \sum_{i=1}^N \frac{1}{\lambda-\lambda_i-i \varepsilon} } = \frac{1}{\pi N} \lim_{\varepsilon \to 0^+} \Im \pdv{\lambda} \expval{ \sum_{i=1}^N \ln(\lambda_i -\lambda +i\varepsilon) } \, ,
\end{equation}
where the average is taken over the distribution of the entries of $\cor{H}$. In the last step we indicated by $\ln(z)$ the principal branch of the complex logarithm. Using the properties of Gaussian integrals, we finally obtain the E-J formula~\cite{Edwards_1976}
\begin{align}
   \rho(\lambda) &\equiv \expval{\rho_N(\lambda)} = -\frac{2}{\pi N} \lim_{\varepsilon \to 0^+} \Im \dv{\lambda} \expval{\ln \cor{Z}(\lambda_\varepsilon)  } \, ,
    \label{eq:edwards_jones} \\
    \cor{Z}(\lambda) &\equiv \det ( \cor{H}- \lambda \mathbb{1} )^{-1/2} =  (2\pi i)^{-N/2}\int_{\mathbb{R}^N} \dd{\vb{r}} e^{-\frac{i}{2} \vb{r}^T (\lambda \mathbb{1} -\cor{H} ) \vb{r} } \, , \label{eq:partition_function}
\end{align}
where $\lambda_\varepsilon \equiv \lambda -i \varepsilon$ with $\varepsilon>0$. Note that the negative imaginary part of $\lambda_\varepsilon$ ensures the convergence of the integral in \cref{eq:partition_function}.
The expectation value of the logarithm can then be handled by using the replica trick \cite{Mezard_1987}
\begin{equation}
    \expval{\ln \cor{Z}(\lambda) } = \lim_{n\to 0} \frac{1}{n} \ln \expval{ \cor{Z}^n(\lambda) } \, \rev{,}
    \label{eq:replica_trick}
\end{equation}
\rev{which allows us to trade the \textit{quenched} average on the left-hand-side for the \textit{annealed} average on the right-hand-side. The latter can be evaluated} by standard methods (see \cref{app:replica}) \rev{to give}
\begin{equation}
    \expval{ \cor{Z}^n(\lambda) }  \propto \int \cor{D}\mu \, \cor{D}\hat \mu \, e^{N \cor{S}_n[\mu , \hat \mu; \lambda]  } \, ,
    \label{eq:average_Zn}
\end{equation}
where the proportionality holds up to an irrelevant numerical constant, and where we introduced the action
\begin{align}
    \cor{S}_n[\mu , \hat \mu; \lambda] \equiv& - i \int \dd{\vec{y}} \mu(\vec{y}) \hat \mu(\vec{y}) - \frac{\eta}{2}  \int \dd{\vec{y}} \dd{\vec{w}} \mu(\vec{y}) \mu(\vec{w}) \left( \vec{y} \cdot \vec{w} \right)^2 \n\\
    &+ \ln \int \dd{\vec{y}} \int \dd{a} p_a (a) \exp[-\frac{i}{2} (\lambda -a) |\vec{y}|^2+i  \hat \mu(\vec{y}) ] \, .
    \label{eq:action}
\end{align}
The parameter $\eta$ is defined in \cref{eq:eta},
while $\vec{y},\vec{w}$ are $n$-dimensional vectors (one component for each of the replicas).

The strategy to obtain the finite-$N$ averaged $\rho(\lambda)$ is the following:
\begin{enumerate}
    \item For large $N$, we look for a saddle-point estimate of the path integral in \cref{eq:average_Zn} in the form
    \begin{equation}
        \expval{ \cor{Z}^n(\lambda) } \propto  e^{N \cor{S}_n[\mu^* , \hat \mu^*; \lambda] +\order{1} } \, ,
    \end{equation}
    where the proportionality holds up to a $\lambda$-independent (even though possibly $N$-dependent) prefactor.
    \item Using \cref{eq:edwards_jones}, we recover the spectral density via
    \begin{align}
        &\rho(\lambda) \simeq -\frac{2}{\pi} \lim_{\varepsilon \to 0^+} \Im  \lim_{n\to 0} \frac{1}{n} \dv{\lambda} \cor{S}_n[\mu^* , \hat \mu^*; \lambda_\varepsilon] \label{eq:strategy_level} \\
        & \qquad\, = \frac{1}{\pi} \lim_{\varepsilon \to 0^+} \Im{  \lim_{n\to 0} \frac{i}{n} \frac{ \int \dd{\vec{y}} |\vec{y}|^2 \int \dd{a} p_a (a) \exp[-\frac{i}{2} (\lambda_\varepsilon -a) |\vec{y}|^2 +i  \hat \mu^*(\vec{y}) ]}{ \int \dd{\vec{y}} \int \dd{a} p_a (a) \exp[-\frac{i}{2} (\lambda_\varepsilon -a) |\vec{y}|^2 +i  \hat \mu^*(\vec{y}) ]}} \, . \n
    \end{align}
    Indeed, only the third term in the action of \cref{eq:action} will contribute, because the dependence on $\lambda$ in the first two terms is only implicit,
    \begin{equation}
        \dv{\lambda} \cor{S}_n[\mu , \hat \mu; \lambda_\varepsilon] = \partial_\lambda  \cor{S}_n + \int \dd{\vec{y}} \left[ \fdv{ \cor{S}_n}{\mu(\vec{y})} \dv{\mu(\vec{y})}{\lambda} +\fdv{ \cor{S}_n}{\hat\mu(\vec{y})} \dv{\hat\mu(\vec{y})}{\lambda}  \right] \, ,
    \end{equation}    
    and the term under the integral vanishes at the saddle-point (where the action is stationary by construction).
    In turn this implies that, to compute $\rho(\lambda)$ from Eq. \eqref{eq:strategy_level}, we do not need to determine $\mu^*(\vec{y})$, but only $\hat \mu^*(\vec{y})$. Finally, by the $\simeq$ symbol in \cref{eq:strategy_level} we mean that the corrections are \textit{at most} \rev{of} $\order{1/N}$.
\end{enumerate}

\subsection{Saddle-point equations and rotationally-invariant Ansatz}
\label{par:saddle_point_density}
The leading contribution to \cref{eq:average_Zn} for large $N$ can be found by minimizing the action in \cref{eq:action}. Omitting (to ease the notations) the superscript $*$ from $\mu^*(\vec{y})$ and $\hat \mu^*(\vec{y})$, and understanding the dependencies on $\lambda$ as computed in correspondence of $\lambda_\varepsilon$, the saddle-point equations read
\begin{align}
    0 &\equiv \fdv{\cor{S}_n}{\mu(\vec{x})} = - i \hat \mu(\vec{x}) -  \eta \int \dd{\vec{w}} \mu(\vec{w}) \left( \vec{x} \cdot \vec{w} \right)^2 \, , \label{eq:saddle_mu} \\
    0 &\equiv \fdv{\cor{S}_n}{\hat \mu(\vec{x})} = - i \mu(\vec{x})+i  \frac{ \int \dd{a} p_a (a) \exp[-\frac{i}{2} (\lambda -a) |\vec{x}|^2  +i \hat \mu(\vec{x}) ] }{\int \dd{\vec{y}} \int \dd{a} p_a (a) \exp[-\frac{i}{2} (\lambda -a) |\vec{y}|^2  +i \hat \mu(\vec{y}) ]} \, . \label{eq:saddle_muhat}
\end{align}
Substituting the expression for $\mu(\vec{x})$ obtained from the second equation \eqref{eq:saddle_muhat} in the first one \eqref{eq:saddle_mu}, one obtains a closed equation for $\hat \mu(\vec{x})$ which reads
\begin{equation}
    \hat \mu(\vec{x}) = i\eta \frac{ \int \dd{\vec{y}} \int \dd{a} p_a (a) \exp[-\frac{i}{2} (\lambda -a) |\vec{y}|^2  +i \hat \mu(\vec{y}) ] \left( \vec{x} \cdot \vec{y} \right)^2 }{\int \dd{\vec{y}} \int \dd{a} p_a (a) \exp[-\frac{i}{2} (\lambda -a) |\vec{y}|^2  +i \hat \mu(\vec{y}) ]} \, .
    \label{eq:saddle_tot}
\end{equation}
To make progress, we plug in the Ansatz $\hat \mu(\vec{x})=\hat \mu(x)$, with $x\equiv |\vec{x}|$,
which is rotationally symmetric in the space of replicas (i.e., it is invariant under $O(n)$ transformations). Note that requiring invariance under $O(n)$ is a stronger request than the mere replica-symmetry (RS): indeed, the exchange between any pair of components of $\vec{x}$ can be obtained by means of a $O(n)$ transformation\footnote{\rev{In the literature, multifractality has sometimes been associated with the breaking of replica-symmetry \cite{Kravtsov_2018}. As we stressed in \cref{subsec:RPmodel}, the intermediate phase of the GRP model is \textit{fractal}, but not \textit{multifractal} \cite{vonSoosten_2019}: it is then natural to look for, and remain with, a replica-symmetric solution, as we do here.}}.
Stepping to spherical coordinates and using the identity
\begin{equation}
    \int \dd{\Omega_n} \left( \vec{x} \cdot \vec{y} \right)^2 = \frac{(xy)^2}{n} \int \dd{\Omega_n} \, ,
\end{equation}
where $\dd{\Omega_n}$ is the differential of the $n$-dimensional solid angle around $\vec{y}$, we find
\begin{equation}
    \hat \mu(x) = i\frac{\eta }{n} \frac{ \int_0^\infty \dd{y} y^{n-1} \int \dd{a} p_a (a) \exp[-\frac{i}{2} (\lambda -a) y^2  +i \hat \mu(y) ] \left( xy \right)^2 }{\int_0^\infty \dd{y} y^{n-1} \int \dd{a} p_a (a) \exp[-\frac{i}{2} (\lambda -a) y^2  +i \hat \mu(y) ] } \, .
    \label{eq:replica_symmetrization}
\end{equation}
Let us now introduce the auxiliary function
\begin{equation}
    G(y;\lambda) \equiv \int \dd{a} p_a (a) \exp[-\frac{i}{2} (\lambda -a) y^2  +i \hat \mu(y) ]  = \exp[-\frac{i}{2} \lambda y^2  +i \hat \mu(y) ] \psi_a(-y^2/2) \, ,
    \label{eq:G(y)}
\end{equation}
where in the second equality we recognized the characteristic function of $p_a(a)$ (\ie, its Fourier transform), namely
\begin{equation}
    \psi_a(k) = \int \dd{a} p_a (a) e^{-ika} \, .
    \label{eq:characteristic}
\end{equation}
Equation \eqref{eq:replica_symmetrization} can readily be expressed in terms of $G(y;\lambda)$. We then integrate by parts in the denominator of \cref{eq:replica_symmetrization}, finding that boundary terms disappear at least as long as $\varepsilon >0$ in $G(y;\lambda_\varepsilon)$ -- we will check \textit{a posteriori} that the presence of $\hat \mu(y)$ does not spoil the convergence. We thus get
\begin{equation}
    \hat \mu(x) = -i\eta x^2 \frac{ \int_0^\infty \dd{y} y^{n+1} G(y;\lambda) }{\int_0^\infty \dd{y} y^{n} G'(y;\lambda) } \, ,
\end{equation}
where $G'(y;\lambda) = \partial_y G(y;\lambda)$. Under this form, we can now take the limit $n\to 0$, which yields
\begin{equation}
    i \hat \mu(x) = x^2 \eta \frac{ \int_0^\infty \dd{y} y \, G(y;\lambda) }{\int_0^\infty \dd{y} G'(y;\lambda) } \equiv x^2 C(\lambda) \, .
    \label{eq:C(l)}
\end{equation}
The function $C(\lambda)$ must now be determined self-consistently via
\begin{equation}
    C(\lambda) =\eta \frac{ \int_0^\infty \dd{y} y \, G(y;\lambda) }{\int_0^\infty \dd{y} G'(y;\lambda) } \, ,
    \label{eq:C(l)_defining}
\end{equation}
where $G(y;\lambda)$ can be read from \cref{eq:G(y)} upon setting $i\hat \mu(y)=y^2 C(\lambda)$.

\subsection{General result}
We now repeat the same steps at the level of the saddle-point action in \cref{eq:strategy_level}: we plug in the rotationally symmetric Ansatz, we integrate by parts in the denominator and we finally take the limit $n\to 0$. This gives
\begin{equation}
     \rho(\lambda) = -\frac{1}{\pi} \lim_{\varepsilon \to 0^+} \Re \frac{ \int_0^\infty \dd{y}y G(y;\lambda_\varepsilon)}{ \int_0^\infty \dd{y} G'(y;\lambda_\varepsilon)} +\order{1/N} =  -\frac{1}{\pi \eta} \lim_{\varepsilon \to 0^+} \Re C(\lambda_\varepsilon) +\order{1/N}
    \, .
    \label{eq:rho_general}
\end{equation}
We then go back to \cref{eq:C(l)_defining}, which contains an exact differential in its denominator. If $\Re C(\lambda_\varepsilon) \leq 0$, then $G(y\to \infty;\lambda_\varepsilon)=0$ and $G(0;\lambda_\varepsilon)=1$, and we obtain a self-consistency equation for $C(\lambda_\varepsilon)$,
\begin{equation}
    C(\lambda)= -\eta \int_0^\infty \dd{z} \psi_a(-z) \, e^{-i\lambda z +2z C(\lambda)} \, ,
    \label{eq:C(l)_self_consistent}
\end{equation}
where we recall that $\psi_a(x)$ is the characteristic function associated with $p_a(a)$ (see \cref{eq:characteristic}). This determines $C(\lambda)$ implicitly. Equations~\eqref{eq:rho_general} and \eqref{eq:C(l)_self_consistent} represent the main result of this section.

\subsubsection{Limiting cases}

Here we briefly check the consistency of the result we just obtained in the limiting cases in which only one of the two matrices $A$ or $B$ in \cref{eq:hamiltonian} are retained.

First, when $\nu=\eta=0$ we find $\cor{H}=A$, and the level statistics can only be determined by $p_a (a)$. Indeed, from \cref{eq:saddle_mu} we read $\hat \mu(\vec{x})=0$, and even without invoking a particular Ansatz we get from \cref{eq:strategy_level}, upon sending $N \to \infty$,
\begin{equation}
    \rho(\lambda)
    = -\frac{1}{\pi} \lim_{\varepsilon \to 0^+} \Re \frac{ \int_0^\infty \dd{y}y G(y;\lambda_\varepsilon)}{ \int_0^\infty \dd{y} G'(y;\lambda_\varepsilon)} \, ,
\end{equation}
where $G(y;\lambda)$ is given by \cref{eq:G(y)} with $\hat \mu=0$. For any $\varepsilon >0$ the function $G(y;\lambda_\varepsilon)$ is well behaved at $y\to+\infty$, so that
\begin{equation}
    \rho(\lambda)
    = \frac{1}{\pi} \lim_{\varepsilon \to 0^+} \Re \int_0^\infty \dd{y}y G(y;\lambda_\varepsilon) = \frac{1}{\pi} \lim_{\varepsilon \to 0^+} \Re \int_0^\infty \dd{z} \psi_a(-z) e^{-i\lambda_\varepsilon z} = p_a (\lambda) \, ,
\end{equation}
where we changed variable to $z=y^2/2$ in the first step, and in the last one we used $\psi_a^*(-z)=\psi_a(z)$.

Conversely, the GOE case is formally recovered by setting $\gamma=1$ and $a_i=0$ identically.\footnote{The way we constructed the action in \cref{eq:action} does not allow us to explore the region $\gamma<1$ where $A$ becomes subleading for large $N$ (although this could of course be achieved with minor modifications of the calculation presented in \cref{app:replica}).} With this choice, the limiting spectrum of ${\mathcal H}$ is supported within $\lambda \in [-\sqrt{2}\nu,\sqrt{2}\nu ]$, and we set for simplicity $\nu=1$ (corresponding to $\eta=1/4$). Using \cref{eq:C(l)_defining} (without integrating by parts in the denominator, but really computing $G'(y;\lambda)$) one easily finds
\begin{equation}
    C(\lambda) = \frac14 \left( i\lambda \pm \sqrt{2-\lambda^2} \right) \, .
\end{equation}
We choose the branch with the minus sign, as we assumed above that $\Re C(\lambda_\varepsilon)\leq 0$. Using \cref{eq:rho_general} then immediately renders, in the limit $N\to \infty$, the semicircle law
\begin{equation}
    \rho(\lambda) \to \rho_\T{GOE}(\lambda) = \frac{\sqrt{2-\lambda^2}}{\pi}\, \Theta(\sqrt{2}-|\lambda|) \, ,
    \label{eq:rho_GOE}
\end{equation}
where $\Theta(x)$ is the Heaviside step function, with $\Theta(x) = 1$ if $x>0$ and $\Theta(x) = 0$ if $x \leq 0$. 

\subsection{Resolvent formulation and connection with the Zee formula}
\label{par:resolvent}
One can observe that, under the same convergence hypotheses as above, the self-consistency equation \eqref{eq:C(l)_self_consistent} which determines $C(\lambda)$ can be rewritten as
\begin{equation}
    C(\lambda) = i \eta \, \cor{G}_a\left[\lambda+2iC(\lambda)\right] \, ,
    \label{eq:self_consistence_resolvent}
\end{equation}
where
\begin{equation}
    \cor{G}_a(z) = \int \dd{a} \frac{p_a(a)}{z-a}
    \label{eq:resolvent}
\end{equation}
is the resolvent (or Cauchy-Stieltjes transform) of the distribution $p_a(a)$. The resolvent can be inverted to give back
\begin{equation}
    p_a(x) = \frac{1}{\pi} \lim_{\varepsilon \to 0^+} \Im \cor{G}_a (x-i\varepsilon) \, ,
    \label{eq:inversion_resolvent}
\end{equation}
which can be easily proved by using the Plemelj-Sokhotski formula in \cref{eq:plemelj}.

By comparing \cref{eq:inversion_resolvent} with \cref{eq:rho_general}, one immediately realizes that our function $C(\lambda)$ is nothing but
\begin{equation}
    C(\lambda) = i \eta \, \cor{G}(\lambda) + \order{1/N} \, ,
    \label{eq:connection_zee}
\end{equation}
where we denoted by $\cor{G}(\lambda)$ the resolvent of the spectral density $\rho(\lambda)$ of our model. Choosing $\gamma=1$ (so that $\eta$ in \cref{eq:eta} becomes $N$-independent) and taking the limit $N\to \infty$, the correction in \cref{eq:connection_zee} vanishes, and \cref{eq:self_consistence_resolvent} takes the form
\begin{equation}
    \cor{G}(\lambda_\varepsilon) = \cor{G}_a (\lambda_\varepsilon -2\eta \cor{G}(\lambda_\varepsilon)) \, . 
    \label{eq:resolvent_ledoussal}
\end{equation}
This is analogous to Eq.~(148) in Ref. \cite{Krajenbrink_2021}, which was derived in the case in which the matrix $B$ belongs to the GUE (and not the GOE) ensemble, and it is consistent with the results of \ccite{Bouchaud_Les_Houches}. Moreover, in \cref{app:zee} we show how \cref{eq:resolvent_ledoussal} can be recovered by direct application of the Zee formula for the addition of two random matrices derived in \ccite{Zee_1996}.

One may legitimately wonder, at this point, whether our calculation is solely another way of obtaining Zee's result \cite{Zee_1996} for the particular case in which one of the two matrices being summed is a GOE matrix.
In \cref{par:exactly_solvable} we will analyze a few cases in which the self-consistency equation \eqref{eq:C(l)_self_consistent} (or \eqref{eq:self_consistence_resolvent}) can be solved exactly, so that from \cref{eq:rho_general} one can find an expression for $\rho(\lambda)$ which is correct, up to $\order{1/N}$, for any value of $\eta$. These cases are in fact the very same which could be cracked by applying the Zee formula to the deformed GOE matrix.
The advantage of our framework is that, whenever $\eta$ is $N$-dependent and decays slower than $1/N$ (as it happens, for instance, in \cref{eq:eta} for $1<\gamma<2$), then we are still able to keep track of the finite-$N$ corrections. Moreover, in \cref{par:approximate_solutions} we will provide approximate solutions which can be used whenever the resolvent $\cor{G}_a$ is not available in closed form, so that the Zee route is not viable.

\subsection{Exactly solvable cases}
\label{par:exactly_solvable}
In some particular cases, the self-consistency equation \eqref{eq:C(l)_self_consistent} admits an analytic solution, and we can access the limiting distribution $\rho(\lambda)$ for any value of $\eta$ (\ie, not necessarily small). This happens whenever the following conditions are met:
\begin{enumerate}[(i)]
    \item{the resolvent $\cor{G}_a(z)$ associated with $p_a(a)$ is known analytically, and}
    \item{the self-consistency equation for $C(\lambda)$ resulting from \cref{eq:C(l)_self_consistent} or \cref{eq:self_consistence_resolvent} is not transcendental, so that we can solve for $C(\lambda)$.}
\end{enumerate}
Below we present two such examples, which will also prove useful in our discussion of the level compressibility presented later in \cref{par:comp_exactly_solvable}.

\begin{figure}[t]
    \centering
    \subfloat[]{
    \includegraphics[width=0.47\textwidth]{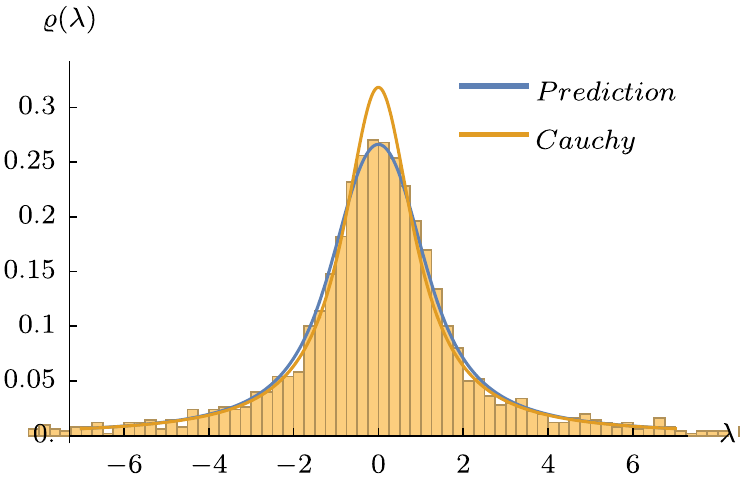}
    \label{fig:cauchy_numeric}
    }
    \subfloat[]{
     \includegraphics[width=0.47\textwidth]{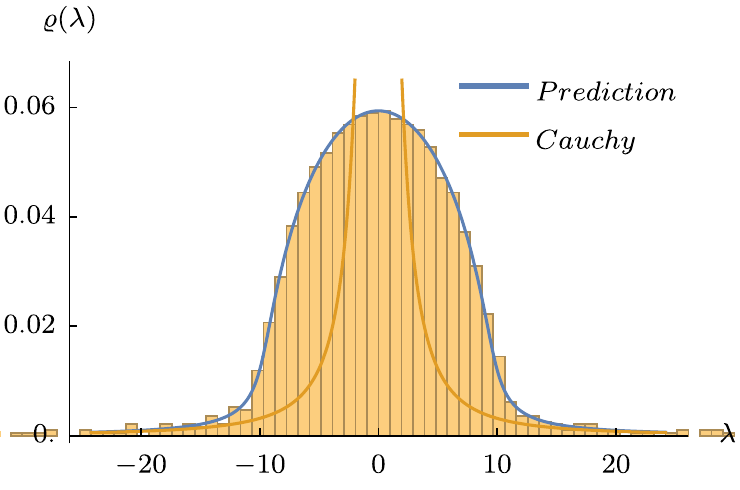}
    \label{fig:cauchy_numeric_large_eta}
    }
    \caption{Distribution of the eigenvalues $\rho(\lambda)$ in the case in which $p_a(a)$ is the Cauchy distribution (see \cref{par:cauchy}). We show a numerical check of \cref{eq:density_cauchy} in the regime of (a) small $\eta$ (here $\eta=0.12$) and (b) large $\eta$ (here $\eta=11.7$). The histogram was built using $\mu=0$, $\omega=1$, $\gamma=1.1$ and $N=2000$, with $\nu=1$ in (a) and $\nu=10$ in (b).}
    \label{fig:cauchy}
\end{figure}

\subsubsection{Cauchy distributed $a_i$}
\label{par:cauchy}
Let us choose $p_a(a)$ to be a Lorentzian of width $\omega$ and centered at $\mu$,
\begin{equation}
    p_a(a) = \frac{1}{\pi \omega}\left[\frac{\omega^2}{(a-\mu)^2+\omega^2} \right] \, .
    \label{eq:lorentzian}
\end{equation}
Its characteristic function is an exponential, $\psi_a(z) = \exp(-i\mu z -\omega |z|)$, and then by using \cref{eq:C(l)_self_consistent} we can compute $C(\lambda)$ in closed form:
\begin{equation}
    C(\lambda) = \frac14 \left\lbrace \omega +i(\lambda-\mu) \pm \sqrt{\left[\omega +i(\lambda-\mu)\right]^2+8\eta } \right\rbrace \, .
    \label{eq:C(l)_cauchy}
\end{equation}
We choose the branch with the minus sign for which $\Re C(\lambda)\leq \omega/2$. Using \cref{eq:rho_general} then yields
\begin{equation}
    \rho(\lambda) = -\frac{\omega -\left\lbrace 4\omega^2 (\lambda-\mu)^2 + \left[ 8\eta +\omega^2 -(\lambda-\mu)^2 \right]^2    \right\rbrace^{1/4}\cos(\theta_\lambda /2)}{4\pi \eta}  +\order{1/N} \, ,
    \label{eq:density_cauchy}
\end{equation}
with
\begin{equation}
    \theta_\lambda \equiv \arg \left\lbrace \left[\omega +i(\lambda-\mu)\right]^2+8\eta \right \rbrace \, .
\end{equation}
In \cref{fig:cauchy_numeric} we plot \cref{eq:density_cauchy} against numerical results in the small-$\eta$ region: we find a good agreement with the theoretical prediction, as well as visible departures from the Cauchy distribution, especially in the bulk.

\begin{figure}[t]
    \centering
    \subfloat[]{
    \includegraphics[width=0.47\textwidth]{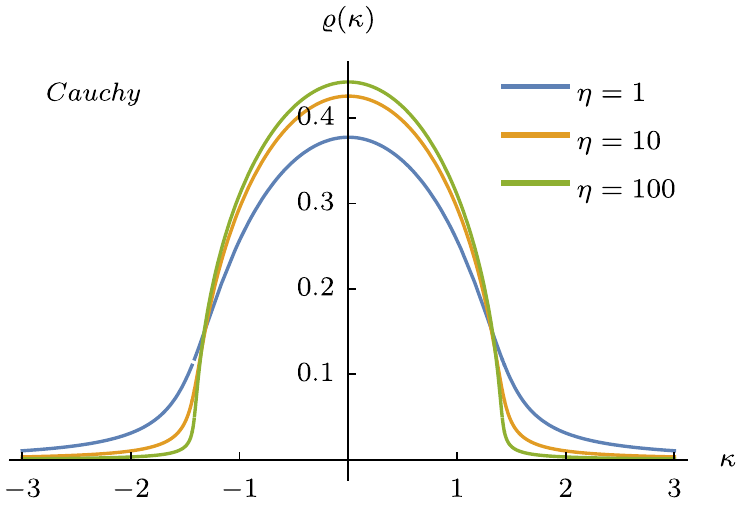}
    \label{fig:cauchy_large_eta}
    }
    \subfloat[]{
     \includegraphics[width=0.47\textwidth]{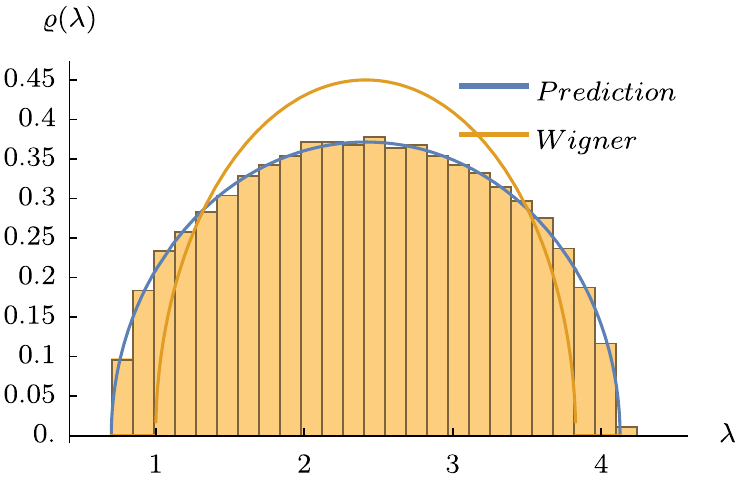}
    \label{fig:wigner_density}
    }
    \caption{\rev{(a)} Limiting distribution of the rescaled eigenvalues $\kappa=\lambda/\sqrt{4\eta}$, with $\rho(\lambda)$ given in \cref{eq:density_cauchy} for the Cauchy case (see \cref{par:cauchy}). In this plot we used $\mu=0$ and $\omega=0.5$. \rev{(b) Distribution of the eigenvalues $\rho(\lambda)$ in the case in which $p_a(a)$ is the Wigner semicircle distribution (see \cref{par:wigner}), and numerical check of the prediction in \cref{eq:wigner_deformed}. The histogram was built using $r=\sqrt{2}$, $\mu=1+\sqrt{2}$, $\nu=1$, $\gamma=1.1$ and $N=2000$.} }
\end{figure}

Another interesting limit is the one of large $\eta$, \ie, $\gamma=1$ and $\nu$ large. It has been shown in \ccite{Krajenbrink_2021} that, whenever $p_a(a)$ is rapidly decaying close to the edge of its finite support, then the spectral density $\rho(\lambda)$ interpolates between $p_a(a)$ and $\rho_\T{GOE}(\lambda)$ as the value of $\nu$ is increased. Note, however, that in the present case $p_a(a)$ decays algebraically and its support is not compact, so the outcome is less clear. The correct way of taking this limit is to rescale the eigenvalues as $\kappa \equiv\lambda/\sqrt{4\eta}$ and look for the distribution $\rho_\kappa(\kappa) = \sqrt{4\eta}\rho(\sqrt{4\eta} \kappa)$. From \cref{eq:C(l)_cauchy} we see that, if $\eta\gg \omega,\mu$, then
\begin{equation}
    C(\sqrt{4\eta}\kappa) = \frac{\sqrt{\eta}}{2}\left[i \kappa- \sqrt{2-\kappa^2} \right] +\order{\eta^0,1/N} \, ,
\end{equation}
and for large $N$ we get from \cref{eq:rho_general} that $\rho_\kappa(\kappa) \to \rho_\T{GOE}(\kappa)$ (see \cref{eq:rho_GOE}). For large but finite $\eta$ and $N\to \infty$, on the other hand, the bulk distribution of $\kappa$ looks like a semicircle (as in the GOE ensemble), but with fat tails whose width grows with $\omega$ (see \cref{fig:cauchy_large_eta}). Moreover, the whole distribution shifts rigidly by changing its center $\mu/\sqrt{4\eta}$.
Numerical results in the large-$\eta$ region are again nicely reproduced by \cref{eq:density_cauchy}, as shown in \cref{fig:cauchy_numeric_large_eta}.

Note that one can equivalently get to \cref{eq:C(l)_cauchy} by first computing the resolvent associated with the Lorentzian distribution in \cref{eq:lorentzian}, \ie,
\begin{equation}
    \cor{G}_\T{Cauchy}(\lambda) = \frac{1}{\lambda - \mu \pm i\omega} \, ,
    \label{eq:cauchy_resolvent}
\end{equation}
where the $\pm$ branches correspond to $\Im{\lambda}>0$ or $\Im{\lambda}<0$, respectively. This can be obtained by explicitly performing the complex integral in \cref{eq:resolvent}, which only entails simple poles for a Cauchy distribution. One can then easily solve the self-consistency equation \eqref{eq:self_consistence_resolvent}, which turns out to be quadratic, and thus recover \cref{eq:C(l)_cauchy}.

\subsubsection{Wigner distributed $a_i$}
\label{par:wigner}
Another simple case is the one in which $p_a(a)$ itself is chosen as the ($\mu$-centered) Wigner distribution
\begin{equation}
    p_a(a) \equiv \frac{2}{\pi \sigma^2} \sqrt{\sigma^2-(a-\mu)^2} \, ,
    \label{eq:wigner}
\end{equation}
whose corresponding resolvent is
\begin{equation}
    \cor{G}_a(z) = \frac{2}{\sigma^2} \left[z-\mu-\sqrt{(z-\mu)^2-\sigma^2} \right] \, .
    \label{eq:wigner_resolvent}
\end{equation}
The self-consistency equation \eqref{eq:self_consistence_resolvent} is again quadratic and it yields
\begin{equation}
    C(\lambda) = 2\eta \frac{i(\lambda-\mu) \pm \sqrt{\sigma^2+8\eta-(\lambda-\mu)^2}}{\sigma^2+8\eta} \, ,
\end{equation}
so that by choosing the branch with the minus sign and using \cref{eq:rho_general} we find
\begin{equation}
    \rho(\lambda) = \frac{2\sqrt{\sigma^2+8\eta-(\lambda-\mu)^2}}{\pi(\sigma^2+8\eta)} \Theta\left(\sigma^2+8\eta-(\lambda-\mu)^2 \right) + \order{1/N}\, .
    \label{eq:wigner_deformed}
\end{equation}
As expected, this is still a Wigner distribution centered in $\lambda=\mu$, but its width gets corrected as $\sigma^2 \to \sigma^2+8\eta$.

\subsection{Approximate solutions}
\label{par:approximate_solutions}
For most choices of the diagonal disorder distribution $p_a(a)$, \cref{eq:C(l)_self_consistent} cannot be solved exactly. However, since $\eta$ is a small parameter for large $N$ and $\gamma>1$ (which is the case we are mostly interested in), it may be sufficient to note that $C(\lambda) \sim \order{\eta} \sim \order{N^{1-\gamma}}$: this can be taken as a starting point for constructing self-consistent approximations for large $N$.

By expanding recursively the exponential in \cref{eq:C(l)_self_consistent} and using that $C(\lambda) \sim \order{\eta}$, one gets
\begin{align}
    C(\lambda) = &
    -\eta \int_0^\infty \dd{z} \psi_a(-z) e^{-i\lambda z} 
    \nonumber\\
    &  + 2\eta^2 \int_0^\infty \dd{z} \int_0^\infty \dd{s} z \psi_a(-z) \psi_a(-s) e^{-i\lambda (z+s)} + \order{\eta^3} \, .
    \label{eq:C(l)_largeN}
\end{align}
Note that
\begin{equation}
    \Re C(\lambda_\varepsilon) = -\eta \, p_a(\lambda_\varepsilon) +\order{\eta^2} \xrightarrow[\varepsilon \to 0]{N\gg 1} -\eta\, p_a(\lambda) \leq 0 \, , 
\end{equation}
in such a way that $G(y)$ is indeed well-behaved at $y\to +\infty$, as we had assumed in \cref{par:saddle_point_density}.
Plugging \cref{eq:C(l)_largeN} into \cref{eq:rho_general} now gives
\begin{align}
    \rho(\lambda) = p_a (\lambda) - \frac{2\eta}{\pi} \Re \int_0^\infty \dd{z} \int_0^\infty \dd{s} z \psi_a(-z) \psi_a(-s) e^{-i\lambda (z+s)} + \order{\eta^2,N^{-1}} \, .
    \label{eq:rho_first_order}
\end{align}
Of course this approximation becomes meaningless if $\gamma \geq 2$, where the higher order corrections we neglected when we took the saddle-point approximation mingle with 
the $\mathcal{O}(\eta)$ correction.

Alternatively, one can expand the exponential in \cref{eq:C(l)_self_consistent} as
\begin{equation}
    C(\lambda) \simeq -\eta \int_0^\infty \dd{z} \psi_a(-z) e^{-i\lambda z} - 2\eta C(\lambda) \int_0^\infty \dd{z}  z \psi_a(-z) e^{-i\lambda} \, ,
    \label{eq:resummed}
\end{equation}
corresponding to a partial resummation of the perturbative series in $\eta$: this is the so-called self-consistent Hartree-Fock approximation \cite{parisi_statistical_1988}. From \cref{eq:resummed} we then obtain
\begin{equation}
    C(\lambda) \simeq \frac{-\eta \int_0^\infty \dd{z} \psi_a(-z) e^{-i\lambda z}}{1+2\eta \Sigma(\lambda)} \, ,
    \label{eq:C(l)_hartree_fock}
\end{equation}
with the \textit{self-energy}
\begin{equation}
    \Sigma(\lambda) = \int_0^\infty \dd{z}  z \psi_a(-z) e^{-i\lambda z} \, .
\end{equation}

\begin{figure}[t]
    \centering
    \includegraphics[width=0.5\textwidth]{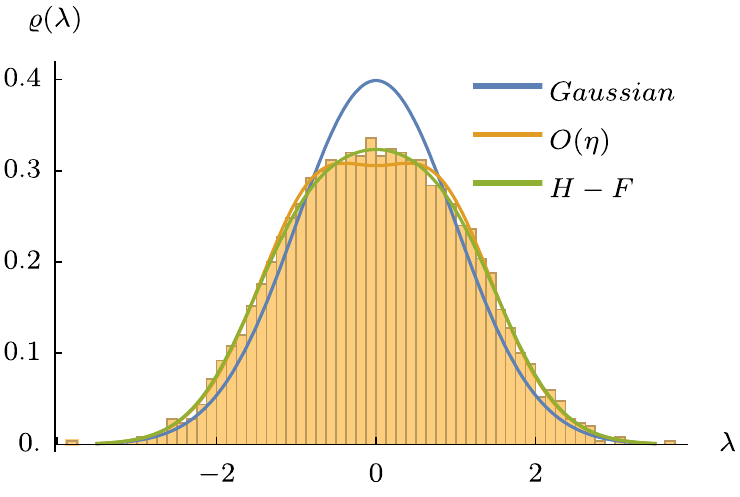}
    \caption{Distribution of the eigenvalues $\rho(\lambda)$ in the case in which $p_a(a)$ is Gaussian with unit variance, and $B$ is a GOE matrix as in \cref{eq:hamiltonian}. Numerical results are compared to the first $\order{\eta}$ correction given in \cref{eq:gaussian_correction}, and to the Hartree-Fock approximation given in \cref{eq:C(l)_hartree_fock}. The histogram was built using $N=2000$, $\nu=1$ and $\gamma=1.1$, corresponding to $\eta=0.12$.}
    \label{fig:gaussian_HF}
\end{figure}

\noindent\textbf{An explicit example: the Gaussian case.} We finally test both approximations in the case where $p_a(a)$ is chosen to be Gaussian (with zero mean and variance $\sigma^2$), as in the original Rosenzweig-Porter model \cite{RP_1960}. By using $\psi_a(z)=\exp(-z^2 \sigma^2/2)$ in \cref{eq:rho_first_order}, one finds
\begin{equation}
    \rho(\lambda) = p_a (\lambda) + \sqrt{\frac{2}{\pi}}\frac{\eta}{ \sigma^3} e^{-\lambda^2/(2\sigma^2)}\left[ \frac{2\sqrt{2}\lambda}{\sigma}F\left(\frac{\lambda}{\sqrt{2}\sigma}\right)-1 \right]+ \order{\eta^2,N^{-1}} \, ,
    \label{eq:gaussian_correction}
\end{equation}
where $F(z)$ is the Dawson function defined as
$
    F(z) \equiv e^{-z^2} \int_0^z  e^{y^2} \dd{y} \, .
$
The corrected spectral density is plotted in \cref{fig:gaussian_HF}, and comparison with the eigenvalue spectrum computed numerically shows a fair agreement.
Using instead the Hartree-Fock approximation for $C(\lambda)$ given in \cref{eq:C(l)_hartree_fock} results in a better agreement with the numerical data in the central region of the spectrum.

\section{Number of eigenvalues in an interval and level compressibility}
\label{par:level_compress}
In this Section we consider the number of eigenvalues $I_N[\alpha,\beta]$ in a finite interval $[\alpha,\beta]$, as given by \cref{eq:levels_number}, and compute its cumulant generating function. This, in turn, can be used to access the level compressibility $\chi(E)$ defined in \cref{eq:level_compress}.
As we explained in the Introduction, $\chi(E)$ represents a simple measure of the rigidity of the spectrum, which in turn allows us to distinguish between the phases of the model.

This program can be achieved by following the replica-based procedure introduced and exploited in \ccite{Metz_2016,Metz_2017}, which we briefly outline here. Starting from the definition of the spectral density in \cref{eq:def_density}, we first rewrite \cref{eq:levels_number} as
\begin{equation}
    I_N[\alpha,\beta] = \sum_{i=1}^N \left[ \Theta(\beta-\lambda_i) - \Theta(\alpha-\lambda_i) \right] \, .
\end{equation}
Now we recall that the Heaviside function can be represented in terms of the discontinuity of the complex logarithm, 
\begin{equation}
    \Theta(-x) = \frac{1}{2\pi i}\lim_{\varepsilon \to 0^+} \left[ \ln(x+i\varepsilon) - \ln(x-i\varepsilon) \right] \, ,
\end{equation}
so that we interpret
\begin{equation}
    \sum_{i=1}^N \Theta(\alpha-\lambda_i) = \frac{1}{2\pi i}\lim_{\varepsilon \to 0^+} \left[ \ln \det ( \cor{H}- \alpha_\varepsilon \mathbb{1} ) - \ln\det ( \cor{H}- \alpha_\varepsilon^* \mathbb{1} ) \right] \, ,
\end{equation}
where we called as before $\alpha_\varepsilon \equiv \alpha-i\varepsilon$ with $\varepsilon>0$. This allows us to express $I_N[\alpha,\beta]$ in terms of the partition function given in \cref{eq:partition_function}, \rev{which leads to}\footnote{\rev{It should be noted that \cref{eq:I_N_complex_representation} was obtained by naively adopting the identity $\ln (ab) = \ln a + \ln b$, which is however not satisfied in general by the complex logarithm (whose principal branch is bounded within $(-\pi,\pi]$ -- see \ccite{Vivo_2020}). As a result, the right-hand-side of \cref{eq:I_N_complex_representation} is not extensive, and thus seemingly unfit to count the number of eigenvalues in an interval for a single realization of $\cor{H}$ \cite{Metz_2017}. The introduction of replicas (see \cref{eq:Q_ab} and \cref{app:replica_level_compress}) is essential in order to restore the extensivity of the ensemble-averaged moments of $\cor{I}_N$. This remarkable fact was dubbed \textit{folding-unfolding mechanism} in \ccite{Castillo_2018}.
}}
\begin{equation}
    I_N[\alpha,\beta] = -\frac{1}{\pi i}\lim_{\varepsilon \to 0^+} \ln \left[\frac{\cor{Z}(\beta_\varepsilon)\cor{Z}(\alpha_\varepsilon^*) }{\cor{Z}(\beta_\varepsilon^*)\cor{Z}(\alpha_\varepsilon)} \right] \, .
    \label{eq:I_N_complex_representation}
\end{equation}
In order to compute the moments of $I_N$, we first address its cumulant generating function
\begin{align}
    \cor{F}_{[\alpha,\beta]}(s) \equiv  
    \frac{1}{N} \ln \expval{e^{-s I_N[\alpha,\beta]}} 
    =
    \frac{1}{N} \ln 
    \langle 
    e^{\frac{s}{\pi i} \lim_{\varepsilon \to 0^+} 
    \ln \{ 
    {\mathcal Z}(\beta_\epsilon) 
    {\mathcal Z}(\alpha_\epsilon^*) 
    [{\mathcal Z}(\beta^*_\epsilon) {\mathcal Z}(\alpha_\epsilon)]^{-1}
    \}
    }
    \rangle
    \,. 
    \end{align}
    Assuming now that one can move the limit $\varepsilon \to 0^+$ at the front of this expression, we obtain
    \begin{align}
    \cor{F}_{[\alpha,\beta]}(s) 
    =
    \lim_{\varepsilon \to 0^+}  \frac{1}{N} \ln Q_{[\alpha,\beta]}(s) \, ,
    \label{eq:cgf}
\end{align}
where we introduced
\begin{equation}
    Q_{[\alpha,\beta]}(s) \equiv \expval{\left[ \cor{Z}(\beta_\varepsilon^*)\cor{Z}(\alpha_\varepsilon)\right]^{is/\pi}  \left[\cor{Z}(\beta_\varepsilon) \cor{Z}(\alpha_\varepsilon^*)\right]^{-is/\pi}   } \, .
\end{equation}
The latter can be accessed by first evaluating
\begin{equation}
    Q_{[\alpha,\beta]}(n_\pm) \equiv \expval{\left[ \cor{Z}(\beta_\varepsilon^*)\cor{Z}(\alpha_\varepsilon)\right]^{n_+}   \left[\cor{Z}(\beta_\varepsilon) \cor{Z}(\alpha_\varepsilon^*)\right]^{n_-}  }
    \label{eq:Q_ab}
\end{equation}
within the replica formalism with $n_\pm$ integer, and then performing its analytic continuation to 
\begin{equation}
    Q_{[\alpha,\beta]}(s) = \lim_{n_\pm \to \pm is/\pi} Q_{[\alpha,\beta]}(n_\pm) \, .
\end{equation}
To obtain the level compressibility in \cref{eq:level_compress}, we finally compute the cumulants
\begin{equation}
    \frac{\kappa_j[\alpha,\beta]}{N} = (-1)^{j} \eval{\partial_s^j \cor{F}_{[\alpha,\beta]}(s)}_{s=0} \, ,
    \label{eq:cumulants}
\end{equation}
and we evaluate them at $\alpha=-E$, $\beta=E$.

\rev{We remark that the average spectral density $\rho(E)$ is formally proportional to the derivative with respect to $E$ of the first cumulant $\kappa_1(E)=\left\langle \mathcal{I}[-E,E] \right\rangle$ (see \cref{eq:levels_number}), thus providing 
an alternative way to compute $\rho(E)$ which does not rely on the Edwards-Jones formula in \cref{eq:edwards_jones}.}

\subsection{Replica action and saddle-point equations}
The details of this derivation are reported in \cref{app:replica_level_compress}. In analogy with \cref{par:spectral_replica_approach}, here we express
\begin{align}
    Q_{[\alpha,\beta]}(n_\pm) \propto & \int  \cor{D}(i\varphi) \, \exp{N \cor{S}_{n_\pm}[\varphi; \hat \Lambda]  } \, , \label{eq:Q_n_varphi} \\
    \cor{S}_{n_\pm}[\varphi ;\hat \Lambda] \equiv&  \frac{1}{2\eta}  \int \dd{\vec{\tau}} \dd{\vec{\tau}'} \varphi(\vec{\tau}) M(\vec{\tau},\vec{\tau}') \varphi(\vec{\tau}')  
    \nonumber\\
    &+ \ln \int \dd{\vec{\tau}} \exp[-\frac{i}{2}\vec{\tau}\, \hat \Lambda \,\vec{\tau} -\int \dd{\vec{\tau}'} M(\vec{\tau},\vec{\tau}') \varphi(\vec{\tau}') ] \, \psi_a\left(-\frac{1}{2} \vec{\tau} \, \hat L \, \vec{\tau}\right) \, ,
    \label{eq:action_comp_varphi}
\end{align}
where the replica vectors $\vec{\tau}\in \mathbb{R}^n$ live in dimension $n=2(n_+ + n_-)$. Each of the four replica sets corresponds to one sector in the block matrices
\begin{equation}
    \hat{\Lambda} \equiv \mqty(\dmat{\alpha_\varepsilon \mathbb{1}_{n_+}, \bar{\beta}_\varepsilon \mathbb{1}_{n_+}, \beta_\varepsilon \mathbb{1}_{n_-},\bar{\alpha}_\varepsilon \mathbb{1}_{n_-}}) \, ,
    \;\;\;\qquad \hat{L} \equiv \mqty(\dmat{\mathbb{1}_{n_+}, -\mathbb{1}_{n_+}, \mathbb{1}_{n_-},- \mathbb{1}_{n_-}}) \, ,
    \label{eq:lambda_L_def}
\end{equation}
where we called $\bar{\alpha}_\varepsilon=-\alpha^*_\varepsilon$, and similarly for $\bar\beta_\epsilon$. Finally, in \cref{eq:action_comp_varphi} we introduced the function
\begin{equation}
    M(\vec{\tau},\vec{\tau}') \equiv \left( \vec{\tau} \, \hat L \, \vec{\tau}' \right)^2\, ,
    \label{eq:M_def}
\end{equation}
while we recall that $\psi_a(z)$ was given in \eqref{eq:characteristic}. Next, we need to compute the functional integration over $\cor{D}(i\varphi)$ in \cref{eq:Q_n_varphi} in the limit of large $N$.
From \cref{eq:action_comp_varphi}, the saddle-point equation follows simply as
\begin{equation}
    \varphi_0(\vec{\tau}) =  \frac{\eta}{Z_\varphi} \exp[-\frac{i}{2}\vec{\tau}\, \hat \Lambda \,\vec{\tau} -\int \dd{\vec{\tau}'} M(\vec{\tau},\vec{\tau}') \varphi_0(\vec{\tau}') ] \, \psi_a\left(-\frac{1}{2} \vec{\tau} \, \hat L \, \vec{\tau}\right) \, , 
    \label{eq:saddle_point_comp}
\end{equation}
with
\begin{equation}
    Z_\varphi \equiv \int \dd{\vec{\tau}} \exp[-\frac{i}{2}\vec{\tau}\, \hat \Lambda \,\vec{\tau} -\int \dd{\vec{\tau}'} M(\vec{\tau},\vec{\tau}') \varphi_0(\vec{\tau}') ] \, \psi_a\left(-\frac{1}{2} \vec{\tau} \, \hat L \, \vec{\tau}\right) \, .
\end{equation}
In order to better quantify the finite-size corrections and to make contact with the calculation performed in \ccite{Metz_2017} for the pure GOE case, in \cref{app:gaussian_fluctuations} we also compute the Gaussian fluctuations around the saddle-point $\varphi_0(\vec{\tau})$, leading to
\begin{equation}
    Q_{[\alpha,\beta]}(n_\pm) = \exp{N\cor{S}_{n_\pm}[\varphi_0 ; \hat \Lambda]+\frac12 \sum_{k=1}^\infty \frac{(-1)^k}{k} \Tr T^k}+\order{1/N^2} \, ,
    \label{eq:Q_n_gaussian}
\end{equation}
with
\begin{equation}
   T (\vec{\tau}_1,\vec{\tau}_2) \equiv \varphi_0(\vec{\tau}_1) \left[ M (\vec{\tau}_1,\vec{\tau}_2) - \frac{1}{\eta} \int \dd{\vec{\tau}'} M (\vec{\tau}_2,\vec{\tau}') \varphi_0(\vec{\tau}') \right] \, .
   \label{eq:T_def}
\end{equation}
Note that $\varphi_0(\va{\tau}) \sim \order{\eta}$ (see \cref{eq:saddle_point_comp}), so that the function $T (\vec{\tau}_1,\vec{\tau}_2)$ itself is also of $\order{\eta}$.

\subsection{Rotationally-invariant Ansatz}
In the pure GOE case (which is recovered, for instance, by setting $\psi_a(z)=1$ identically in the expressions above), the saddle-point equation \eqref{eq:saddle_point_comp} suggests to look for a replica-symmetric solution in the form of a Gaussian, \ie,
\begin{equation}
    \varphi_0(\vec{\tau}) = \cor{N} \exp(-\frac{1}{2} \vec{\tau} \, \hat C^{-1} \, \vec{\tau}) \, ,
    \label{eq:ansatz}
\end{equation}
where $\hat C$ is another block-diagonal matrix of the form
\begin{equation}
    \hat C \equiv \mqty(\dmat{\Delta_\alpha \mathbb{1}_{n_+}, \bar{\Delta}_\beta \mathbb{1}_{n_+}, \Delta_\beta \mathbb{1}_{n_-},\bar{\Delta}_\alpha \mathbb{1}_{n_-}}) \, .
    \label{eq:hat_C_def}
\end{equation}
The four parameters $\Delta_\alpha, \bar\Delta_\alpha, \Delta_\beta$ and $\bar\Delta_\beta$ and  are to be fixed, together with the prefactor $\cor{N}$ in \cref{eq:ansatz}, by substituting the Ansatz in \cref{eq:ansatz} back into the saddle-point equation \eqref{eq:saddle_point_comp}.

This strategy has proven effective in \ccite{Metz_2017} and, for completeness, in \cref{app:GOE} we sketch  the entire calculation for the GOE case. However, in the presence of a nontrivial $\psi_a(z)$ in \cref{eq:saddle_point_comp}, there is no reason why the Ansatz in \cref{eq:ansatz} should work\rev{: the structure of \cref{eq:saddle_point_comp} suggests instead} to extend it to the form
\begin{equation}
    \varphi_0(\vec{\tau}) = \cor{N} \exp(-\frac{1}{2} \vec{\tau} \, \hat C^{-1} \, \vec{\tau}) \psi_a\left(-\frac{1}{2} \vec{\tau} \, \hat L \, \vec{\tau}\right) \, .
    \label{eq:ansatz_new}
\end{equation}
This can be plugged into \cref{eq:saddle_point_comp} to first obtain $\cor{N} = \eta / Z_\varphi $, where 
\begin{equation}
    Z_\varphi = \int \dd{\vec{\tau}} \varphi_0(\vec{\tau}) \, .
    \label{eq:Z_varphi}
\end{equation}
By using Gaussian integration one can then show that
\begin{equation}
    \int \dd{\vec{\tau}'} M(\vec{\tau},\vec{\tau}') \varphi_0(\vec{\tau}') = \eta \vec{\tau} \, \hat K \, \vec{\tau} \, ,
    \label{eq:property}
\end{equation}
where $\hat K$ is a diagonal matrix given by
\begin{equation}
    \hat K = -i \hat L \, \cor{G}_a \left( (i\hat L \hat C)^{-1} \right)\, ,
    \label{eq:K_def}
\end{equation}
$\cor{G}_a$ is the resolvent in \cref{eq:resolvent}, and $\hat{L}$ is given in \cref{eq:lambda_L_def}.
The remaining free parameters in \cref{eq:ansatz} can be determined by solving the set of four self-consistency equations which follow from \cref{eq:saddle_point_comp}, i.e., 
\begin{equation}
    \hat C^{-1} \,=  2\eta \, \hat K +i \, \hat \Lambda \, .
     \label{eq:four_set_p(a)}
\end{equation}
Finally, both the action and its Gaussian fluctuation matrix $T$ (see \cref{eq:T_def}) can be computed 
by using the saddle-point solution in \cref{eq:ansatz_new}. First, notice that the action in \cref{eq:action_comp_varphi} takes the form
\begin{align}
    \cor{S}_{n_\pm}[\varphi_0 ; \hat \Lambda] &= \frac12 \int \dd{\vec{\tau}} \left( \vec{\tau} \, \hat K \, \vec{\tau} \right) \varphi_0(\vec{\tau})+ \ln Z_\varphi \n\\
    &=  \frac{\eta}{2} \sum_i K_{ii}^2 + \ln \int_{-\infty}^\infty \dd{a} p_a(a) \exp{ -\frac12 \Tr \ln[ (i\hat L \hat C)^{-1} - a ] }  + \T{const.} \, .
\end{align}
The constant term vanishes upon taking the analytic continuation $n_\pm \to \pm is/\pi$, yielding
\begin{align}
    \cor{S}_{\pm \frac{is}{\pi}}[\varphi_0 ; \hat \Lambda]=& \frac{is \eta}{2\pi} \left( k_\alpha^{2} + \bar{k}_\beta^{2} - k_\beta^{2} - \bar{k}_\alpha^{2} \right) \n\\
    &+ \ln \int_{-\infty}^\infty \dd{a} p_a(a) \exp{ -\frac{is}{2\pi} \ln \left[ \frac{(\bar{\Delta}_\beta^{-1}+ia)(\Delta_\alpha^{-1}-ia)}{(\bar{\Delta}_\alpha^{-1}+ia)(\Delta_\beta^{-1}-ia)} \right] }\, ,
    \label{eq:action_prolungata}
\end{align}
where we introduced for brevity $\hat{K} \equiv \T{diag}(k_\alpha \mathbb{1}_{n_+}, \bar{k}_\beta \mathbb{1}_{n_+}, k_\beta \mathbb{1}_{n_-},\bar{k}_\alpha \mathbb{1}_{n_-})$. Note that this action coincides at leading order with the cumulant generating function in \cref{eq:cgf}, \ie,
\begin{align}
    \cor{F}_{[\alpha,\beta]}(s) = \lim_{\varepsilon \to 0^+}  \cor{S}_{\pm \frac{is}{\pi}}[\varphi_0 ; \hat \Lambda] +\order{\eta/N} \, ,
    \label{eq:cgf_general_solution}
\end{align}
where we used \cref{eq:Q_n_gaussian} (the estimate of the large-$N$ correction will soon be justified). In the next Section we will make this result more explicit in the case of an interval which is symmetric around the origin.

The Gaussian fluctuations around the saddle-point are studied in \cref{app:gaussian_fluctuations}. A closed-form result is not available in this case (in contrast to the GOE case, see \cref{app:GOE}), because the calculation involves increasingly complex generalizations of the resolvent $\cor{G}_a(z)$ which encode higher order correlations (see \cref{eq:gaussian_fluc_first_term}). However, we can show that the Gaussian fluctuations add to the leading order term in \cref{eq:cgf_general_solution} a correction of $\order{\eta/N}=\order{N^{-\gamma}}$, which is strongly suppressed for large $N$ in the region $\gamma>1$ which we focus on here. 

\subsection{General result in the case of a symmetric interval}
\label{par:symmetric_interval}
We consider now the simpler case in which $\alpha=-E$ and $\beta=E$, and we take a symmetric distribution $p_a(a)$. From \cref{eq:K_def,eq:four_set_p(a)} one can deduce that the entries of the matrix $\hat{C}$ are related by the following symmetries:
\begin{equation}
    \Delta_\alpha \equiv \Delta \equiv r e^{i\theta}\,, \;\;\;\;\; \bar{\Delta}_\alpha=\Delta_\alpha^* = \Delta_\beta \,, \;\;\;\;\; \bar{\Delta}_\beta = \Delta_\beta^* \, .
    \label{eq:symmetries}
\end{equation}
The same holds for the entries of $\hat{K}$ (see \cref{eq:K_def}), hence we will simply call $k_\alpha \equiv k$. The problem is then reduced to computing one single unknown, namely $\Delta$: from \cref{eq:K_def,eq:four_set_p(a)}, this amounts to solving the self-consistency equations 
\begin{equation}
\begin{dcases}
    \Delta^{-1}= \varepsilon-iE +2\eta k \, , \\
    k = -i\cor{G}_a \left(-i \Delta^{-1} \right) \, .
    \label{eq:self_consistency_p(a)_symm}
\end{dcases}
\end{equation}
The action in \cref{eq:action_prolungata} then takes the form
\begin{align}
    \cor{S}[\varphi_0 ; \hat \Lambda]= -\frac{2\eta s}{\pi} \Im{k^2} + \ln \int_{-\infty}^\infty \dd{a} p_a(a) \exp[ -\frac{s}{\pi} \arctan \left( \frac{ \sin 2\theta}{a^2r^2 + \cos 2\theta} \right) ]\, ,
    \label{eq:action_comp_symmetric_interval}
\end{align}
where the branch of the $\arctan$ is chosen so that it returns an angle in $[0,\pi]$. From \cref{eq:cgf} we can then read the leading order contribution to the rate function, namely
\begin{align}
    \cor{F}_{[-E,E]}(s) = \cor{S}[\varphi_0 ; \hat \Lambda] + \order{\eta/N} = -m s + \ln \expval{e^{-s f(a) }}_a + \order{\eta/N} \, .
    \label{eq:cgf_GRP}
\end{align}
Here we used the notation $\expval{\bullet}_a$ to indicate the average over $p_a(a)$, and we introduced for brevity
\begin{align}
    m&\equiv \frac{2\eta}{\pi} \Im{k^{2}} = - \frac{2\eta}{\pi} \Im \left[ \cor{G}_a \left(-i\Delta^{-1} \right) \right]^2  \, , \label{eq:m_def} \\
    f(a) &\equiv \frac{1}{\pi} \arctan \left( \frac{ \sin 2\theta}{a^2r^2 + \cos 2\theta} \right) \in [0,1] \, ,
    \label{eq:f_def}
\end{align}
where in the first line we used \cref{eq:self_consistency_p(a)_symm}.
The cumulant generating function in \cref{eq:cgf_GRP}, together with the self-consistency equations \eqref{eq:self_consistency_p(a)_symm}, represent our second main result. As we stressed above, when $\eta$ is given by \cref{eq:eta}, the correction to \cref{eq:cgf_GRP} is of $\order{\eta/N}=\order{N^{-\gamma}}$, which is strongly suppressed for large $N$ in the region $\gamma>1$.

Expanding \cref{eq:cgf_GRP} in powers of $s$ we get
\begin{align}
    \cor{F}_{[-E,E]}(s) \simeq -s \left[ m+ \expval{f(a)}_a \right] + \frac{s^2}{2} \left[ \expval{f^2(a)}_a-\expval{f(a)}_a^2 \right]+ \order{s^3,\eta/N}\, ,
\end{align}
and by comparison with \cref{eq:cumulants} we can identify the first two cumulants
\begin{align}
    \frac{\kappa_1}{N} &= m +\expval{f(a)}_a +  \order{\eta/N}\, , \label{eq:first_cumulant} \\
    \frac{\kappa_2}{N} &= \expval{f^2(a)}_a - \expval{f(a)}_a^2+ \order{\eta/N} \, .
\end{align}
From \cref{eq:level_compress}, we finally obtain the level compressibility
\begin{equation}
    \chi(E) =  \frac{\expval{f^2(a)}_a - \expval{f(a)}_a^2}{ m +\expval{f(a)}_a} +  \order{\eta/N}\, .
    \label{eq:comp_general_prediction}
\end{equation}
We remark that not only the level compressibility, but actually all the moments of $I_N[-E,E]$ can be simply computed starting from Eq. \eqref{eq:cgf_GRP}: they read (at leading order for large $N$)
\begin{eqnarray}
\langle (I_N[-E,E])^m\rangle \simeq N \langle [f(a)]^m \rangle_a \, , \quad m \geq 2 \;.
\end{eqnarray}

\subsubsection{Limit of a pure diagonal matrix with random i.i.d. entries}
It is instructive, at this point, to consider the limit $\eta\to 0$. In this case the GOE part of \cref{eq:hamiltonian} is neglected, and the spectral properties are completely determined by the matrix $A$, whose entries are independent and identically distributed according to $p_a(a)$. The self-consistency equation \eqref{eq:four_set_p(a)} then reduces to
\begin{equation}
    \hat C \,= (i \, \hat \Lambda)^{-1} \, ,
    \label{eq:SCE_iid}
\end{equation}
whence
\begin{equation}
    \Delta = 
    \eval{\Delta_\alpha}_{\alpha=-E} = \frac{iE+\varepsilon}{E^2+\varepsilon^2} \equiv re^{i\theta} \, , 
\end{equation}
and there is no need to determine the entries of $\hat K$ since it does not enter the expression of the saddle-point action \eqref{eq:action_comp_symmetric_interval} for $\eta = 0$. From \cref{eq:action_comp_symmetric_interval} we obtain, for $\varepsilon \to 0^+$,
\begin{align}
    \cor{S}[\varphi_0 ; \hat \Lambda] &= \ln \int_{-\infty}^\infty \dd{a} p_a(a) \exp{ -\frac{s}{\pi} \arctan \left[ \frac{ 0^+}{(a/E)^2 -1} \right] } \n\\
    &= \ln \left[ 1+(e^{-s}-1) \int_{-E}^E \dd{a} p_a(a)  \right] \, ,
\end{align}
where we used that the branch of $\arctan (z)\in [0,\pi]$ has a discontinuity in $z=0$, \ie, it jumps from $\pi$ to $0$ as $z$ becomes positive.
From \cref{eq:cgf_GRP,eq:levels_number}, we can then read out the cumulant generating function
\begin{equation}
    \cor{F}_{[-E,E]}(s) = \ln \left[ 1+(e^{-s}-1) \frac{\expval*{I_N[-E,E]}_a}{N} \right] \, .
    \label{eq:cgf_iid}
\end{equation}
In this way we recover the standard textbook result for the cumulant generating function in the case of i.i.d. random variables, which we sketch for completeness in \cref{app:iid}. In particular, the level compressibility in \cref{eq:level_compress} reads in this case 
\begin{equation}
    \chi(E) = 1 - \frac{\expval{I_N[-E,E]}_a}{N} \, ,
    \label{eq:comp_Poisson}
\end{equation}
so that in general $\chi(E)\sim 1$ for small $E$, and $\chi(E)\to 0$ for large $E$.

Finally, we note that \cref{eq:cgf_iid} is exact, because we have stressed above that the Gaussian fluctuations around the saddle-point are at least of $\order{\eta}$ and so they vanish in the limit $\eta \to 0$.

\subsubsection{Limit of a pure GOE matrix}
\label{par:comp_pureGOE}
The opposite limit in which $A$ is neglected in \cref{eq:hamiltonian} is obtained by setting $p_a(a)=\delta(a)$, whose corresponding resolvent is $\cor{G}_a(z) = 1/z$ (indeed, summing zeros to the matrix $B$ in \cref{eq:hamiltonian} does not change its spectrum). Equation \eqref{eq:K_def} then implies $\hat K = \hat C$, where we used $\hat L^2 = \mathbb{1}$ (see \cref{eq:lambda_L_def}). The self-consistency equations \eqref{eq:four_set_p(a)} are then seen to coincide with \cref{eq:four_set_GOE}, corresponding to the GOE case studied in \ccite{Metz_2017} and here revisited in \cref{app:GOE}. From \cref{eq:action_comp_symmetric_interval} we obtain, in terms of $r$ and $\theta$ introduced in \cref{eq:symmetries},
\begin{align}
    \cor{S}[\varphi_0 ; \hat \Lambda]= -\frac{2\eta r^2 s}{\pi} \sin 2\theta -  \frac{2s\theta}{\pi}  \, ,
    \label{eq:action_GOE}
\end{align}
which is linear in $s$: we deduce that the cumulants higher than the average $\expval{I_N[-E,E]}$ are subleading for large $N$, and they are only accessible by explicitly computing the Gaussian fluctuations around the saddle-point (see \cref{app:gaussian_fluctuations}).
Still, the leading order term in the first cumulant (see \cref{eq:k1_GOE}) is correctly reproduced by \cref{eq:action_GOE}.

\subsection{Exactly solvable cases}
\label{par:comp_exactly_solvable}

The cumulant generating function we found in \cref{eq:cgf_GRP} is formal, in that $\Delta = re^{i\theta}$ must first be determined by solving the self-consistency equation \eqref{eq:four_set_p(a)}. In the following, we will present two cases in which the result can be expressed in closed form, namely those in which $p_a(a)$ is the Cauchy or the Wigner distribution, respectively. While the former presents fat tails and is thus slowly decaying, the latter has a compact support and a well-defined edge.

\subsubsection{Cauchy distributed $a_i$}

Let us start from the case in which $p_a(a)$ is the Cauchy distribution, see \cref{eq:lorentzian}.
We set $\mu=0$, so that the problem remains symmetric around the origin. Using the expression of $\cor{G}_a(z)$ in \cref{eq:cauchy_resolvent}, $\hat K$ in \cref{eq:K_def} reduces to
\begin{equation}
    \hat K = \frac{\hat C}{1+\omega \hat C} \, .
    \label{eq:K_cauchy}
\end{equation}
Solving the self-consistency equation \eqref{eq:four_set_p(a)} then yields
\begin{equation}
    \Delta=\eval{\Delta_\alpha}_{\alpha=-E}  = 2 \left[ \varepsilon -iE -\omega + \sqrt{8\eta+\left(\varepsilon -iE +\omega \right)^2 } \right]^{-1} \equiv re^{i \theta} \, ,
\end{equation}
where we chose the positive branch of the square root so that $\Re \Delta_\alpha >0$. After this choice, the constant $\varepsilon$ can be safely sent to zero. The resulting cumulant generating function is given in \cref{eq:cgf_GRP}, where the averages are taken over the Cauchy distribution in \cref{eq:lorentzian}, and the coefficient of the term linear in $s$ reads
\begin{equation}
    m= \frac{4\eta r^2 \sin \theta \left( r \omega +\cos\theta \right)}{\pi \left[ 1+(r \omega)^2+2 r \omega \cos\theta  \right]^2}\, .
\end{equation}

\begin{figure}[t]
    \centering
    \subfloat[]{
    \includegraphics[width=0.47\textwidth]{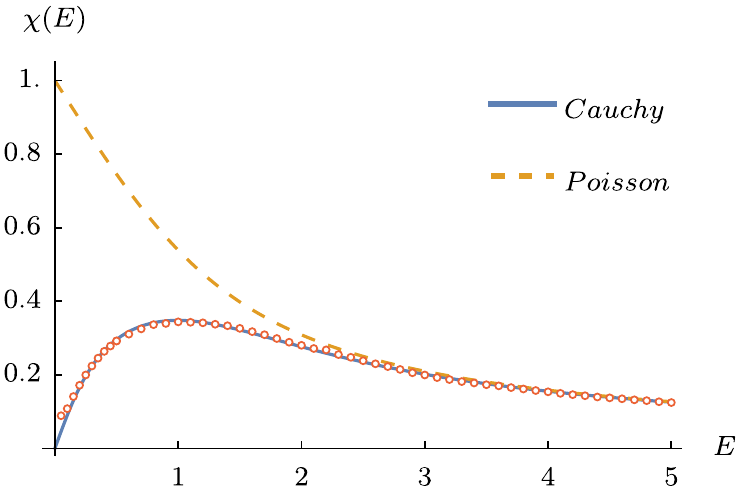}
    \label{fig:cauchy_comp}
    }
    \subfloat[]{
     \includegraphics[width=0.47\textwidth]{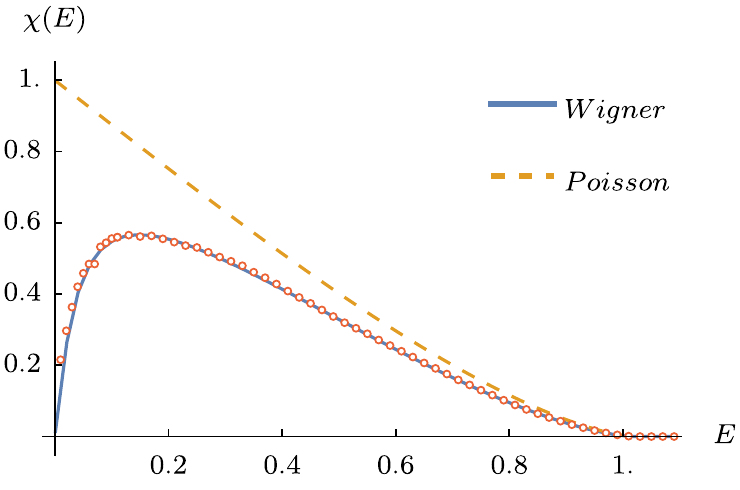}
    \label{fig:wigner_comp}
    }
    \caption{Numerical check of the level compressibility $\chi(E)$ predicted in \cref{eq:comp_general_prediction} (solid blue line), for the case in which the elements of the diagonal matrix $A$ belong to the (a) Cauchy, or (b) Wigner distribution. Numerical data (symbols) are obtained from the numerical diagonalization of $N_\T{tot}=1000$ random matrices of size $N=5000$. The (dashed yellow) line denoted by \textit{Poisson} shows the level compressibility as it would be in the absence of level repulsion -- see the main text. 
    In both plots we used the parameters $\mu=0$, $\gamma=1.1$. In (a) we chose $\omega=1$ and $\nu=1$, corresponding to $\eta=0.125$, while in (b) we set $\sigma=1$ and $\nu=0.2$, yielding $\eta=0.005$.}
    \label{fig:comp_numerics}
\end{figure}

The resulting level compressibility $\chi(E)$ (see \cref{eq:comp_general_prediction}) is plotted in \cref{fig:cauchy_comp}, and it is compared to numerical results showing excellent agreement. We show in the same plot the level compressibility under the hypothesis that the level statistics is of the Poisson type, \ie, that the energy levels do not repel each other. This is given by \cref{eq:comp_Poisson} upon interpreting the average $\expval{\bullet}$ as taken over the average eigenvalue density $\rho(\lambda)$ for the case in which the matrix $A$ has Cauchy-distributed entries: this has been found previously in \cref{eq:density_cauchy}. Even for very small values of $\eta$, the behavior at low energies $E$ of the compressibility $\chi (E)$ is qualitatively very different: in the Poisson case we have $\chi (E)\sim 1$, while in the GRP model it is $\chi (E) \sim 0$. The latter $\chi(E)$ increases up to a maximum, whose position $E_\T{max}(\eta)$ grows monotonically (and sublinearly) with $\eta$. 

\subsubsection{Wigner distributed $a_i$}
Let us now consider the case in which $p_a(a)$ is the Wigner distribution, see \cref{eq:wigner}.
Again we set $\mu=0$, so that the problem remains symmetric around the origin. Using the expression of $\cor{G}_a(z)$ in \cref{eq:wigner_resolvent}, the quantity $\hat K$ in \cref{eq:K_def} becomes
\begin{equation}
    \hat K = - \frac{2}{\sigma^2 \hat C}\left[  1-\sqrt{1+(\sigma \hat C)^2} \right] \, .
    \label{eq:K_wigner}
\end{equation}
Solving the self-consistency equation \eqref{eq:four_set_p(a)} yields, after choosing the positive branch of the square root so that $\Re \Delta_\alpha >0$ and letting $\varepsilon\to 0$,
\begin{equation}
    \Delta = \eval{\Delta_\alpha}_{\alpha=-E}  = \frac{iE(\sigma^2+4\eta) + 4\eta \sqrt{\sigma^2+8\eta-E^2}}{16\eta^2+\sigma^2E^2} \equiv re^{i \theta} \, .
\end{equation}
The resulting cumulant generating function is again given in \cref{eq:cgf_GRP}, where the averages are taken over the Wigner distribution in \cref{eq:wigner}. The analytical expression of the quantity $m$ in \cref{eq:m_def} is cumbersome, but it follows readily from \cref{eq:K_wigner}.
In \cref{fig:wigner_comp} we plot the level compressibility $\chi (E)$ as we did for the Cauchy case, finding a qualitatively similar behavior.

\subsection{Scaling limit and Thouless energy}
\label{par:scaling}
In this Section we focus on the limit in which $E=x \eta^\delta$ and $\eta \ll 1$, while $x\sim \order{1}$. We can envision a different behavior depending on whether the exponent $\delta>1$, $\delta<1$ or $\delta=1$. The latter case turns out to be particularly interesting: we will show that the level compressibility computed in $\chi(E=x\eta^\delta)$ assumes for $\delta=1$ a \textit{universal} scaling form, which is independent of the particular choice of the distribution $p_a(a)$ of the entries of the diagonal matrix $A$.

To this end, let us go back to the self-consistency equations \eqref{eq:self_consistency_p(a)_symm}, which we can rewrite for $\delta=1$ and $\eta \ll 1$ as
\begin{align}
    \Delta^{-1} &= \varepsilon-ix \eta -2i \eta \cor{G}_a\left(-i \Delta^{-1} \right) \n\\
    &\simeq \varepsilon-ix \eta -2i \eta \cor{G}_a\left(-i \varepsilon + \order{\eta} \right) \, ,
    \label{eq:SCE_scaling}
\end{align}
where again $\Delta=r e^{i\theta}$.
By taking the complex conjugate of \cref{eq:SCE_scaling} and by summing and subtracting the two equations, we obtain the two conditions
\begin{align}
    r^{-1}  \cos\theta &= \varepsilon + 2 \eta \Im \cor{G}_a\left(-i \varepsilon + \order{\eta} \right) \xrightarrow[\varepsilon\to 0^+]{\eta \ll 1} 2 \pi \eta p_a(0) \, , \\
    r^{-1}  \sin\theta &= x \eta + 2 \eta \Re \cor{G}_a\left(-i \varepsilon + \order{\eta} \right) \xrightarrow[\varepsilon\to 0^+]{\eta \ll 1} x \eta \, , 
\end{align}
where we used the Plemelj-Sokhotski formula recalled in \cref{eq:plemelj}, and the fact that for a symmetric distribution $p_a(a)$ one has
\begin{equation}
    \pv{\int \dd{a} \frac{p_a(a)}{a}} = 0 \, .
\end{equation}
One then easily obtains, at leading order for small $\eta$ (and with $x=E/\eta$),
\begin{align}
    \tan\theta &\simeq \frac{x}{2\pi p_a(0)} \equiv y \, , \\
    r^{-1}  &\simeq 2\pi \eta p_a(0) \sqrt{1+y^2} \, . \label{eq:rho_scaling}
\end{align}
\rev{Note that these manipulations are only possible under the additional assumption that the distribution $p_a(a)$ behaves regularly close to $a=0$, and that $p_a(0)\neq 0$.} 

\begin{figure}[t]
    \centering
    \subfloat[]{
    \includegraphics[width=0.47\textwidth]{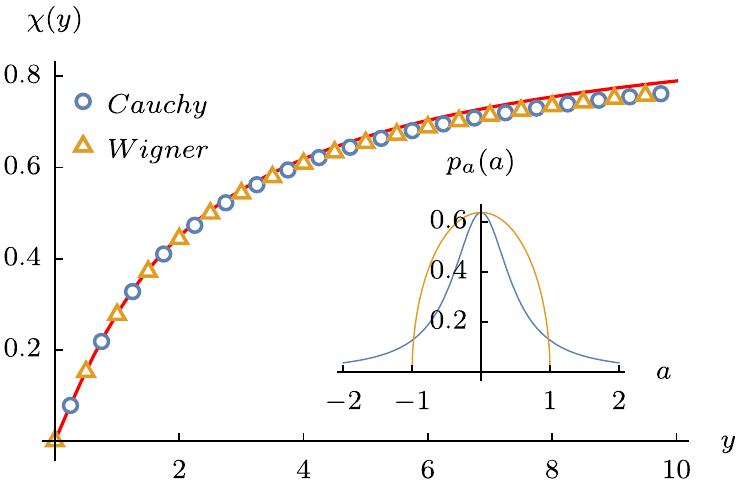}
    \label{fig:scaling}
    }
    \subfloat[]{
     \includegraphics[width=0.47\textwidth]{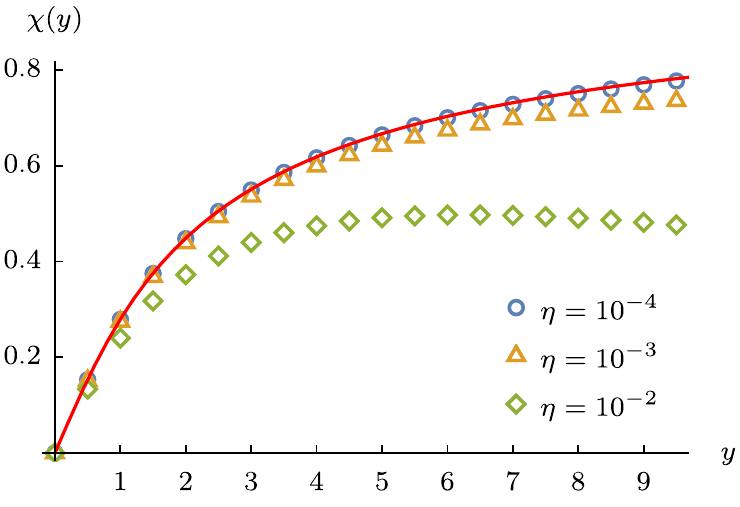}
    \label{fig:collapse_cauchy}
    }
    \caption{Scaling form of the level compressibility, in the limit in which $E\propto y \eta$ and $\eta \ll 1$. In (a) we compare the universal prediction in \cref{eq:comp_universal} (solid red line) with the two exactly solvable cases studied in \cref{par:comp_exactly_solvable} (symbols), showing a good agreement at low energies $y =E/[2\pi p_a(0) \eta]$ (\ie, the condition under which \cref{eq:comp_universal} was derived). We chose $\eta=5\cdot 10^{-4}$ and $\omega=0.5$, $\sigma=1$, so that $p_a(0)$ assumes the same value for the two distributions (see inset). In (b) we exemplify in the Cauchy case how the curves corresponding to different values of $\eta$ collapse onto the same master curve, when plotted as a function of $y\propto E/\eta$ for $\eta\ll 1$.}
    \label{fig:scaling_chi}
\end{figure}

We can now estimate the level compressibility in this limit. First, notice that 
\begin{equation}
    \cor{G}_a(-i\Delta^{-1}) = i\pi p_a(0) +\order{\eta} \, ,
\end{equation}
so that from \cref{eq:m_def} we read
\begin{equation}
    m = - \frac{2\eta}{\pi} \Im \left[ \cor{G}_a(-i\Delta^{-1}) \right]^2 = \order{\eta^2} \, .
\end{equation}
From \cref{eq:first_cumulant}, the first cumulant $\kappa_1$ thus reduces to
\begin{align}
    \frac{\kappa_1}{N} &\simeq \expval{f(a)}_a = \int \dd{a} p_a(a) f(a) \simeq \frac{p_a(0)}{r} \int_{-\infty}^\infty \dd{u} f(u/r) \n\\
    &= \frac{p_a(0)}{r} \frac{2y}{\sqrt{1+y^2}} = 2 p_a(0) x \eta  \, ,
    \label{eq:k1_scaling}
\end{align}
where in the first line we changed variable to $u= r a$ and we used the fact that $r^{-1} \sim \order{\eta}$ (see \cref{eq:rho_scaling}), while in the second line we explicitly computed the integral
\begin{equation}
    \frac{1}{\pi} \int_0^\infty \dd{u} \arctan \left( \frac{\sin 2\theta}{u^2+\cos 2\theta} \right) = \frac{\tan \theta}{\sqrt{1+\tan^2 \theta}} =\sin{\theta} \, , \;\;\;\;\;\;\; \theta \in \left[0,\frac{\pi}{2}\right] \, ,
\end{equation}
and we inserted the expression for $r$ found in \cref{eq:rho_scaling} (note that $f(u/r)$ is actually $r$-independent -- see \cref{eq:f_def}). One could alternatively compute $\kappa_1$ by taking the average of \cref{eq:levels_number}: this leads to the same result upon expanding for small $E$ and $\eta$, since
\begin{equation}
    \rho(\lambda) = p_a(\lambda) + \order{\eta} \, .
\end{equation}
The same steps can be repeated for the second (and possibly any other) cumulant $\kappa_2$, yielding
\begin{equation}
    \frac{\kappa_2}{N} \simeq \expval{f^2(a)}_a  \simeq \frac{p_a(0)}{r} \int_{-\infty}^\infty \dd{u} f^2(u/r) \, .
    \label{eq:k2_approx}
\end{equation}
From \cref{eq:comp_general_prediction} we thus obtain the leading order estimate for the level compressibility when $\eta \ll 1$, which takes the universal scaling form
\rev{
\begin{eqnarray}
    \chi(E)  
    &\simeq& \chi_T\left( y= \frac{E}{2\pi p_a(0) \eta} \right) \, ,
    \nonumber\\
    \chi_T(y)
    & \equiv & \frac{\sqrt{1+y^2}}{\pi^2 y} \int_0^\infty \dd{u} \left\lbrace \arctan \left[\frac{2y}{u^2(1+y^2)+1-y^2 }\right] \right\rbrace^2  ,
    \label{eq:comp_scaling_calculation}
\end{eqnarray}
}
where we stress that we have chosen the branch $\arctan(z) \in [0,\pi]$. Upon integrating by parts and performing some algebra \cite{stack,table}, the integral over $u$ can be computed explicitly to give
\begin{equation}
    \chi_T(y) = \frac{1}{\pi y}\left[2y \arctan(y) -\ln (1+y^2)   \right] \, .
    \label{eq:comp_universal}
\end{equation}
The function $\chi_T(y)$ grows monotonically from $0$ to $1$ as we increase $y$, and it is plotted in \cref{fig:scaling} together with the level compressibility for the two cases explicitly solved above, \ie, Cauchy and Wigner. We find a good agreement at low energies $y$, while we observe a departure at large energies: here the scaling prediction keeps growing, while the actual compressibility must hit a maximum and start decreasing -- see \cref{fig:comp_numerics}. The same trend can be observed in \cref{fig:collapse_cauchy}, where we evaluate the level compressibility for different values of $\eta$, and show that they collapse on a common master curve when they are plotted as a function of $y \propto E/\eta$.

The other two cases ($\delta>1$ or $\delta<1$) can be easily addressed by the same token. When $\delta>1$, by studying the self-consistency equations as in \cref{eq:SCE_scaling} we obtain at leading order
\begin{align}
    &\tan\theta \simeq  y \eta^{\delta-1} \, , \\
    &r^{-1}  \simeq 2\pi \eta p_a(0)  \, .
\end{align}
It can be readily seen that $\kappa_1/N \simeq 2p_a(0)x\eta^\delta$ and $\kappa_2/N \sim \order{\eta^{2\delta-1}}$, so that in this limit
\begin{equation}
    \chi(E=x\eta^\delta) \sim \order{\eta^{\delta-1}} \, , \;\;\;\;\;\;\; \delta >1 \, , \; \eta \ll 1 \, .
    \label{eq:comp_lowE}
\end{equation}
This resembles the behavior of $\chi(E)$ in the case of a pure GOE matrix, see \cref{par:comp_pureGOE,app:GOE}.
Conversely, for $\delta<1$ the self-consistency equations \eqref{eq:self_consistency_p(a)_symm} reduce to
\begin{equation}
    \Delta^{-1} \simeq \varepsilon -ix\eta^\delta \, ,
\end{equation}
and, by comparison with \cref{eq:SCE_iid}, we identify this limit as that in which the eigenvalues behave as i.i.d. random variables. In particular (compare with \cref{eq:comp_Poisson}), 
\begin{equation}
    \chi(E=x\eta^\delta) \simeq 1- \frac{\expval{I_N[-E,E]}_a}{N} \, , \;\;\;\;\;\;\; \delta <1 \, , \; \eta \ll 1 \, .
    \label{eq:comp_largeE}
\end{equation}

The above analysis suggests to interpret the quantity $E_T \sim 2\pi p_a(0) \eta $ as the \textit{Thouless energy} of the system. Indeed, consider again the limit in which $\eta\ll 1$, and let $E\propto \eta^\delta$. For $E\ll E_T$ (\ie, $\delta>1$), the eigenvalues organize in multiplets (or mini-bands \cite{Kravtsov_2015}) and they repel each other as in the GOE ensemble -- as a result, the level compressibility is zero at leading order (see \cref{eq:comp_lowE}). For $E\gg E_T$ ($\delta<1$), on the other hand, the various multiplets no longer interact, and we recover the Poisson statistics -- see \cref{eq:comp_largeE}. Finally, the case $\delta=1$ marks a crossover in which the level compressibility $\chi(E/E_T)$ assumes the universal scaling form given in \cref{eq:comp_universal}.
Indeed, the asymptotics of the function $\chi_T(y)$ in \cref{eq:comp_universal} can be checked to give
\begin{equation}
    \chi_T(y) \simeq
    \begin{dcases}
        \frac{y}{\pi} \, , & y \ll 1 \, , \\
        1- \frac{2(1+\ln y)}{\pi y} \, , & y \gg 1 \, ,
    \end{dcases}
\end{equation}
showing that $\chi_T(y)$ interpolates between Wigner-Dyson statistics at low energy, and Poisson statistics at higher energy.

\rev{We finally note that a close relative of the level compressibility, namely the \textit{two-level spectral correlation function} $C(t,t')$ -- see Eqs.~(\ref{eq:C1})-(\ref{eq:C3}) for its definition -- was computed in \ccite{Kravtsov_2015} for the Hermitian GRP model. In the latter, the GOE matrix $B$ in \cref{eq:hamiltonian} is replaced by a GUE matrix with complex entries, so that additional analytical techniques (notably the Harish-Chandra-Itzykson-Zuber
integral \cite{Mehta_2004_book}) are available. In particular, $C(t,t')$ is shown in \ccite{Kravtsov_2015} to assume a universal scaling form within the fractal region $1<\gamma<2$, and for large $N$. 
In \cref{app:Kravtsov} we show that the corresponding scaling form of the level compressibility coincides, in the crossover regime in which $E\sim E_T$, with $\chi_T(y)$ in \cref{eq:comp_universal}. 
This is quite remarkable, since these are in fact two distinct random matrix ensembles -- and indeed their level compressibilities do not coincide for $E \gg E_T$ or $E\ll E_T$.
This identification suggests that $\chi_T(y)$ originates from the structural properties of the model, rather than from the specific choice of the matrix $B$ (e.g., GOE, GUE, but also possibly Wishart or sparse random matrices).}

\begin{figure}[t]
    \centering
        \subfloat[]{
    \includegraphics[width=0.47\textwidth]{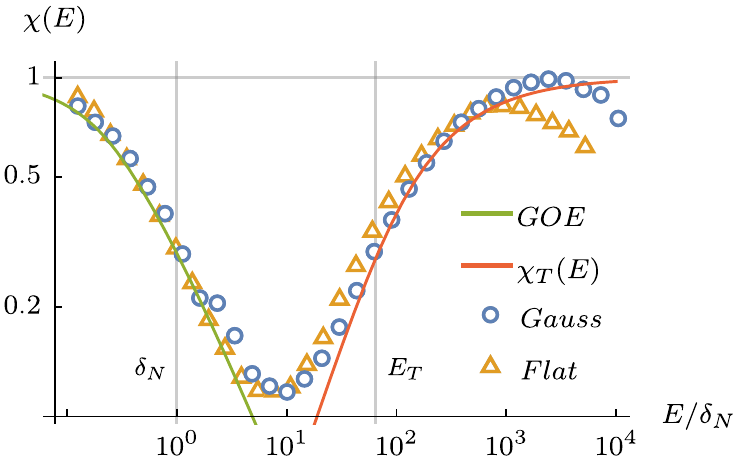}
    \label{fig:numerics_universal}
    }
    \subfloat[]{
     \includegraphics[width=0.47\textwidth]{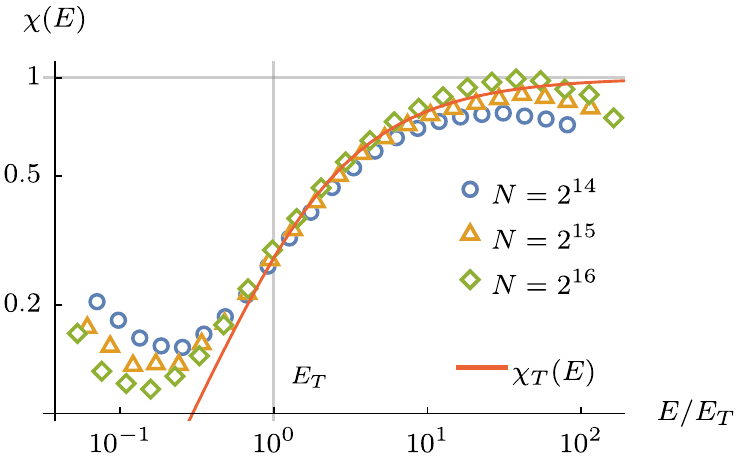}
    \label{fig:numerics_N}
    }
    \caption{Behavior of the level compressibility $\chi(E)$ at low energies. The symbols correspond to numerical results, and we indicated with vertical lines the mean level spacing $\delta_N$ (see \cref{eq:mean_level_spacing}) and the Thouless energy $E_T \sim N^{1-\gamma}$. In panel (a), $p_a(a)$ is chosen Gaussian with unit variance or uniform, and $N=2^{16}$. The region $E \lesssim \delta_N$ is described by the GOE prediction in \cref{chi_GOE_final}, while the crossover region $E\sim E_T$ is described by the universal scaling form in \cref{eq:comp_universal}. In panel (b), $p_a(a)$ is Gaussian, and we show the approach to the universal curve $\chi_T(E)$ for increasing values of the matrix size $N$. We used $\gamma=1.5$ throughout, and the simulations with $N=2^{14},2^{15},$ or $2^{16}$ are averaged over $N_\T{tot}=256,64,$ or $32$ samples, respectively.}
    \label{fig:numerics_scaling}
\end{figure}

\subsection{Behavior \rev{for small $E$}}
\label{par:low_energy}
In this final Section we use extensive numerical exact diagonalization of large random matrices in order to inspect the low-energy behavior of the level compressibility $\chi(E)$. Indeed, our prediction of \cref{par:scaling} is expected to break down for energies of the order of the mean level spacing $\delta_N$ of the finite-sized matrix $\cor{H}$, which is given by $\delta_N \simeq [N p_a(0)]^{-1}$, Eq.~\eqref{eq:mean_level_spacing}.
Equivalently, we expect that \rev{the leading order term in the saddle-point approximation adopted in our} replica calculation \rev{(see \cref{eq:cgf,eq:Q_n_gaussian})} provides the correct result in the $N \to \infty$ limit, while for a matrix of size $N$ and for sufficiently small $E$ we should eventually recover the exact GOE result $\chi_\T{GOE}(E)$ \cite{Dyson_1962_I,Dyson_63_IV,Mehta_2004_book,Forrester_2010_book,Mirlin_2000,marino2014phase,Marino_2016, Tikhonov_2019}, whose derivation is reported in \cref{app_chi_GOE} -- see \cref{chi_GOE_final}.
The overall picture is thus the one we present in \cref{fig:numerics_scaling}, and which we support by numerical results. The region $E \lesssim \delta_N$ is described by the GOE prediction in \cref{chi_GOE_final}. For $1< \gamma < 2$ the Thouless energy $E_T \sim N^{1-\gamma}$ is such that $\delta_N \ll E_T \ll 1$, so that the crossover region with $E \sim E_T$ is described by the universal function $\chi_T(E)$ given in \cref{eq:comp_universal}. For larger $E \gtrsim \order{1}$, the level compressibility becomes model-dependent and it is described by \cref{eq:comp_general_prediction} (see also \cref{fig:comp_numerics}).

The datapoints\footnote{Although the scaling function in \cref{eq:comp_universal} has been derived under the assumption that the spectral density is symmetric, the numerical results are obtained by averaging the cumulants of the number of eigenvalues within many energy windows across the whole bandwidth, for which the symmetry with respect to the center of the window is lost. However, on the scale of the Thouless energy $E \simeq E_T \propto \eta$, the corrections due to the fact that $p_a(-E) \neq p_a(E)$ are of $\order{\eta}$, and therefore yield a contribution which is of the same order as the finite-size corrections, and which can be neglected for sufficiently large $N$.} presented in panel (a) of \cref{fig:numerics_scaling} correspond to the choices of $p_a(a)$ Gaussian or uniform, which supports our claim of universality of $\chi_T(E)$ in the region $E\sim E_T$. Note, in fact, that the data follow the predicted curves (up to finite-size effects) with no adjustable parameters (\ie, no fitting was needed). The datapoints are eventually observed to deviate from the scaling prediction, as they reach a maximum \rev{in correspondence of $E_\text{max}(\eta)\gg E_\text{T}$} and start decaying to zero as in \cref{fig:comp_numerics}. By increasing $N$, however, this maximum is observed to shift to\rev{wards} larger values of $E$, and the plateau around $\chi(E)\sim 1$ broadens accordingly.

\section{Conclusions}

In this work we used the replica method to study the average spectral density (\cref{par:spectral_replica_approach}) and the local level statistics (\cref{par:level_compress}) of a deformed GOE random matrix ensemble known as the generalized RP model. We focused on its fractal intermediate phase with $1 < \gamma < 2$ \cite{Kravtsov_2015}, which is conveniently characterized in terms of the level compressibility $\chi(E)$ (see \cref{eq:level_compress}). We  showed that $\chi(E)$ assumes a universal form \rev{$\chi_T(E/E_T)$} independent\rev{ly} of the character $p_a(a)$ of the deformation matrix $A$ (see \cref{eq:hamiltonian}), provided that the system is probed over energy scales of the order of the Thouless energy $E_T$ (see \cref{par:scaling}).

It is natural to conjecture that this universal regime should persist in structurally similar random matrix ensembles \rev{(even more so, since we showed that the \textit{same} scaling function $\chi_T(y)$ can be recovered for the Hermitian GRP model -- see \cref{app:Kravtsov})}. For instance, one could numerically inspect the case in which the GOE matrix is replaced by a Wigner matrix (\ie, any real symmetric matrix with i.i.d. random entries taken from a probability distribution with finite variance), whose limiting average spectral density is still given by the semi-circle law \cite{arous2011wigner}. Similarly, it would be interesting to study the effect of the diagonal deformation matrix $A$ on a Wishart matrix \cite{potters2020first}, or else on a sparse (rather than dense) matrix $B$, such as those describing the Erdös-Rényi random graph \cite{Semerjian_2002,Rogers_2008,Metz_2014},
which can still be treated analytically (at least to some extent); 
if the average connectivity is chosen
\rev{to be} finite, the spectral density of Erdös-Rényi is no longer a semi-circle, but the local statistics is still of the Wigner-Dyson type. 

\rev{Along our derivation, we heavily relied on the \textit{independence} of the elements $a_i$ characterizing the diagonal disorder (as usually assumed in standard formulations of the GRP model). However, the introduction of short-ranged correlations between the levels $a_i$ seems within reach of the replica method. In particular, the analysis of \ccite{DeTomasi_2022} suggests that changing the power $d$ at which the distance between sorted diagonal elements $a_n<a_{n+1}$ grows, e.g., $(a_{n+k}-a_n)\propto k^d$, has important implications on the phase diagram. Moreover, in \cref{par:scaling} we assumed a regular character of $p_a(a)$ in correspondence of $a\to 0$, while it would be interesting to check the fate of the universal scaling form $\chi_T(E)$ upon choosing a singular (but normalizable) distribution $p_a(a)$. Similarly, the analysis in \cref{par:scaling} suggests that the particular choice of a distribution $p_a(a)$ such that $p_a(0)=0$ may not be innocent.}

Finally, it is worth mentioning that the spectral properties of the intermediate phase of the generalized RP model are particularly simple: differently from realistic interacting quantum systems, the mini-bands are compact and the eigenvectors are fractal but not \textit{multi}fractal (meaning that all the moments of the wave-functions' amplitudes are described by the same fractal exponent $D_\gamma = 2 - \gamma$). In order to overcome some of these issues, several extensions  of the RP model have been proposed in the last few years, featuring either log-normal~\cite{kravtsov2020localization,khaymovich2020fragile} or power-law~\cite{biroli2021levy} distributed off-diagonal matrix elements. \rev{Recent developments suggest, however, that the only way to obtain multifractality is to introduce correlations between the matrix elements of $\cor{H}$ (either the diagonal or the off-diagonal ones \cite{Kutlin_2023}).} 
It would therefore be illuminating to study the behavior of the level compressibility at small energy \rev{in these generalizations of the RP model}, and check whether or not the universal form discussed here is robust with respect to these modifications.

\section*{Acknowledgements}
We thank I. Khaymovich, F. Metz, M. Potters, and P. Vivo for illuminating insights. DV would like to thank F. Ares and L. Capizzi for interesting discussions, and the LPTHE for the kind hospitality during a substantial part of the preparation of this work.

\paragraph{Funding information}
DV acknowledges partial financial support from Erasmus+ 2020-1-IT02-KA103-078180.
LFC acknowledges partial financial support from FCM2b ANR-19-CE30-0014. GS acknowledges partial financial support from ANR Grant No. ANR-17-CE30-0027-01 RaMaTraF.

\begin{appendix}
\numberwithin{equation}{section}

\section{Number of i.i.d. variables in an interval}
\label{app:iid}
In this Appendix we revise the standard textbook result for the statistics of the number of eigenvalues in a finite interval, when such eigenvalues behave as independent and identically distributed random variables. This will help clarifying the behavior of the level compressibility $\chi(E)$ in the case of Poisson level statistics.

Given $N$ random variables $a_i$ distributed according to $p_a(a_i)$ (\eg, the eigenvalues of the matrix $A$ in \cref{eq:hamiltonian}), the number of variables contained in the interval $[\alpha,\beta]$ can be written as
\begin{equation}
    I_N[\alpha,\beta] = \sum_{i=1}^N \theta_i \, ,
\end{equation}
where we introduced the indicator function
\begin{equation}
    \theta_i \equiv \mathbb{1}_{[\alpha,\beta]}(a_i) \, , \;\;\;\;\;\qquad \mathbb{1}_{[\alpha,\beta]}(x) =
    \begin{cases}
        1 & x\in [\alpha,\beta] \, , \\
        0 & x \not\in [\alpha,\beta] \, .
    \end{cases}
\end{equation}
Its cumulant generating function can be constructed by noting that
\begin{equation}
    e^{-s I_N[\alpha,\beta]} = \prod_{i=1}^N e^{-s \theta_i} = \prod_{i=1}^N \left[1+(e^{-s}-1) \theta_i \right] \, ,
\end{equation}
and then
\begin{equation}
    \ln \expval{e^{-s I_N[\alpha,\beta]}} =  \sum_{i=1}^N \ln  \left[1+(e^{-s}-1) \int_\alpha^\beta \dd{a} p_a(a) \right] = N \ln \left[1+(e^{-s}-1) \frac{\expval{I_N[\alpha,\beta]}}{N} \right] \, .
    \label{eq:cgf_iid_textbook}
\end{equation}
Note that this coincides with the limiting case in \cref{eq:cgf_iid} of our general result.

By expanding in powers of $s$ the two sides of \cref{eq:cgf_iid_textbook} and comparing with \cref{eq:cgf,eq:cumulants}, we can in particular extract the first two cumulants
\begin{equation}
    \kappa_1 =  \expval{I_N[\alpha,\beta]} \, , \;\;\;\;\;\qquad \kappa_2 = \expval{I_N[\alpha,\beta]} \left(1 - \frac{\expval{I_N[\alpha,\beta]}}{N}\right) \, ,
\end{equation}
and from \cref{eq:level_compress} we obtain the level compressibility
\begin{equation}
    \chi(E) = 1 - \frac{\expval{I_N[-E,E]}}{N} \, .
\end{equation}
We thus generically expect $\chi(E)\sim 1$ for small $E$, and $\chi(E)\to 0$ for large $E$.

\section{Details of the replica calculation of the spectral density}
\label{app:replica}

In this Appendix we fill in the missing steps which lead from \cref{eq:replica_trick} to \cref{eq:average_Zn} in \cref{par:spectral_replica_approach}. A replica-based calculation for the pure GOE ensemble can be found in \ccite{Livan_2018}, from which we partially adopt the notation. We start by expressing the average of the replicated partition function as
\begin{equation}
    \expval{ \cor{Z}^n(\lambda) } \propto \expval{ \int_{\mathbb{R}^{Nn}} \left( \prod_{\alpha=1}^n \dd{\vb{r}_\alpha}\right) \exp[-\frac{i}{2} \sum_{i,j=1}^N \sum_{\alpha=1}^n r_{i\alpha} (\lambda_\varepsilon \delta_{ij} -h_{ij} )r_{j\alpha} ]}_{A,B},
    \label{eq:representation_Zn}
\end{equation}
where we indicated by $h_{ij}\equiv a_i \delta_{ij}+J b_{ij}$ the elements of the random matrix ${\mathcal H}$ given in \cref{eq:hamiltonian}, and $J= J(N) \equiv \nu N^{-\gamma/2}$. The average symbol means
\begin{equation}
    \expval{\bullet}_{A,B} \equiv \int_{\mathbb{R}^N} \left(\prod_{i\leq j}^N \dd{b_{ij}}\right) p_B(\{b_{ij}\}) \left( \prod_{i=1}^N \int \dd{a_i} p_a(a_i) \right) \left( \bullet \right) \, ,
    \label{eq:averages_def}
\end{equation}
where the probability distribution of the elements $b_{ij}$ of the GOE matrix $B$ reads
\begin{equation}
    p_B(\{b_{ij}\}) = \prod_{i=1}^N \frac{e^{-b_{ii}^2/2}}{\sqrt{2\pi}} \prod_{\rev{i<j} } \frac{e^{-b_{ij}^2}}{\sqrt{\pi}} \, .
\end{equation}
Here and henceforth, Latin indices run up to $N$ in real space, while Greek indices run up to $n$ in replica space.
Computing the Gaussian integrals over $b_{ij}$ gives (up to a numerical constant)
\begin{align}
    \expval{ \cor{Z}^n(\lambda) } \propto & \int_{\mathbb{R}^{Nn}} \left( \prod_{\alpha=1}^n \dd{\vb{r}_\alpha}\right) \Bigg\langle \exp[-\frac{i}{2} \sum_{i=1}^N \sum_{\alpha=1}^n (\lambda_\varepsilon -a_i )r_{i\alpha}^2 ] \Bigg\rangle_{A} \n \\
    & \times \exp{-\frac{J^2}{4} \left[\frac12 \sum_{i=1}^N \left( \sum_{\alpha=1}^n r_{i\alpha}^2 \right)^2 + \sum_{\rev{i<j} } \left( \sum_{\alpha=1}^n r_{i\alpha} r_{j\alpha} \right)^2  \right] } \, ,
    \label{eq:after_gaussian_integrals}
\end{align}
where $\expval{\bullet}_{A}$ indicates the reduced averaged over the entries of $A$ -- see \cref{eq:averages_def}. The interacting term in the second line can be usually decoupled by means of the Hubbard-Stratonovich transformation \cite{Mezard_1987}, which is however ineffective in our case, for a generic choice of $p_a(a)$.
We introduce instead the normalized density
\begin{equation}
    \mu(\vec{y}) \equiv \frac{1}{N} \sum_{i=1}^N \prod_{\alpha=1}^n \delta (y_\alpha-r_{i\alpha}) \, ,
\end{equation}
where $\vec{y}\in \mathbb{R}^n$ has components $y_a\in \mathbb{R}$, and we insert into \cref{eq:after_gaussian_integrals} the functional integral representation of the identity
\begin{equation}
    1 = N^{\T{dim}(\mu)}\int \cor{D}\mu \, \cor{D}\hat \mu \, \exp{-i \int \dd{\vec{y}} \hat \mu(\vec{y}) \left[ N\mu(\vec{y}) - \sum_{i=1}^N \prod_{\alpha=1}^n \delta (y_\alpha-r_{i\alpha}) \right] } \, .
    \label{eq:identity}
\end{equation}
Here $\T{dim}(\mu)$ is the dimension of the field $\mu$, which renders the prefactor on the right hand side formally infinite -- this will be of no consequence in the following calculation, since this prefactor is $\lambda$-independent. Equation~\eqref{eq:identity} is useful because it allows us to rewrite
\begin{align}
    &\frac12 \sum_{i=1}^N \left( \sum_{\alpha=1}^n r_{i\alpha}^2 \right)^2 + \sum_{i\leq j} \left( \sum_{\alpha=1}^n r_{i\alpha} r_{j\alpha} \right)^2 = \frac12 \sum_{i,j=1}^N \left( \sum_{\alpha=1}^n r_{i\alpha} r_{j\alpha} \right)^2 \\
    &= \frac{N^2}{2}\int \dd{\vec{y}} \dd{\vec{w}} \mu(\vec{y}) \mu(\vec{w}) \left( \sum_{\alpha=1}^n y_\alpha w_\alpha \right)^2 = \frac{N^2}{2}\int \dd{\vec{y}} \dd{\vec{w}} \mu(\vec{y}) \mu(\vec{w}) \left( \vec{y} \cdot \vec{w} \right)^2 \n \, ,
\end{align}
so that inserting the identity in \cref{eq:identity} into \cref{eq:after_gaussian_integrals} leads to
\begin{align}
    &\expval{ \cor{Z}^n(\lambda) } \propto 
    \int \cor{D}\mu \, \cor{D}\hat \mu \, \exp{-i N \int \dd{\vec{y}} \hat \mu(\vec{y}) \mu(\vec{y}) -\frac{(JN)^2}{8}\int \dd{\vec{y}} \dd{\vec{w}} \mu(\vec{y}) \mu(\vec{w}) \left( \vec{y} \cdot \vec{w} \right)^2 }\n \\
    &\times \int_{\mathbb{R}^{Nn}} \left( \prod_{\alpha=1}^n \dd{\vb{r}_\alpha}\right) \expval{ \exp[-\frac{i}{2} \sum_{i=1}^N \sum_{\alpha=1}^n (\lambda_\varepsilon -a_i )r_{i\alpha}^2 +i\sum_{i=1}^N \int \dd{\vec{w}} \hat \mu(\vec{w})  \prod_{\alpha=1}^n \delta (w_\alpha-r_{i\alpha}) ]}_A . \label{eq:huge}
\end{align}
Staring at \cref{eq:huge} for long enough, one realizes that the second line contains $N$ copies of the same integral,
\begin{align}
    &\int_{\mathbb{R}^{Nn}} \left( \prod_{\alpha=1}^n \dd{\vb{r}_\alpha}\right) \expval{  \exp[-\frac{i}{2} \sum_{i=1}^N \sum_{\alpha=1}^n (\lambda_\varepsilon -a_i )r_{i\alpha}^2 +i\sum_{i=1}^N \int \dd{\vec{w}} \hat \mu(\vec{w})  \prod_{\alpha=1}^n \delta (w_\alpha-r_{i\alpha}) ]}_A \n\\
    &=\left\lbrace \int_{\mathbb{R}^n} \dd{\vec{y}} \int \dd{a} p_a (a)  \exp[-\frac{i}{2} \sum_{\alpha=1}^n (\lambda_\varepsilon -a )y_\alpha^2 +i \int \dd{\vec{w}} \hat \mu(\vec{w})  \prod_{\alpha=1}^n \delta (w_\alpha-y_{\alpha}) ] \right\rbrace^N \n\\
    &= \left\lbrace \int_{\mathbb{R}^n} \dd{\vec{y}} \int \dd{a} p_a (a)  \exp[-\frac{i}{2} (\lambda_\varepsilon -a )|\vec{y}|^2 +i \hat \mu(\vec{y}) ] \right\rbrace^N \, .
\end{align}
Note that it was crucial to assume independent entries $a_i$, so that their distribution in \cref{eq:averages_def} is factorized.
Plugging this expression back into \cref{eq:huge} allows us to rewrite $\expval{ \cor{Z}^n(\lambda) }$ as reported in \cref{eq:average_Zn} of the main text.

\section{Connection with the Zee formula}
\label{app:zee}
In this Appendix we show why \cref{eq:resolvent_ledoussal} is hiddenly the Zee formula. In Ref. \cite{Zee_1996}, the recipe for computing the spectrum $\rho_{1+2}(\lambda)$ of the sum of two random matrices $M_1+M_2$ is given as follows:
\begin{enumerate}[(i)]
    \item Compute the resolvents (or Green's functions) associated to $\rho_1(\lambda)$ and $\rho_2(\lambda)$, \ie, $\cor{G}_1(z)$ and $\cor{G}_2(z)$.
    \item Compute their functional inverses $B_1(z)$ and $B_2(z)$, or Blue's functions, defined by
    \begin{equation}
        B(\cor{G}(z))=z \, .
    \end{equation}
    \item Apply the sum rule 
    \begin{equation}
        B_{1+2}(z) = B_1(z) + B_2(z) -1/z \, .
    \end{equation}
    \item Invert the result back (see \cref{eq:inversion_resolvent}) to find
    \begin{equation}
        B_{1+2}(z) \to \cor{G}_{1+2}(z) \to \rho_{1+2}(\lambda) \, .
    \end{equation}
\end{enumerate}
Another interesting object is however the $R$-function, which is simply defined as
\begin{equation}
    R(z) \equiv B(z) - 1/z \, ,
\end{equation}
and which is easily seen to satisfy the free-sum rule \cite{Livan_2018,Voiculescu_1991,voiculescu1992free}
\begin{equation}
    R_{1+2}(z) = R_1(z) + R_2(z) \, .
\end{equation}
It follows that
\begin{equation}
    B_1(x) = B_{1+2}(x) - R_2(x) \, ,
\end{equation}
which we can choose to apply in particular on $x=\cor{G}_{1+2}(z)$, yielding by construction
\begin{equation}
    B_1(\cor{G}_{1+2}(z)) = z - R_2(\cor{G}_{1+2}(z)) \, .
\end{equation}
Applying $\cor{G}_1$ on both sides finally yields
\begin{equation}
    \cor{G}_{1+2}(z) = \cor{G}_1(z-R_2(\cor{G}_{1+2}(z))) \, .
\end{equation}
The analogy with \cref{eq:resolvent_ledoussal} is readily established once we recall that, if $M_2$ is a GOE matrix, then its $R$-function is simply $R_2(z) = z$ \cite{Biane_97}.

\section{Details of the replica calculation of the level compressibility}

In this Appendix we provide the technical steps for the derivation of the cumulant generating function and the level compressibility presented in \cref{par:level_compress}. A similar calculation for the pure GOE/GUE ensemble can be found in \ccite{Metz_2017}, while the derivation in the case of the Erdös-Rényi graph and the Anderson model on a random regular graph was reported in \ccite{Metz_2016}. 

\subsection{Functional representation}
\label{app:replica_level_compress}
The target of this Section is to express $Q_{[\alpha,\beta]}(n_\pm)$ given in \cref{eq:Q_ab} within the replica formalism, as we did in \cref{app:replica}. The first step is to choose a suitable representation for the partition function $\cor{Z}(z)$ which appears in \cref{eq:Q_ab}: indeed, the one we introduced in \cref{eq:partition_function} is only appropriate if $\Im{z}<0$, being the integral not convergent otherwise. If on the contrary $\Im{z}>0$, then one should choose instead
\begin{equation}
    \cor{Z}_+(z) \equiv \left(\frac{i}{2\pi}\right)^{N/2}\int_{\mathbb{R}^N} \dd{\vb{r}} e^{\frac{i}{2} \vb{r}^T (z \mathbb{1} -\cor{H} ) \vb{r} } \, . 
    \label{eq:partition_function_+}
\end{equation}
Since the various prefactors in front of the integral in \cref{eq:partition_function} will cancel out in \cref{eq:Q_ab} after we take the analytic continuation to $n_\pm \to \pm i s/\pi$, we will not need to keep track of them in the following. 

In analogy with the representation in \cref{eq:representation_Zn}, we can still write
\begin{equation}
    Q_{[\alpha,\beta]}(n_\pm) \propto \expval{ \int_{\mathbb{R}^{Nn}} \left( \prod_{\sigma=1}^n \dd{\vb{r}_\sigma}\right) \exp[-\frac{i}{2} \sum_{i,j=1}^N \sum_{\sigma=1}^n r_{i\sigma} (\Lambda_{\sigma \sigma} \delta_{ij} -L_{\sigma \sigma}h_{ij} )r_{j\sigma} ]}_{A,B},
\end{equation}
but now we interpret
\begin{equation}
    n=2(n_+ + n_-) ,
    \label{eq:n}
\end{equation}
because each of the four partition functions in \cref{eq:Q_ab} requires its own set of replicas (here labelled by the Greek index $\sigma$, to avoid confusion with the left boundary $\alpha$ of the interval). We have also replaced the eigenvalue $\lambda_\varepsilon$ by the block matrix $\hat \Lambda$, which is defined in \cref{eq:lambda_L_def} together with the block matrix $\hat{L}$. Notice that the elements $\bar{\alpha}_\varepsilon=-\alpha_\varepsilon^*$ of the matrix $\hat \Lambda$ follow from the representation in \cref{eq:partition_function_+}.

The very same steps which in \cref{app:replica} led us to \cref{eq:average_Zn} of the main text now give
\begin{equation}
    Q_{[\alpha,\beta]}(n_\pm) \propto \int \cor{D}\mu \, \cor{D}\hat \mu \, \exp{N \cor{S}_{n_\pm}[\mu , \hat \mu;\hat \Lambda]  } \, ,
    \label{eq:average_Qn}
\end{equation}
with the action
\begin{align}
    \cor{S}_{n_\pm}[\mu , \hat \mu;\hat \Lambda] \equiv& - i \int \dd{\vec{\tau}} \mu(\vec{\tau}) \hat \mu(\vec{\tau}) - \frac{\eta}{2}  \int \dd{\vec{\tau}} \dd{\vec{\tau}'} \mu(\vec{\tau}) \mu(\vec{\tau}') \left( \vec{\tau} \, \hat L \, \vec{\tau}' \right)^2  \label{eq:action_comp_mu_muhat}\\
    &+ \ln \int \dd{\vec{\tau}} \exp[-\frac{i}{2}\vec{\tau}\, \hat \Lambda \,\vec{\tau} +i  \hat \mu(\vec{\tau}) ] \int \dd{a} p_a (a) \exp(\frac{i}{2} a \vec{\tau} \, \hat L \, \vec{\tau}) \, . \n
\end{align}
This generalizes the action in \cref{eq:action}, and the vector $\vec{\tau}\in \mathbb{R}^n$ plays the same role as the vector $\vec{y}$ but in an extended replica space, with $n$ given in \cref{eq:n}.
By noting that the action in \cref{eq:action_comp_mu_muhat} is quadratic in $\mu$, we can evaluate the Gaussian functional integral in $\cor{D}\mu$ to obtain
\begin{align}
    Q_{[\alpha,\beta]}(n_\pm) \propto & \int  \cor{D}\hat \mu \, \exp{N \cor{S}_{n_\pm}[\hat \mu;\hat \Lambda]  } \, , \label{eq:Q_n_muhat} \\
    \cor{S}_{n_\pm}[\hat \mu;\hat \Lambda] \equiv&  - \frac{1}{2\eta}  \int \dd{\vec{\tau}} \dd{\vec{\tau}'} \hat\mu(\vec{\tau}) M^{-1}(\vec{\tau},\vec{\tau}') \hat\mu(\vec{\tau}')  \label{eq:action_comp_muhat}\\
    &+ \ln \int \dd{\vec{\tau}} \exp[-\frac{i}{2}\vec{\tau}\, \hat \Lambda \,\vec{\tau} +i  \hat \mu(\vec{\tau}) ] \psi_a\left(-\frac{1}{2} \vec{\tau} \, \hat L \, \vec{\tau}\right) \, , \n
\end{align}
where $\psi_a(z)$ was given in \cref{eq:characteristic}, and we introduced the function $M(\vec{\tau},\vec{\tau}')$ as in \cref{eq:M_def}. We omitted from \cref{eq:Q_n_muhat} a $\hat \Lambda-$independent prefactor coming from the Gaussian integration, which will in general depend on $n$. However, one can check that all the prefactors cancel out smoothly by including the Jacobian of the variable transformations (see later), and after taking the functional integral over the Gaussian fluctuations in \cref{app:gaussian_fluctuations}. We will thus avoid reporting these prefactors, so as to lighten the notation.

The saddle-point equation follows simply from \cref{eq:action_comp_muhat} as 
\begin{equation}
    \hat\mu(\vec{\tau}) = i \eta \frac{\int \dd{\vec{\tau}'} M(\vec{\tau},\vec{\tau}') \exp[-\frac{i}{2}\vec{\tau}'\, \hat \Lambda \,\vec{\tau}' +i  \hat \mu(\vec{\tau}') ] \psi_a\left(-\frac{1}{2} \vec{\tau}' \, \hat L \, \vec{\tau}'\right) }{ \int \dd{\vec{\tau}'} \exp[-\frac{i}{2}\vec{\tau}'\, \hat \Lambda \,\vec{\tau}' +i  \hat \mu(\vec{\tau}') ] \psi_a\left(-\frac{1}{2} \vec{\tau}' \, \hat L \, \vec{\tau}'\right)  } \, ,
     \label{eq:saddle_comp_muhat}
\end{equation}
which is analogous to \cref{eq:saddle_tot}. In the following, we will look for an explicit rotationally-invariant solution: to this end, it is useful to introduce the new variable $\varphi(\vec{\tau})$ defined via
\begin{equation}
    \hat\mu(\vec{\tau}) = i \int \dd{\vec{\tau}'} M(\vec{\tau},\vec{\tau}')\varphi(\vec{\tau}') \, .
\end{equation}
Changing variables from $\hat \mu$ to $\varphi$ in \cref{eq:Q_n_muhat} leads to the expression reported in \cref{eq:action_comp_varphi}.

\subsection{Gaussian fluctuations around the saddle-point}
\label{app:gaussian_fluctuations}

In order to go beyond the saddle-point approximation, we introduce the fluctuation $\phi (\vec{\tau})$ around the saddle-point solution $\varphi_0 (\vec{\tau})$ in the form
\begin{equation}
    \varphi(\vec{\tau}) = \varphi_0(\vec{\tau}) + \phi (\vec{\tau}) \, .
\end{equation}
Calling for brevity $\cor{S}_{n_\pm}[\varphi; \hat \Lambda]\equiv \cor{S}[\varphi]$, we then have up to $\order{N^{-2}}$
\begin{equation}
    Q_{[\alpha,\beta]}(n_\pm) \propto e^{N\cor{S}[\varphi_0]} \int  \cor{D}(i\phi)  \exp{\frac{N}{2} \int  \dd{\vec{\tau}_1} \dd{\vec{\tau}_2} \phi(\vec{\tau}_1) \eval{\frac{\delta^2 \cor{S}[\varphi]}{\delta \varphi(\vec{\tau}_1) \delta \varphi(\vec{\tau}_2) }}_{\varphi=\varphi_0} \phi(\vec{\tau}_2) } \, ,
    \label{eq:gaussian_fluc}
\end{equation}
and one can check that we can express
\begin{equation}
    \eval{\frac{\delta^2 \cor{S}[\varphi]}{\delta \varphi(\vec{\tau}_1) \delta \varphi(\vec{\tau}_2) }}_{\varphi=\varphi_0} = \frac{1}{\eta}  M(\vec{\tau}_1,\vec{\tau}_2) \left[ \mathbb{1}(\vec{\tau}_1,\vec{\tau}_2) + T (\vec{\tau}_1,\vec{\tau}_2)  \right] 
    \label{eq:fluctuation_matrix}
\end{equation}
in terms of the functions $M$ and $T$ given in \cref{eq:M_def,eq:T_def}, respectively.
Computing the Gaussian integral in \cref{eq:gaussian_fluc} we thus find
\begin{equation}
    Q_{[\alpha,\beta]}(n_\pm) = \exp{N\cor{S}[\varphi_0]-\frac12 \ln \det (\mathbb{1}+T)}+\order{1/N^2} \, .
\end{equation}
Expanding the logarithm in series as
\begin{equation}
    \ln(1+x) = -\sum_{k=1}^\infty \frac{(-x)^k}{k}
\end{equation}
we finally get \cref{eq:Q_n_gaussian}, where the trace and matrix operations are intended over the replica vectors as
\begin{equation}
    \Tr T = \int \dd{\vec{\tau}} \, T(\vec{\tau},-\vec{\tau}) \, , \;\;\;\;\;\;\;\;\;\; 
    T^2(\vec{\tau}_1,\vec{\tau}_2) = \int \dd{\vec{\tau}} \,T(\vec{\tau}_1,\vec{\tau})T(\vec{\tau},\vec{\tau}_2) \, .
    \label{eq:trace}
\end{equation}
We may try to specialize \cref{eq:Q_n_gaussian} to the rotationally invariant Ansatz in \cref{eq:ansatz_new}. The fluctuation matrix in \cref{eq:fluctuation_matrix} becomes
\begin{equation}
    T (\vec{\tau}_1,\vec{\tau}_2) = \varphi_0(\vec{\tau}_1) \left[ \left( \vec{\tau}_1 \, \hat L \, \vec{\tau}_2 \right)^2 - \left( \vec{\tau}_2 \, \hat K \, \vec{\tau}_2 \right) \right] \, ,
\end{equation}
and by using Wick's theorem together with \cref{eq:property} we obtain, for instance,
\begin{equation}
    \Tr T = \int \dd{\vec{\tau}} T (\vec{\tau},\vec{\tau}) = \eta \left[ \sum_i \left( 2K_{ii}^{(2)}-K_{ii}^2  \right) + \sum_{ij}L_{ii}L_{jj} K_{ij}^{(2)}\right] \, ,
\end{equation}
where we introduced the matrix
\begin{equation}
    K_{ij}^{(2)} \equiv \int \dd{a} p_a(a) \left( \frac{\hat C}{1-ia\hat L \hat C}  \right)_{ii} \left( \frac{\hat C}{1-ia\hat L \hat C}  \right)_{jj} \, .
    \label{eq:gaussian_fluc_first_term}
\end{equation}
We recognize in the last expression a generalization of the resolvent (see \cref{eq:resolvent}) which encodes higher order correlations. The next terms $\Tr T^k$ with $k>1$ in the series of \cref{eq:Q_n_gaussian} will involve some matrices $K_{ij}^{(k+1)}$ with increasingly higher order correlations, which are nontrivial to compute in general. However, it is straightforward to show that $\Tr T^k = \order{\eta^k}$, so that when $\eta$ is small the series in \cref{eq:Q_n_gaussian} is dominated by its first few terms. To the best of our efforts, it has not been possible to resum the whole series in \cref{eq:Q_n_gaussian}, as it happens instead in the pure GOE case -- see \cref{app:GOE} and \ccite{Metz_2017}.

Specializing to the Cauchy distribution in \cref{eq:lorentzian} with $\mu=0$, one finds for instance that $K_{ii}$ is given in \cref{eq:K_cauchy} in terms of the matrix elements of $\hat C$, while using complex integration in \cref{eq:gaussian_fluc_first_term} yields
\rev{
\begin{equation}
    K_{ij}^{(2)} =
        \dfrac{C_{ii}C_{jj}}{(\omega C_{ii}-1)(\omega C_{jj}-1)} \left[ 1- \dfrac{2\omega C_{ii}C_{jj}}{ C_{ii}+ C_{jj}} \Theta(- L_{ii}\cdot L_{jj} )\right] \, .    
\end{equation}
}

\section{Level compressibility in the pure GOE case}
\label{app:GOE}
In this Appendix we recover the results of \ccite{Metz_2017} concerning the level compressibility for a GOE matrix. This allows us to inspect the similarities and the differences with respect to the GRP case analyzed in this manuscript. In the final section \ref{app_chi_GOE}, we will repeat the derivation of $\chi_\T{GOE}(E)$ using more standard techniques in order to address the low-energy region $E\ll \delta_N$ (see \cref{eq:mean_level_spacing} and \cref{par:low_energy}).

The pure GOE case can be formally obtained from \cref{eq:hamiltonian} by letting the distribution $p_a(a)$ of the diagonal elements of the matrix $A$ tend to a delta function, so that $\psi_a(z) \to 1$.
The calculation then becomes analogous to that reported in \ccite{Metz_2017}, whose main steps we detail here for completeness. By replacing the replica-symmetric Ansatz of \cref{eq:ansatz} into the saddle-point equation \eqref{eq:saddle_point_comp} one first obtains $\cor{N} = \eta / Z_\varphi $, where $Z_\varphi$ is given in \cref{eq:Z_varphi}. By using Gaussian integration one can then show that
\begin{equation}
    \int \dd{\vec{\tau}'} M(\vec{\tau},\vec{\tau}') \varphi_0(\vec{\tau}') = \eta \vec{\tau} \, \hat C \, \vec{\tau} \, ,
    \label{eq:property_GOE}
\end{equation}
and thus the remaining free parameters in \cref{eq:ansatz} can be determined by solving the set of four self-consistency equations which follow from \cref{eq:saddle_point_comp} as
\begin{equation}
    \hat C^{-1} = 2\eta \, \hat C +i \, \hat \Lambda.
    \label{eq:four_set_GOE}
\end{equation}
Note that this can be recovered from \cref{eq:K_def,eq:four_set_p(a)} by using the fact that the resolvent corresponding to $p_a(a)=\delta(a)$ is $\cor{G}_a(z) = 1/z$.

\subsection{Action and fluctuations around the saddle-point}
Both the action and its Gaussian fluctuation matrix $T$ can now be computed in correspondence of the saddle-point solution in \cref{eq:ansatz}. By using the definition of the action given in \cref{eq:action_comp_varphi} together with the saddle-point equation \eqref{eq:saddle_point_comp} and the property in \cref{eq:property_GOE}, one can easily deduce
\begin{align}
    \cor{S}_{n_\pm}[\varphi_0 ; \hat \Lambda]= &\; \frac{\eta}{2}\left[ n_+ \left( \Delta_\alpha^{2} + \bar{\Delta}_\beta^{2}\right) +n_- \left( \Delta_\beta^{2} + \bar{\Delta}_\alpha^{2} \right) \right] + \frac12 (n_+ + n_-) \ln(2\pi) \n\\
    & + \frac12 n_+ \ln(\Delta_\alpha \bar{\Delta}_\beta) + \frac12 n_- \ln(\Delta_\beta \bar{\Delta}_\alpha) \, .
    \label{eq:S_saddlepoint_first}
\end{align}
The computation of $\Tr T^k$ in \cref{eq:Q_n_gaussian} requires more work. First, we rewrite in correspondence of the Ansatz in \cref{eq:ansatz}
\begin{equation}
    T (\vec{\tau}_1,\vec{\tau}_2) = \varphi_0(\vec{\tau}_1) \left[ \left( \vec{\tau}_1 \, \hat L \, \vec{\tau}_2 \right)^2 - \left( \vec{\tau}_2 \, \hat C \, \vec{\tau}_2 \right) \right] \, ,
\end{equation}
where we have used the definition of the function $M(\vec{\tau}_1,\vec{\tau}_2)$ in \cref{eq:M_def} and the property in \cref{eq:property_GOE}. The first few powers of $T$ can then be computed by applying Wick's theorem: by introducing the notation
\begin{equation}
    \expval{\bullet}_1 \equiv \frac{1}{\eta} \int \dd{\vec{\tau}_1} (\bullet) \varphi_0(\vec{\tau}_1) \, ,
\end{equation}
it is sufficient to note that
\begin{equation}
    \expval{\tau_{1i}\tau_{1j}}_1 = C_{ij} \, ,
\end{equation}
so that more complicated averages can be handled as
\begin{equation}
    \expval{\tau_{1i}\tau_{1j}\tau_{1k}\tau_{1l}}_1 = C_{ij} C_{kl} + C_{ik} C_{jl}+C_{il} C_{jk} \, .
\end{equation}
Upon noting that $\hat L^2 = \mathbb{1}_n$, one can then prove by induction the relation
\begin{equation}
    T^{k+1} (\vec{\tau}_1,\vec{\tau}_2) = (2\eta)^k \varphi_0(\vec{\tau}_1) \left[ \left( \vec{\tau}_1 \, \hat L^{k+1}  \, \hat C^k \, \vec{\tau}_2 \right)^2 - \left( \vec{\tau}_2 \, \hat C^{2k+1} \, \vec{\tau}_2 \right) \right] \, .
\end{equation}
Using \cref{eq:trace} now yields
\begin{equation}
    \Tr T^k = 2^{k-1}\eta^k \left[ \left( \sum_i L_{ii}^k C_{ii}^k \right)^2 + \sum_i C_{ii}^{2k} \right] \, ,
\end{equation}
and inserting the definition of the matrices $\hat L$ and $\hat C$ given in \cref{eq:lambda_L_def,eq:hat_C_def} gives
\begin{align}
    \Tr T^k = & 
    \; 
    2^{k-1}\eta^k \Big\lbrace \left[ n_+\left( \Delta_\alpha^{k} + (-\bar{\Delta}_\beta)^{k} \right) + n_- \left( \Delta_\beta^{k} + (-\bar{\Delta}_\alpha)^{k}\right) \right]^2 \n\\
    & + n_+ \left( \Delta_\alpha^{2k} + \bar{\Delta}_\beta^{2k} \right) + n_- \left( \Delta_\beta^{2k} + \bar{\Delta}_\alpha^{2k} \right) \Big\rbrace \, .
    \label{eq:Tk_first}
\end{align}
Taking the limit $n_\pm \to \pm is/\pi $ in \cref{eq:S_saddlepoint_first,eq:Tk_first} then results in
\begin{align}
    \cor{S}_{\pm \frac{is}{\pi}}[\varphi_0 ; \hat \Lambda]=& \frac{is}{2\pi}\Bigg[ \eta \left( \Delta_\alpha^{2} + \bar{\Delta}_\beta^{2} - \Delta_\beta^{2} - \bar{\Delta}_\alpha^{2} \right) + \ln(\frac{\Delta_\alpha \bar{\Delta}_\beta}{\Delta_\beta \bar{\Delta}_\alpha}) \Bigg]\, , \label{eq:S_saddlepoint} \\
    \eval{\Tr T^k}_{\varphi=\varphi_0} \!\!\!\!\!\!\! = & \;
    2^{k-1}\eta^k \Big\lbrace \frac{is}{\pi} \left( \Delta_\alpha^{2k} + \bar{\Delta}_\beta^{2k} - \Delta_\beta^{2k} - \bar{\Delta}_\alpha^{2k} \right) - \frac{s^2}{\pi^2}  \left[ \Delta_\alpha^{k} + (-\bar{\Delta}_\beta)^{k} - \Delta_\beta^{k} - (-\bar{\Delta}_\alpha)^{k} \right]^2 \Big\rbrace \, .
    \label{eq:Tk}
\end{align}
Comparing with the definitions of the cumulant generating function and the cumulants in \cref{eq:cgf,eq:cumulants}, respectively, we can finally identify
\begin{align}
    \frac{\kappa_1}{N}=& -\frac{i}{2\pi}\Bigg[ \eta \left( \Delta_\alpha^{2} + \bar{\Delta}_\beta^{2} - \Delta_\beta^{2} - \bar{\Delta}_\alpha^{2} \right) + \ln(\frac{\Delta_\alpha \bar{\Delta}_\beta}{\Delta_\beta \bar{\Delta}_\alpha}) \Bigg] \n\\
    &- \frac{i}{4\pi N } \sum_{k=1}^\infty \frac{(-2\eta)^k}{k} \left( \Delta_\alpha^{2k} + \bar{\Delta}_\beta^{2k} - \Delta_\beta^{2k} - \bar{\Delta}_\alpha^{2k} \right) \, , \label{eq:k1} \\
    \frac{\kappa_2}{N} =&  - \frac{i}{2\pi^2 N } \sum_{k=1}^\infty \frac{(-2\eta)^k}{k}
    \left[ \Delta_\alpha^{k} + (-\bar{\Delta}_\beta)^{k} - \Delta_\beta^{k} - (-\bar{\Delta}_\alpha)^{k} \right]^2 \, . \label{eq:k2}
\end{align}
In the next Section we will specialize these results to the case in which the interval $[\alpha,\beta]$ is chosen symmetric.

\subsection{Case of a symmetric interval}
We consider here the case in which $\alpha=-E$ and $\beta=E$. With this choice, solving \cref{eq:four_set_GOE} gives\footnote{The angle $\theta$ should be compared with $\theta_L$ in Ref. \cite{Metz_2017}.} 
\begin{equation}
    \eval{\Delta_\alpha}_{a=-E}  = \frac{1}{4\eta} \left[ -(\varepsilon-iE) \pm \sqrt{8\eta+\left(\varepsilon -iE\right)^2 } \right] \equiv r e^{i \theta} \, ,
\end{equation}
where we choose the positive branch of the square root so that $\Re \Delta_\alpha \geq 0$ for any positive $\eta$ (recall that $\Delta_\alpha$ represents the variance of a Gaussian distribution, see \cref{eq:ansatz}). Similarly, from \cref{eq:four_set_GOE} one finds for the entries of $\hat C$ the same symmetries as in \cref{eq:symmetries}. The first two cumulants in \cref{eq:k1,eq:k2} are then found to yield
\begin{align}
    \frac{\kappa_1}{N} &= \frac{x}{\pi} \sin 2\theta +\frac{2\theta}{\pi} +\frac{i}{2\pi N} \ln( \frac{1+x e^{2i\theta}}{1+x e^{-2i\theta}}  )\, , \label{eq:k1_GOE} \\
    \frac{\kappa_2}{N} &= \frac{1}{\pi^2 N} \ln \left[ 1+ \left( \frac{2x \sin2 \theta}{1-x^2} \right)^2 \right]\, , \label{eq:k2_GOE}
\end{align}
where we called $x\equiv 2\eta r^2 $. As a first check, one can easily verify that both $\kappa_1$, $\kappa_2\to 0$ in the limit of a vanishing interval $E \to 0$. 

By choosing $2\eta = 1$, we obtain in the $N\to \infty$ limit an eigenvalue spectrum distributed within the interval $[-2,2]$. Sending $\varepsilon \to 0^+$ as prescribed by \cref{eq:cgf}, one can check that $r \to 1$ (hence $x\to 1$), while
\begin{equation}
    \theta \xrightarrow[\varepsilon \to 0^+]{} \theta_0 \equiv \arctan( \frac{E}{\sqrt{4-E^2}} ) \in \left[-\frac{\pi}{2},\frac{\pi}{2}\right] \, .
\end{equation}
This concludes the calculation of $\kappa_1$ (see \cref{eq:k1_GOE}), which includes both the leading order term and its $\order{N^0}$ correction (note that the latter is actually real-valued). However, the second cumulant $\kappa_2$ is seen to diverge in the limit $\varepsilon\to 0^+$; the problem is addressed in \ccite{Metz_2017} by introducing a $N$-dependent regularization of the infinite sum which appears in \cref{eq:k2}.
Nonetheless, we have shown that such infinite sum (and hence $\kappa_2$ itself in \cref{eq:k2_GOE}) does \textit{not} diverge for any finite value of $\varepsilon$. This hints at the well-known fact that the limit $\varepsilon \to 0^+$ and that for $N\to \infty$ in \cref{eq:cgf} are not interchangeable. It is then useful to expand for small $\varepsilon$
\begin{equation}
    x = r^2\simeq 1-\frac{2\varepsilon}{\sqrt{4-E^2}} \, ,
\end{equation}
so as to rewrite
\begin{equation}
    \kappa_2 \simeq \frac{2}{\pi^2} \ln \left[ \frac{\sqrt{4-E^2} \sin 2 \theta}{2\varepsilon} \right] \, .
\end{equation}
In order to recover the leading order result $\kappa_2 \sim \ln N$ found in previous literature \cite{Dyson_62,Cavagna_2000,Majumdar_2009,Majumdar_2011,Castillo_2014}, one has to assume some type of functional relation $\varepsilon = \varepsilon(N)$, so that the limit $\varepsilon \to 0^+$ is taken by controlling the product $\varepsilon N$ \cite{Cavagna_2000}. This goes however beyond the scope of the present paper.

\subsection{Derivation of the scaling function $\chi_{\rm GOE}(y)$} 
\label{app_chi_GOE}
The prediction for $\chi(E)$ which we have pursued in the previous Section using the replica method is in any case expected to fail if $E$ lies below the mean level spacing, \ie, $E\ll \delta_N$ (as discussed in \cref{par:low_energy}).
In this Appendix, we thus derive in this limit an explicit expression for the level compressibility for a pure $N \times N$ GOE matrix (such as the matrix $B$ in Eq. \eqref{eq:hamiltonian}) using an alternative and more standard technique. We denote by $\rho_B(\lambda)$ (defined as in Eq. \eqref{eq:def_density}) the average density of eigenvalues normalized to unity. In the large $N$ limit, $\rho_B(\lambda)$ has a finite support over $[-\sqrt{2N}, \sqrt{2N}]$ and it is given by the standard Wigner semi-circle. We focus here on the number of eigenvalues $I_N[-E,E]$ in a symmetric interval $[-E,E]$ in the bulk of the spectrum, i.e., we choose $E$ of the order of the mean inter-particle spacing $[N \rho_B(0)]^{-1} = O(N^{-1/2})$.    
The mean number $\langle I_N[-E,E]\rangle$ is easily obtained as
\begin{eqnarray} \label{av_app_GOE}
\langle I_N[-E,E]\rangle = N \int_{-E}^E \rho_B(\lambda) \, \dd \lambda \approx 2 N \rho_B(0) E = 2 \tilde E \, , \quad\qquad \tilde E = N \rho_B(0) E \;.
\end{eqnarray}
The variance of $I_N[-E,E]$ has been studied since the pioneering works of Dyson and Mehta \cite{Dyson_63_IV}. However, an explicit expression for it, valid for any value of $E$ in the bulk, seems hard to find in the literature. This observable was recently revisited in the context of full counting statistics of interacting fermions in Ref. \cite{smith2021full}, which provides a useful starting point, namely (see Eqs.~(31)-(37) therein)
\begin{align} 
    \langle (I_N[-E,E])^2\rangle - \langle I_N[-E,E]\rangle^2  &= - N \rho_B(0)^2 \int_{-E}^E \dd x \left(\int_{-\infty}^{-E} \dd y + \int_{E}^{+ \infty} \dd y \right) C(x,y) \, , \n \\
    C(x,y) &= - N \rho_B(0)^2 Y_{21}(N \rho_B(0)|x-y|) \;,
    \label{var_app_GOE}
\end{align}
where the ``cluster'' function $Y_{21}(r)$ is given by \cite{pandey1979statistical}
\begin{eqnarray} \label{cluster}
Y_{21}(r) = \left(\frac{\sin(\pi r)}{\pi r} \right)^2 - \left(\frac{{\rm Si}(\pi r)}{\pi} - \frac{1}{2} \right)\left(\frac{\pi r \cos(\pi r) - \sin{(\pi r)}}{\pi r^2} \right) \, ,
\end{eqnarray}
with ${\rm Si}(z) = \int_0^z \sin(t)/t\, \dd t$ denoting the sine-integral function. By inserting this explicit expression \eqref{cluster} in \cref{var_app_GOE} and performing explicitly the integrals over $x$ and $y$, one obtains for the level compressibility \begin{eqnarray} \label{com_app_GOE}
\chi(E) = \frac{\langle (I_N[-E,E])^2\rangle - \langle I_N[-E,E]\rangle^2}{\langle I_N[-E,E]\rangle} \sim \chi_{\rm GOE}(y = N \rho_B(0)\,E) \, ,
\end{eqnarray} 
where the function $\chi_{\rm GOE}(y)$ is given by
\begin{align}
\label{chi_GOE_final}
\chi_{\rm GOE}(y) = & \frac{1}{2 \pi^2 y}
 \big\lbrace [\text{Si}(2 \pi  y)]^2 -2\, \text{Ci}(4 \pi  y) -\pi \, \text{Si}(2 \pi  y)\\
 &+2 \left[-4
   \pi  y \, \text{Si}(4 \pi  y)+2 \pi ^2 y+\log (4 \pi  y)-\cos (4 \pi  y)+\gamma_E
   +1\right]\big\rbrace \, , \n
\end{align}
where ${\rm Ci}(z) = - \int_z^\infty \cos(t)/t\, \dd t$ is the cosine integral function and $\gamma_E = 0.577216\ldots$ is the Euler-gamma constant. Its asymptotic behaviors are given by
\begin{eqnarray} \label{chi_GOE_asympt}
\chi_{\rm GOE}(y) =
\begin{dcases}
&1 - 2y + O(y^2) \, , \quad y \to 0 \\
&\frac{\ln y}{\pi^2 y} + O(1/y) \, , \quad y \to \infty \;.
\end{dcases}
\end{eqnarray}

\section{\rev{Scaling function for the Hermitian GRP model}}
\label{app:Kravtsov}
\rev{In this Appendix we consider the case in which the matrix $B$ has complex (rather than real) entries, \ie, it belongs to the GUE ensemble. In this case, powerful analytical tools such as the Harish-Chandra-Itzykson-Zuber
integral are available \cite{Mehta_2004_book}. Following the method introduced in \ccite{Kunz_1998}, the authors of \ccite{Kravtsov_2018} demonstrated that the two-level spectral correlation function assumes a universal form in the fractal regime $1<\gamma<2$, and for large $N$. Our aim here is to link their result to the level compressibility $\chi(E)$, and to show that the latter assumes in the fractal regime the \textit{same} universal form as in the real symmetric case (\ie, in the deformed GOE ensemble studied in this manuscript).}

\rev{Let us begin by defining, as in \ccite{Kunz_1998} (see Eqs.~(2.9) and (3.1) therein),
\begin{align}
    C_1(t) &\equiv \sum_n e^{i t \lambda_n} \, , 
    \label{eq:C1}\\
    C_2(t,t') &\equiv \sum_{n\neq m} e^{i t \lambda_m +it' \lambda_n}\, , \label{eq:C2} \\
    NC(t,t') &\equiv \expval*{C_2(t,t')}-\expval*{C_1(t)}\expval*{C_1(t')} \, ,
    \label{eq:C3}
\end{align}
where $C(t,t')$ is the \textit{spectral form factor}, and the average is intended over the entries of the matrix $\cor{H}$. 
Inserting the identity in the form of $1=\int\dd{\lambda} \delta(\lambda-\lambda_n)$ and using \cref{eq:def_density}, it is simple to derive
\begin{align}
    \expval{C_1(t)} &= N \int \dd{\lambda} e^{i t \lambda } \expval{\rho_N(\lambda)} \, , \\
    \expval{C_2(t,t')} &= N^2 \int \dd{\lambda} \int \dd{\lambda'} e^{i t \lambda +it' \lambda'} \expval{\rho_N(\lambda)\rho_N(\lambda')} - N \int \dd{\lambda} e^{i (t+t') \lambda } \expval{\rho_N(\lambda)}\, .
\end{align}
We now introduce the Fourier transform of the spectral form factor
\begin{align}
    \hat C(\omega,\omega') &\equiv \int \frac{\dd{t}}{2\pi}\int \frac{\dd{t'}}{2\pi} e^{-i\omega t-i\omega' t'} C(t,t') \n \\
    &= N \expval{\rho_N(\omega)\rho_N(\omega')}_c - \expval{\rho_N(\omega)} \delta(\omega -\omega') \, ,
    \label{eq:spectral_ft}
\end{align}
so that using \cref{eq:levels_number} we can express
\begin{equation}
    N \int_{-E}^E \dd{\omega} \int_{-E}^E \dd{\omega'} \hat C(\omega,\omega')  = \expval{I_N^2[-E,E]}_c- \expval{I_N[-E,E]} \, .
    \label{eq:variance_from_spectral_ft}
\end{equation}
Comparing \cref{eq:spectral_ft,eq:variance_from_spectral_ft}, we deduce that the \textit{non-singular} part of $\hat C(\omega,\omega')$ determines the variance of the number of eigenvalues $I_N[-E,E]$ lying within the interval $[-E,E]$.}

\rev{The function $C(t,t')$ was computed in \ccite{Kravtsov_2015} for the Hermitian GRP model with 
$\expval{|\cor{H}_{i\neq j}|^2}=\nu^2/(4N^\gamma)$,
and it reads\footnote{Note that in \ccite{Kravtsov_2015} this function was instead identified with $C_2(t,t')$ given in \cref{eq:C2}. A factor of $p_a(0)$ was also missing.}
\begin{equation}
    C(t,t') = 2\pi p_a(0) \delta(t+t') \left[ S\left( \frac{t-t'}{2} E_\T{Th} \right) -1 \right] \, ,
    \label{eq:C_krav}
\end{equation}
where in the large $N$ limit and for $1<\gamma <2$ the function $S(u)$ assumes the simple form
\begin{equation}
    S(u) = e^{-2\pi \Lambda^2 |u|} \, .
    \label{eq:S(u)}
\end{equation}
We have introduced (as in \cite{Kravtsov_2015}) the quantities
\begin{align}
    E_\T{Th} &\equiv \delta_N N^{2-\gamma} = \frac{2 E_T}{\pi [\nu p_a(0) ]^2} = \frac{E_T}{\pi \Lambda^2} \, , \label{eq:E_th_krav} \\
    \Lambda &\equiv
    \nu p_a(0) /\sqrt{2}\label{eq:lambda_krav} \, ,
\end{align}
where $\delta_N \simeq [N p_a(0)]^{-1}$ is the mean level spacing (see \cref{par:low_energy}), while $E_T \simeq 2\pi p_a(0) \eta$ is the Thouless energy as we introduced it in \cref{par:scaling} (with $\eta$ given in \cref{eq:eta}). 
It follows that
\begin{equation}
    \hat C(\omega,\omega') = \frac{p_a(0)}{E_\T{Th}} \hat S \left(\frac{\omega-\omega'}{E_\T{Th}}\right) - p_a(0)\, \delta (\omega-\omega') \, ,
\end{equation}
where
\begin{equation}
    \hat S(\omega) = \frac{1}{\pi} \frac{2\pi\Lambda^2}{\omega^2 +(2\pi\Lambda^2)^2}
    \label{eq:S(omega)_ft}
\end{equation}
is the Fourier transform of $S(u)$ in \cref{eq:S(u)}. Using \cref{eq:variance_from_spectral_ft}, we thus obtain
\begin{align}
    \expval{I_N^2[-E,E]}_c &= \frac{N p_a(0)}{E_\T{Th}} \int_{-E}^E \dd{\omega} \int_{-E}^E \dd{\omega'} \hat S \left(\frac{\omega-\omega'}{E_\T{Th}}\right) \n\\
    &=  \frac{Np_a(0)}{E_\T{Th}} \int_{-2E}^{2E} \dd{x} (2E-|x|)\; \hat S \left(\frac{x}{E_\T{Th}}\right) \n\\
    &=  2NE p_a(0) \cdot \chi_T\left(\frac{E}{\pi \Lambda^2 E_\T{Th}}\right) = 2NE p_a(0) \cdot \chi_T\left(\frac{E}{E_T}\right) \, ,
    \label{eq:GUE_variance}
\end{align}
where in the second line we changed variables to $x=(\omega-\omega')$, $z=(\omega+\omega'+2E)$, and we integrated out $z$, while in the third line we used \cref{eq:E_th_krav,eq:lambda_krav,eq:S(omega)_ft} and we recognized the scaling function $\chi_T(y)$ given in \cref{eq:comp_universal}.}

\rev{The result in \cref{eq:GUE_variance} should be compared with the one we found in \cref{par:scaling} for the real GRP model: using \cref{eq:k1_scaling,eq:comp_scaling_calculation} yields in fact
\begin{equation}
    \kappa_2(E) = \expval{I_N^2[-E,E]}_c= 2NE p_a(0) \chi_T\left(\frac{E}{E_T}\right) \, .
\end{equation}
Quite interestingly, the same scaling function $\chi_T$ appears both in the deformed GUE and GOE ensembles.}

\end{appendix}

\bibliography{references}

\begin{thebibliography}{100}
\providecommand{\url}[1]{\texttt{#1}}
\providecommand{\urlprefix}{URL }
\expandafter\ifx\csname urlstyle\endcsname\relax
  \providecommand{\doi}[1]{doi:\discretionary{}{}{}#1}\else
  \providecommand{\doi}{doi:\discretionary{}{}{}\begingroup
  \urlstyle{rm}\Url}\fi
\providecommand{\eprint}[2][]{\url{#2}}

\bibitem{anderson1958absence}
P.~W. Anderson,
\newblock \emph{Absence of diffusion in certain random lattices},
\newblock Phys. Rev. \textbf{109}, 1492 (1958),
\newblock \doi{10.1103/PhysRev.109.1492}.

\bibitem{wegner1980inverse}
F.~Wegner,
\newblock \emph{Inverse participation ratio in 2+ $\varepsilon$ dimensions},
\newblock Z. Phys. B Cond. Mat. \textbf{36}(3), 209 (1980),
\newblock \doi{10.1007/BF01325284}.

\bibitem{rodriguez2011multifractal}
A.~Rodriguez, L.~J. Vasquez, K.~Slevin and R.~A. R\"omer,
\newblock \emph{Multifractal finite-size scaling and universality at the
  {Anderson} transition},
\newblock Phys. Rev. B \textbf{84}, 134209 (2011),
\newblock \doi{10.1103/PhysRevB.84.134209}.

\bibitem{srednicki1994chaos}
M.~Srednicki,
\newblock \emph{Chaos and quantum thermalization},
\newblock Phys. Rev. E \textbf{50}, 888 (1994),
\newblock \doi{10.1103/PhysRevE.50.888}.

\bibitem{rigol2008thermalization}
M.~Rigol, V.~Dunjko and M.~Olshanii,
\newblock \emph{Thermalization and its mechanism for generic isolated quantum
  systems},
\newblock Nature \textbf{452}, 854 (2008),
\newblock \doi{10.1038/nature06838}.

\bibitem{mace2019multifractal}
N.~Mac\'e, F.~Alet and N.~Laflorencie,
\newblock \emph{Multifractal scalings across the many-body localization
  transition},
\newblock Phys. Rev. Lett. \textbf{123}, 180601 (2019),
\newblock \doi{10.1103/PhysRevLett.123.180601}.

\bibitem{de2021rare}
G.~De~Tomasi, I.~M. Khaymovich, F.~Pollmann and S.~Warzel,
\newblock \emph{{Rare thermal bubbles at the many-body localization transition
  from the Fock space point of view}},
\newblock Phys. Rev. B \textbf{104}, 024202 (2021),
\newblock \doi{10.1103/PhysRevB.104.024202}.

\bibitem{gornyi2017spectral}
I.~Gornyi, A.~Mirlin, D.~Polyakov and A.~Burin,
\newblock \emph{Spectral diffusion and scaling of many-body delocalization
  transitions},
\newblock Ann. Phys. \textbf{529}, 1600360 (2017),
\newblock \doi{10.1002/andp.201600360}.

\bibitem{tarzia2020many}
M.~Tarzia,
\newblock \emph{{Many-body localization transition in Hilbert space}},
\newblock Phys. Rev. B \textbf{102}, 014208 (2020),
\newblock \doi{10.1103/PhysRevB.102.014208}.

\bibitem{Luitz_2015}
D.~J. Luitz, N.~Laflorencie and F.~Alet,
\newblock \emph{{Many-body localization edge in the random-field Heisenberg
  chain}},
\newblock Phys. Rev. B \textbf{91}, 081103 (2015),
\newblock \doi{10.1103/PhysRevB.91.081103}.

\bibitem{Serbyn_2017}
M.~Serbyn, Z.~Papi\ifmmode~\acute{c}\else \'{c}\fi{} and D.~A. Abanin,
\newblock \emph{Thouless energy and multifractality across the many-body
  localization transition},
\newblock Phys. Rev. B \textbf{96}, 104201 (2017),
\newblock \doi{10.1103/PhysRevB.96.104201}.

\bibitem{Tikhonov_2018}
K.~S. Tikhonov and A.~D. Mirlin,
\newblock \emph{Many-body localization transition with power-law interactions:
  Statistics of eigenstates},
\newblock Phys. Rev. B \textbf{97}, 214205 (2018),
\newblock \doi{10.1103/PhysRevB.97.214205}.

\bibitem{Luitz_2020}
D.~J. Luitz, I.~M. Khaymovich and Y.~B. Lev,
\newblock \emph{{Multifractality and its role in anomalous transport in the
  disordered XXZ spin-chain}},
\newblock SciPost Phys. Core \textbf{2}, 006 (2020),
\newblock \doi{10.21468/SciPostPhysCore.2.2.006}.

\bibitem{basko2006metal}
D.~M. Basko, I.~L. Aleiner and B.~L. Altshuler,
\newblock \emph{Metal--insulator transition in a weakly interacting
  many-electron system with localized single-particle states},
\newblock Ann. Phys. \textbf{321}, 1126 (2006),
\newblock \doi{10.1016/j.aop.2005.11.014}.

\bibitem{altshuler1997quasiparticle}
B.~L. Altshuler, Y.~Gefen, A.~Kamenev and L.~S. Levitov,
\newblock \emph{Quasiparticle lifetime in a finite system: A nonperturbative
  approach},
\newblock Phys. Rev. Lett. \textbf{78}, 2803 (1997),
\newblock \doi{10.1103/PhysRevLett.78.2803}.

\bibitem{Faoro_2019}
L.~Faoro, M.~V. Feigel’man and L.~Ioffe,
\newblock \emph{Non-ergodic extended phase of the quantum random energy model},
\newblock Ann. Phys. \textbf{409}, 167916 (2019),
\newblock \doi{10.1016/j.aop.2019.167916}.

\bibitem{baldwin2018quantum}
C.~L. Baldwin and C.~R. Laumann,
\newblock \emph{Quantum algorithm for energy matching in hard optimization
  problems},
\newblock Phys. Rev. B \textbf{97}, 224201 (2018),
\newblock \doi{10.1103/PhysRevB.97.224201}.

\bibitem{biroli2021out}
G.~Biroli, D.~Facoetti, M.~Schir\'o, M.~Tarzia and P.~Vivo,
\newblock \emph{Out-of-equilibrium phase diagram of the quantum random energy
  model},
\newblock Phys. Rev. B \textbf{103}, 014204 (2021),
\newblock \doi{10.1103/PhysRevB.103.014204}.

\bibitem{parolini2020multifractal}
T.~Parolini and G.~Mossi,
\newblock \emph{Multifractal dynamics of the {QREM}},
\newblock \doi{10.48550/arxiv.2007.00315} (2020).

\bibitem{kechedzhi2018efficient}
K.~Kechedzhi, V.~Smelyanskiy, J.~R. McClean, V.~S. Denchev, M.~Mohseni,
  S.~Isakov, S.~Boixo, B.~Altshuler and H.~Neven,
\newblock \emph{Efficient population transfer via non-ergodic extended states
  in quantum spin glass},
\newblock \doi{10.48550/arxiv.1807.04792} (2018).

\bibitem{pino2017multifractal}
M.~Pino, V.~E. Kravtsov, B.~L. Altshuler and L.~B. Ioffe,
\newblock \emph{{Multifractal metal in a disordered Josephson junctions
  array}},
\newblock Phys. Rev. B \textbf{96}, 214205 (2017),
\newblock \doi{10.1103/PhysRevB.96.214205}.

\bibitem{pino2016nonergodic}
M.~Pino, L.~B. Ioffe and B.~L. Altshuler,
\newblock \emph{{Nonergodic metallic and insulating phases of Josephson
  junction chains}},
\newblock Proc. Natl. Acad. Sci. USA \textbf{113}, 536 (2016),
\newblock \doi{10.1073/pnas.1520033113}.

\bibitem{Smelyanskiy_2020}
V.~N. Smelyanskiy, K.~Kechedzhi, S.~Boixo, S.~V. Isakov, H.~Neven and
  B.~Altshuler,
\newblock \emph{Nonergodic delocalized states for efficient population transfer
  within a narrow band of the energy landscape},
\newblock Phys. Rev. X \textbf{10}, 011017 (2020),
\newblock \doi{10.1103/PhysRevX.10.011017}.

\bibitem{Atland_2019}
T.~Micklitz, F.~Monteiro and A.~Altland,
\newblock \emph{Nonergodic extended states in the {Sachdev-Ye-Kitaev} model},
\newblock Phys. Rev. Lett. \textbf{123}, 125701 (2019),
\newblock \doi{10.1103/PhysRevLett.123.125701}.

\bibitem{monteiro2021minimal}
F.~Monteiro, T.~Micklitz, M.~Tezuka and A.~Altland,
\newblock \emph{Minimal model of many-body localization},
\newblock Phys. Rev. Research \textbf{3}, 013023 (2021),
\newblock \doi{10.1103/PhysRevResearch.3.013023}.

\bibitem{RP_1960}
N.~Rosenzweig and C.~E. Porter,
\newblock \emph{Repulsion of energy levels in complex atomic spectra},
\newblock Phys. Rev. \textbf{120}, 1698 (1960),
\newblock \doi{10.1103/PhysRev.120.1698}.

\bibitem{Kravtsov_2015}
V.~E. Kravtsov, I.~M. Khaymovich, E.~Cuevas and M.~Amini,
\newblock \emph{A random matrix model with localization and ergodic
  transitions},
\newblock New J. Phys. \textbf{17}, 122002 (2015),
\newblock \doi{10.1088/1367-2630/17/12/122002}.

\bibitem{vonSoosten_2019}
P.~von Soosten and S.~Warzel,
\newblock \emph{{Non-ergodic delocalization in the Rosenzweig--Porter model}},
\newblock Lett. Math. Phys. \textbf{109}, 905 (2019),
\newblock \doi{10.1007/s11005-018-1131-7}.

\bibitem{Facoetti_2016}
D.~Facoetti, P.~Vivo and G.~Biroli,
\newblock \emph{From non-ergodic eigenvectors to local resolvent statistics and
  back: A random matrix perspective},
\newblock Europhys. Lett. \textbf{115}, 47003 (2016),
\newblock \doi{10.1209/0295-5075/115/47003}.

\bibitem{Truong_2016}
K.~Truong and A.~Ossipov,
\newblock \emph{Eigenvectors under a generic perturbation: Non-perturbative
  results from the random matrix approach},
\newblock Europhys. Lett. \textbf{116}, 37002 (2016),
\newblock \doi{10.1209/0295-5075/116/37002}.

\bibitem{Bogomolny_2018}
E.~Bogomolny and M.~Sieber,
\newblock \emph{{Eigenfunction distribution for the Rosenzweig-Porter model}},
\newblock Phys. Rev. E \textbf{98}, 032139 (2018),
\newblock \doi{10.1103/PhysRevE.98.032139}.

\bibitem{DeTomasi_2019}
G.~D. Tomasi, M.~Amini, S.~Bera, I.~M. Khaymovich and V.~E. Kravtsov,
\newblock \emph{{Survival probability in Generalized Rosenzweig-Porter random
  matrix ensemble}},
\newblock SciPost Phys. \textbf{6}, 14 (2019),
\newblock \doi{10.21468/SciPostPhys.6.1.014}.

\bibitem{amini2017spread}
M.~Amini,
\newblock \emph{Spread of wave packets in disordered hierarchical lattices},
\newblock Europhy. Lett. \textbf{117}, 30003 (2017),
\newblock \doi{10.1209/0295-5075/117/30003}.

\bibitem{pino2019ergodic}
M.~Pino, J.~Tabanera and P.~Serna,
\newblock \emph{{From ergodic to non-ergodic chaos in Rosenzweig-Porter
  model}},
\newblock J. Phys. A - Mat. Theor. \textbf{52}, 475101 (2019),
\newblock \doi{10.1088/1751-8121/ab4b76}.

\bibitem{berkovits2020super}
R.~Berkovits,
\newblock \emph{{Super-Poissonian behavior of the Rosenzweig-Porter model in
  the nonergodic extended regime}},
\newblock Phys. Rev. B \textbf{102}, 165140 (2020),
\newblock \doi{10.1103/PhysRevB.102.165140}.

\bibitem{Pandey_1995}
A.~Pandey,
\newblock \emph{Brownian-motion model of discrete spectra},
\newblock Chaos Soliton. Frac. \textbf{5}, 1275 (1995),
\newblock \doi{10.1016/0960-0779(94)E0065-W}.

\bibitem{Guhr_1996}
T.~Guhr,
\newblock \emph{Transition from {Poisson} regularity to chaos in a
  time-reversal noninvariant system},
\newblock Phys. Rev. Lett. \textbf{76}, 2258 (1996),
\newblock \doi{10.1103/PhysRevLett.76.2258}.

\bibitem{Guhr_1997}
T.~Guhr and A.~Müller-Groeling,
\newblock \emph{{Spectral correlations in the crossover between GUE and Poisson
  regularity: on the identification of scales}},
\newblock J. Math. Phys. \textbf{38}, 1870 (1997),
\newblock \doi{10.1063/1.531918}.

\bibitem{Brezin_1996}
E.~Br\'ezin and S.~Hikami,
\newblock \emph{{Correlations of nearby levels induced by a random potential}},
\newblock Nucl. Phys. B \textbf{479}, 697 (1996),
\newblock \doi{10.1016/0550-3213(96)00394-X}.

\bibitem{Kunz_1998}
H.~Kunz and B.~Shapiro,
\newblock \emph{{Transition from Poisson to Gaussian unitary statistics: the
  two-point correlation function}},
\newblock Phys. Rev. E \textbf{58}, 400 (1998),
\newblock \doi{10.1103/PhysRevE.58.400}.

\bibitem{Atland_1997}
A.~Altland, M.~Janssen and B.~Shapiro,
\newblock \emph{{Perturbation theory for the Rosenzweig-Porter matrix model}},
\newblock Phys. Rev. E \textbf{56}, 1471 (1997),
\newblock \doi{10.1103/PhysRevE.56.1471}.

\bibitem{Krajenbrink_2021}
A.~Krajenbrink, P.~Le~Doussal and N.~O'Connell,
\newblock \emph{{Tilted elastic lines with columnar and point disorder,
  non-Hermitian quantum mechanics, and spiked random matrices: pinning and
  localization}},
\newblock Phys. Rev. E \textbf{103}, 042120 (2021),
\newblock \doi{10.1103/PhysRevE.103.042120}.

\bibitem{Mergny_2021}
P.~Mergny and S.~N. Majumdar,
\newblock \emph{Stability of large complex systems with heterogeneous
  relaxation dynamics},
\newblock J. Stat. Mech.: Theor. Exp. \textbf{2021}, 123301 (2021),
\newblock \doi{10.1088/1742-5468/ac3b47}.

\bibitem{Bouchaud_Les_Houches}
J.-P. Bouchaud,
\newblock \emph{{Random matrix theory and (big) data analysis}},
\newblock In \emph{{Stochastic Processes and Random Matrices: Lecture Notes of
  the Les Houches Summer School: Volume 104, July 2015}}. Oxford University
  Press,
\newblock ISBN 9780198797319,
\newblock \doi{10.1093/oso/9780198797319.003.0006} (2017).

\bibitem{Mezard_1987}
M.~M{\'e}zard, G.~Parisi and M.~Virasoro,
\newblock \emph{Spin Glass Theory and Beyond},
\newblock Lecture Notes in Physics Series. World Scientific,
\newblock ISBN 9789971501150 (1987).

\bibitem{Harukuni_2022}
H.~Ikeda,
\newblock \emph{{Bose{\textendash}Einstein-like condensation of deformed random
  matrix: a replica approach}},
\newblock J. Stat. Mech.: Theor. Exp. \textbf{2023}(2), 023302 (2023),
\newblock \doi{10.1088/1742-5468/acb7d6}.

\bibitem{Claeys_2018}
T.~Claeys, A.~B.~J. Kuijlaars, K.~Liechty and D.~Wang,
\newblock \emph{{Propagation of singular behavior for Gaussian perturbations of
  random matrices}},
\newblock Commun. Math. Phys. \textbf{362}, 1 (2018),
\newblock \doi{10.1007/s00220-018-3195-8}.

\bibitem{Livan_2018}
G.~Livan, M.~Novaes and P.~Vivo,
\newblock \emph{Introduction to Random Matrices},
\newblock Springer International Publishing,
\newblock \doi{10.1007/978-3-319-70885-0} (2018).

\bibitem{altshuler1986repulsion}
B.~Altshuler and B.~Shklovskii,
\newblock \emph{Repulsion of energy levels and conductivity of small metal
  samples},
\newblock Sov. Phys. JETP \textbf{64}, 127 (1986),
\newblock \eprint{http://www.jetp.ras.ru/cgi-bin/dn/e_064_01_0127.pdf}.

\bibitem{cuevas2007two}
E.~Cuevas and V.~E. Kravtsov,
\newblock \emph{Two-eigenfunction correlation in a multifractal metal and
  insulator},
\newblock Phys. Rev. B \textbf{76}, 235119 (2007),
\newblock \doi{10.1103/PhysRevB.76.235119}.

\bibitem{kravtsov2020localization}
V.~E. Kravtsov, I.~M. Khaymovich, B.~L. Altshuler and L.~B. Ioffe,
\newblock \emph{{Localization transition on the Random Regular Graph as an
  unstable tricritical point in a log-normal Rosenzweig-Porter random matrix
  ensemble}},
\newblock \doi{10.48550/arxiv.2002.02979} (2020).

\bibitem{khaymovich2020fragile}
I.~M. Khaymovich, V.~E. Kravtsov, B.~L. Altshuler and L.~B. Ioffe,
\newblock \emph{{Fragile extended phases in the log-normal Rosenzweig-Porter
  model}},
\newblock Phys. Rev. Res. \textbf{2}, 043346 (2020),
\newblock \doi{10.1103/PhysRevResearch.2.043346}.

\bibitem{monthus2017multifractality}
C.~Monthus,
\newblock \emph{{Multifractality of eigenstates in the delocalized non-ergodic
  phase of some random matrix models: Wigner--Weisskopf approach}},
\newblock J. Phys. A - Mat. Theor. \textbf{50}, 295101 (2017),
\newblock \doi{10.1088/1751-8121/aa77e1}.

\bibitem{biroli2021levy}
G.~Biroli and M.~Tarzia,
\newblock \emph{{L\'evy-Rosenzweig-Porter random matrix ensemble}},
\newblock Phys. Rev. B \textbf{103}, 104205 (2021),
\newblock \doi{10.1103/PhysRevB.103.104205}.

\bibitem{buijsman2022circular}
W.~Buijsman and Y.~Bar~Lev,
\newblock \emph{{Circular Rosenzweig-Porter random matrix ensemble}},
\newblock SciPost Phys. \textbf{12}, 082 (2022),
\newblock \doi{10.21468/SciPostPhys.12.3.082}.

\bibitem{khaymovich2021dynamical}
I.~M. Khaymovich and V.~E. Kravtsov,
\newblock \emph{{Dynamical phases in a ``multifractal'' Rosenzweig-Porter
  model}},
\newblock SciPost Phys. \textbf{11}, 045 (2021),
\newblock \doi{10.21468/SciPostPhys.11.2.045}.

\bibitem{sarkar2021mobility}
M.~Sarkar, R.~Ghosh, A.~Sen and K.~Sengupta,
\newblock \emph{{Mobility edge and multifractality in a periodically driven
  Aubry-Andr\'e model}},
\newblock Phys. Rev. B \textbf{103}, 184309 (2021),
\newblock \doi{10.1103/PhysRevB.103.184309}.

\bibitem{roy2018multifractality}
S.~Roy, I.~M. Khaymovich, A.~Das and R.~Moessner,
\newblock \emph{{Multifractality without fine-tuning in a Floquet quasiperiodic
  chain}},
\newblock SciPost Phys. \textbf{4}, 025 (2018),
\newblock \doi{10.21468/SciPostPhys.4.5.025}.

\bibitem{wang2016phase}
J.~Wang, X.-J. Liu, G.~Xianlong and H.~Hu,
\newblock \emph{{Phase diagram of a non-Abelian Aubry-Andr\'e-Harper model with
  $p$-wave superfluidity}},
\newblock Phys. Rev. B \textbf{93}, 104504 (2016),
\newblock \doi{10.1103/PhysRevB.93.104504}.

\bibitem{nosov2019correlation}
P.~A. Nosov, I.~M. Khaymovich and V.~E. Kravtsov,
\newblock \emph{Correlation-induced localization},
\newblock Phys. Rev. B \textbf{99}, 104203 (2019),
\newblock \doi{10.1103/PhysRevB.99.104203}.

\bibitem{duthie2022anomalous}
A.~Duthie, S.~Roy and D.~E. Logan,
\newblock \emph{Anomalous multifractality in quantum chains with strongly
  correlated disorder},
\newblock Phys. Rev. B \textbf{106}, L020201 (2022),
\newblock \doi{10.1103/PhysRevB.106.L020201}.

\bibitem{kutlin2021emergent}
A.~G. Kutlin and I.~M. Khaymovich,
\newblock \emph{{Emergent fractal phase in energy stratified random models}},
\newblock SciPost Phys. \textbf{11}, 101 (2021),
\newblock \doi{10.21468/SciPostPhys.11.6.101}.

\bibitem{motamarri2021localization}
V.~R. Motamarri, A.~S. Gorsky and I.~M. Khaymovich,
\newblock \emph{{Localization and fractality in disordered Russian Doll
  model}},
\newblock SciPost Phys. \textbf{13}, 117 (2022),
\newblock \doi{10.21468/SciPostPhys.13.5.117}.

\bibitem{tang2022non}
W.~Tang and I.~M. Khaymovich,
\newblock \emph{{Non-ergodic delocalized phase with Poisson level statistics}},
\newblock Quantum \textbf{6}, 733 (2022),
\newblock \doi{10.22331/q-2022-06-09-733}.

\bibitem{tarzia2022fully}
M.~Tarzia,
\newblock \emph{{Fully localized and partially delocalized states in the tails
  of Erd\"os-R\'enyi graphs in the critical regime}},
\newblock Phys. Rev. B \textbf{105}, 174201 (2022),
\newblock \doi{10.1103/PhysRevB.105.174201}.

\bibitem{Skvortsov_2022}
M.~A. Skvortsov, M.~Amini and V.~E. Kravtsov,
\newblock \emph{{Sensitivity of (multi)fractal eigenstates to a perturbation of
  the Hamiltonian}},
\newblock Phys. Rev. B \textbf{106}, 054208 (2022),
\newblock \doi{10.1103/PhysRevB.106.054208}.

\bibitem{Cai_2013}
X.~Cai, L.-J. Lang, S.~Chen and Y.~Wang,
\newblock \emph{Topological superconductor to {Anderson} localization
  transition in one-dimensional incommensurate lattices},
\newblock Phys. Rev. Lett. \textbf{110}, 176403 (2013),
\newblock \doi{10.1103/PhysRevLett.110.176403}.

\bibitem{DeGottardi_2013}
W.~DeGottardi, D.~Sen and S.~Vishveshwara,
\newblock \emph{Majorana fermions in superconducting 1d systems having
  periodic, quasiperiodic, and disordered potentials},
\newblock Phys. Rev. Lett. \textbf{110}, 146404 (2013),
\newblock \doi{10.1103/PhysRevLett.110.146404}.

\bibitem{Liu_2015}
F.~Liu, S.~Ghosh and Y.~D. Chong,
\newblock \emph{Localization and adiabatic pumping in a generalized
  {Aubry-Andr\'e-Harper} model},
\newblock Phys. Rev. B \textbf{91}, 014108 (2015),
\newblock \doi{10.1103/PhysRevB.91.014108}.

\bibitem{Das_2022}
A.~K. Das and A.~Ghosh,
\newblock \emph{Nonergodic extended states in the $\ensuremath{\beta}$
  ensemble},
\newblock Phys. Rev. E \textbf{105}, 054121 (2022),
\newblock \doi{10.1103/PhysRevE.105.054121}.

\bibitem{Ahmed_2022}
A.~Ahmed, A.~Ramachandran, I.~M. Khaymovich and A.~Sharma,
\newblock \emph{Flat band based multifractality in the all-band-flat diamond
  chain},
\newblock Phys. Rev. B \textbf{106}, 205119 (2022),
\newblock \doi{10.1103/PhysRevB.106.205119}.

\bibitem{Lee_2022}
S.~Lee, A.~Andreanov and S.~Flach,
\newblock \emph{Critical-to-insulator transitions and fractality edges in
  perturbed flatbands},
\newblock \doi{10.48550/arxiv.2208.11930} (2022).

\bibitem{voiculescu1992free}
D.~V. Voiculescu, K.~J. Dykema and A.~Nica,
\newblock \emph{Free random variables},
\newblock American Mathematical Soc. (1992).

\bibitem{Zee_1996}
A.~Zee,
\newblock \emph{Law of addition in random matrix theory},
\newblock Nucl. Phys. B \textbf{474}, 726 (1996),
\newblock \doi{10.1016/0550-3213(96)00276-3}.

\bibitem{fano1947ionization}
U.~Fano,
\newblock \emph{Ionization yield of radiations. {II}. {The} fluctuations of the
  number of ions},
\newblock Phys. Rev. \textbf{72}, 26 (1947),
\newblock \doi{10.1103/PhysRev.72.26}.

\bibitem{majumdar2019exactly}
S.~N. Majumdar, P.~von Bomhard and J.~Krug,
\newblock \emph{Exactly solvable record model for rainfall},
\newblock Phys. Rev. Lett. \textbf{122}, 158702 (2019),
\newblock \doi{10.1103/PhysRevLett.122.158702}.

\bibitem{majumdar2021universal}
S.~N. Majumdar, P.~Mounaix and G.~Schehr,
\newblock \emph{{Universal record statistics for random walks and L{\'e}vy
  flights with a nonzero staying probability}},
\newblock J. Phys. A - Math. Theor. \textbf{54}, 315002 (2021),
\newblock \doi{10.1088/1751-8121/ac0a2f}.

\bibitem{Mirlin_2000}
A.~D. Mirlin,
\newblock \emph{Statistics of energy levels and eigenfunctions in disordered
  systems},
\newblock Phys. Rep. \textbf{326}, 259 (2000),
\newblock \doi{10.1016/S0370-1573(99)00091-5}.

\bibitem{Edwards_1976}
S.~F. Edwards and R.~C. Jones,
\newblock \emph{The eigenvalue spectrum of a large symmetric random matrix},
\newblock J. Phys. A - Math. Gen. \textbf{9}, 1595 (1976),
\newblock \doi{10.1088/0305-4470/9/10/011}.

\bibitem{Kravtsov_2018}
V.~Kravtsov, B.~Altshuler and L.~Ioffe,
\newblock \emph{{Non-ergodic delocalized phase in Anderson model on Bethe
  lattice and regular graph}},
\newblock Ann. Phys. \textbf{389}, 148 (2018),
\newblock \doi{https://doi.org/10.1016/j.aop.2017.12.009}.

\bibitem{parisi_statistical_1988}
G.~Parisi,
\newblock \emph{Statistical {Field} {Theory}},
\newblock Addison-Wesley (1988).

\bibitem{Metz_2016}
F.~L. Metz and I.~{P\'erez Castillo},
\newblock \emph{Large deviation function for the number of eigenvalues of
  sparse random graphs inside an interval},
\newblock Phys. Rev. Lett. \textbf{117}, 104101 (2016),
\newblock \doi{10.1103/PhysRevLett.117.104101}.

\bibitem{Metz_2017}
F.~L. Metz,
\newblock \emph{{Replica-symmetric approach to the typical eigenvalue
  fluctuations of Gaussian random matrices}},
\newblock J. Phys. A - Math. Theor. \textbf{50}, 495002 (2017),
\newblock \doi{10.1088/1751-8121/aa94f8}.

\bibitem{Vivo_2020}
P.~Vivo,
\newblock \emph{{Index of a matrix, complex logarithms, and multidimensional
  Fresnel integrals}},
\newblock J. Phys. A - Math. Theor. \textbf{54}(2), 025002 (2020),
\newblock \doi{10.1088/1751-8121/abccf9}.

\bibitem{Castillo_2018}
I.~P. Castillo and F.~L. Metz,
\newblock \emph{{Large-deviation theory for diluted Wishart random matrices}},
\newblock Phys. Rev. E \textbf{97}, 032124 (2018),
\newblock \doi{10.1103/PhysRevE.97.032124}.

\bibitem{stack}
D.~H. (https://math.stackexchange.com/users/55051/david h),
\newblock \emph{{Asymptotic expansion of integral involving an ArcTan}},
\newblock Mathematics Stack Exchange,
\newblock \eprint{https://math.stackexchange.com/q/4523130}.

\bibitem{table}
I.~S. Gradshteyn and I.~M. Ryzhik,
\newblock \emph{Table of integrals, series, and products},
\newblock Elsevier/Academic Press, Amsterdam, 7th edn.,
\newblock ISBN 978-0-12-373637-6 (2007).

\bibitem{Mehta_2004_book}
M.~L. Mehta,
\newblock \emph{Random Matrices},
\newblock Academic Press, New York, 3rd edn. (2004).

\bibitem{Dyson_1962_I}
F.~J. Dyson,
\newblock \emph{{Statistical Theory of the Energy Levels of Complex Systems.
  I}},
\newblock J. Math. Phys. \textbf{3}, 140 (1962),
\newblock \doi{10.1063/1.1703773}.

\bibitem{Dyson_63_IV}
F.~J. Dyson and M.~L. Mehta,
\newblock \emph{{Statistical Theory of the Energy Levels of Complex Systems.
  IV}},
\newblock J. Math. Phys. \textbf{4}, 701 (1963),
\newblock \doi{10.1063/1.1704008}.

\bibitem{Forrester_2010_book}
P.~J. Forrester,
\newblock \emph{Log-Gases and Random Matrices (LMS-34)},
\newblock London Mathematical Society Monographs. Princeton University Press,
\newblock ISBN 9781400835416 (2010).

\bibitem{marino2014phase}
R.~Marino, S.~N. Majumdar, G.~Schehr and P.~Vivo,
\newblock \emph{Phase transitions and edge scaling of number variance in
  {Gaussian} random matrices},
\newblock Phys. Rev. Lett. \textbf{112}, 254101 (2014),
\newblock \doi{10.1103/PhysRevLett.112.254101}.

\bibitem{Marino_2016}
R.~Marino, S.~N. Majumdar, G.~Schehr and P.~Vivo,
\newblock \emph{Number statistics for $\ensuremath{\beta}$-ensembles of random
  matrices: Applications to trapped fermions at zero temperature},
\newblock Phys. Rev. E \textbf{94}, 032115 (2016),
\newblock \doi{10.1103/PhysRevE.94.032115}.

\bibitem{Tikhonov_2019}
K.~S. Tikhonov and A.~D. Mirlin,
\newblock \emph{Statistics of eigenstates near the localization transition on
  random regular graphs},
\newblock Phys. Rev. B \textbf{99}, 024202 (2019),
\newblock \doi{10.1103/PhysRevB.99.024202}.

\bibitem{arous2011wigner}
G.~B. Arous and A.~Guionnet,
\newblock \emph{Wigner matrices},
\newblock The Oxford Handbook of Random Matrix Theory pp. 433--451 (2011).

\bibitem{potters2020first}
M.~Potters and J.-P. Bouchaud,
\newblock \emph{A First Course in Random Matrix Theory: For Physicists,
  Engineers and Data Scientists},
\newblock Cambridge University Press (2020).

\bibitem{Semerjian_2002}
G.~Semerjian and L.~F. Cugliandolo,
\newblock \emph{Sparse random matrices: the eigenvalue spectrum revisited},
\newblock J. Phys. A - Math. Gen. \textbf{35}, 4837 (2002),
\newblock \doi{10.1088/0305-4470/35/23/303}.

\bibitem{Rogers_2008}
T.~Rogers, I.~{P\'erez Castillo}, R.~K\"uhn and K.~Takeda,
\newblock \emph{Cavity approach to the spectral density of sparse symmetric
  random matrices},
\newblock Phys. Rev. E \textbf{78}, 031116 (2008),
\newblock \doi{10.1103/PhysRevE.78.031116}.

\bibitem{Metz_2014}
F.~L. Metz, G.~Parisi and L.~Leuzzi,
\newblock \emph{Finite-size corrections to the spectrum of regular random
  graphs: An analytical solution},
\newblock Phys. Rev. E \textbf{90}, 052109 (2014),
\newblock \doi{10.1103/PhysRevE.90.052109}.

\bibitem{DeTomasi_2022}
G.~De~Tomasi and I.~M. Khaymovich,
\newblock \emph{{Non-Hermitian Rosenzweig-Porter random-matrix ensemble:
  Obstruction to the fractal phase}},
\newblock Phys. Rev. B \textbf{106}, 094204 (2022),
\newblock \doi{10.1103/PhysRevB.106.094204}.

\bibitem{Kutlin_2023}
A.~G. Kutlin and I.~M. Khaymovich,
\newblock \emph{{to appear}} (2023).

\bibitem{Voiculescu_1991}
D.~Voiculescu,
\newblock \emph{Limit laws for random matrices and free products},
\newblock Invent. Math. \textbf{104}(1), 201 (1991),
\newblock \doi{10.1007/BF01245072}.

\bibitem{Biane_97}
P.~Biane,
\newblock \emph{On the free convolution with a semi-circular distribution},
\newblock Indiana U. Math. J. \textbf{46}, 705 (1997),
\newblock \eprint{https://hal.archives-ouvertes.fr/hal-00536164}.

\bibitem{Dyson_62}
F.~J. Dyson,
\newblock \emph{Statistical theory of the energy levels of complex systems.
  {III}},
\newblock J. Math. Phys. \textbf{3}, 166 (1962),
\newblock \doi{10.1063/1.1703775}.

\bibitem{Cavagna_2000}
A.~Cavagna, J.~P. Garrahan and I.~Giardina,
\newblock \emph{Index distribution of random matrices with an application to
  disordered systems},
\newblock Phys. Rev. B \textbf{61}, 3960 (2000),
\newblock \doi{10.1103/PhysRevB.61.3960}.

\bibitem{Majumdar_2009}
S.~N. Majumdar, C.~Nadal, A.~Scardicchio and P.~Vivo,
\newblock \emph{{Index distribution of Gaussian random matrices}},
\newblock Phys. Rev. Lett. \textbf{103}, 220603 (2009),
\newblock \doi{10.1103/PhysRevLett.103.220603}.

\bibitem{Majumdar_2011}
S.~N. Majumdar, C.~Nadal, A.~Scardicchio and P.~Vivo,
\newblock \emph{{How many eigenvalues of a Gaussian random matrix are
  positive?}},
\newblock Phys. Rev. E \textbf{83}, 041105 (2011),
\newblock \doi{10.1103/PhysRevE.83.041105}.

\bibitem{Castillo_2014}
I.~{P\'erez Castillo},
\newblock \emph{{Spectral order statistics of Gaussian random matrices: Large
  deviations for trapped fermions and associated phase transitions}},
\newblock Phys. Rev. E \textbf{90}, 040102 (2014),
\newblock \doi{10.1103/PhysRevE.90.040102}.

\bibitem{smith2021full}
N.~R. Smith, P.~Le~Doussal, S.~N. Majumdar and G.~Schehr,
\newblock \emph{Full counting statistics for interacting trapped fermions},
\newblock SciPost Phys. \textbf{11}, 110 (2021),
\newblock \doi{10.21468/SciPostPhys.11.6.110}.

\bibitem{pandey1979statistical}
A.~Pandey,
\newblock \emph{{Statistical properties of many-particle spectra: III. Ergodic
  behavior in random-matrix ensembles}},
\newblock Ann. Phys. \textbf{119}, 170 (1979),
\newblock \doi{10.1016/0003-4916(79)90254-9}.

\end{thebibliography}

\nolinenumbers
\end{document}